\documentclass[10pt]{article} %normal 10pt, other options [11pt],[12pt] before {article}
\usepackage{indentfirst,amsmath,latexsym,amssymb,amsbsy,amscd}  %use in arxiv
\newcommand{\be}{\begin{equation}}
\newcommand{\ee}{\end{equation}}
\newcommand{\ra}{\rightarrow}
\newcommand{\la}{\leftarrow}
\newcommand{\ua}{\uparrow}
\newcommand{\da}{\downarrow}
\newcommand{\lra}{\leftrightarrow}
\newcommand{\uda}{\updownarrow}

\newcommand{\qed}{\hfill $\bullet$}
\newcommand{\cast}{\circledast}
\newcommand{\ccirc}{\circledcirc}

\newcommand{\shrtpll}{\shortparallel}

\newtheorem{prop}{Proposition}
\newtheorem{thm}[prop]{Theorem}

\newtheorem{lem}[prop]{Lemma}
\newtheorem{cor}[prop]{Corollary}

\newenvironment{prf}{\trivlist \item[\hskip \labelsep{\bf Proof.}]}{\qed \endtrivlist}

\begin{document}

%HAVE NOT USED OMEGA OR THETA IN CAP VERSION
%have not used \lambda, just used \Lambda
%have not use \Phi 

\newcommand{\je}[1]{j={#1},\ldots,\ell}
\newcommand{\ifof}{if and only if}
\newcommand{\bone}{\mathbf{1}}
\newcommand{\bzero}{\mathbf{0}}

\newcommand{\bsigma}{{\boldsymbol{\sigma}}}

%$\mathsf{G}, {\mathsf{G}}^t, \mathsf{g}$
\newcommand{\msfcpg}{{\mathsf{G}}}
\newcommand{\msfg}{{\mathsf{g}}}
\newcommand{\msfgdt}{{{\dot{\mathsf{g}}}}}
\newcommand{\msfgddt}{{{\ddot{\mathsf{g}}}}}

\newcommand{\bmcpz}{{\mathbf{Z}}}
\newcommand{\bmcpx}{{\mathbf{X}}}
\newcommand{\bmcpy}{{\mathbf{Y}}}
\newcommand{\mbbc}{{\mathbb{C}}}
\newcommand{\bmcpr}{{\mathbf{R}}}
\newcommand{\bmcpu}{{\mathbf{U}}}
\newcommand{\bmcpi}{{\mathbf{I}}}

\newcommand{\bmd}{{\mathbf{d}}}
\newcommand{\bmdbr}{{{\bar{\bmd}}}}
\newcommand{\bmb}{{\mathbf{b}}}
\newcommand{\bma}{{\mathbf{a}}}

\newcommand{\bmy}{{\mathbf{y}}}
\newcommand{\bmm}{{\mathbf{m}}}
\newcommand{\bmz}{{\mathbf{z}}}
\newcommand{\bms}{{\mathbf{s}}}
\newcommand{\bmt}{{\mathbf{t}}}
\newcommand{\bmcpw}{{\mathbf{W}}}

%%%%%%%%%%%%%%%%%
\newcommand{\bmf}{{\mathbf{f}}}
%%%%%%%%%%%%%%%%%%%

\newcommand{\bmc}{{\mathbf{c}}}
\newcommand{\bmu}{{\mathbf{u}}}
\newcommand{\bmr}{{\mathbf{r}}}
\newcommand{\bmg}{{\mathbf{g}}}
\newcommand{\bmi}{{\mathbf{i}}}
\newcommand{\bme}{{\mathbf{e}}}
\newcommand{\bmh}{{\mathbf{h}}}
\newcommand{\bmk}{{\mathbf{k}}}
\newcommand{\bmp}{{\mathbf{p}}}
\newcommand{\bmn}{{\mathbf{n}}}

\newcommand{\bmw}{{\mathbf{w}}}
\newcommand{\bml}{{\mathbf{l}}}
\newcommand{\bmx}{{\mathbf{x}}}
\newcommand{\bmcpq}{{\mathbf{Q}}}
\newcommand{\bmq}{{\mathbf{q}}}
\newcommand{\bmv}{{\mathbf{v}}}
\newcommand{\bmcpv}{{\mathbf{V}}}

\newcommand{\bmbht}{{\hat{\bmb}}}
\newcommand{\bmcht}{{\hat{\bmc}}}
\newcommand{\bmsht}{{\hat{\bms}}}
\newcommand{\bmrht}{{\hat{\bmr}}}
\newcommand{\bmght}{{\hat{\bmg}}}
\newcommand{\bmhht}{{\hat{\bmh}}}
\newcommand{\bmeht}{{\hat{\bme}}}

\newcommand{\bmubv}{{\breve{\bmu}}}
\newcommand{\bmuinv}{{\bmu^{-1}}}

\newcommand{\bmcck}{{\check{\bmc}}}
\newcommand{\bmuck}{{\check{\bmu}}}
\newcommand{\bmrck}{{\check{\bmr}}}
\newcommand{\bmgck}{{\check{\bmg}}}

\newcommand{\bmabr}{{\bar{\bma}}}
\newcommand{\bmbbr}{{\bar{\bmb}}}
\newcommand{\bmcbr}{{\bar{\bmc}}}
\newcommand{\bmubr}{{\bar{\bmu}}}
\newcommand{\bmrbr}{{\bar{\bmr}}}
\newcommand{\bmsbr}{{\bar{\bms}}}
\newcommand{\bmgbr}{{\bar{\bmg}}}
\newcommand{\bmhbr}{{\bar{\bmh}}}
\newcommand{\bmebr}{{\bar{\bme}}}
\newcommand{\bmvbr}{{\bar{\bmv}}}
\newcommand{\bmwbr}{{\bar{\bmw}}}

\newcommand{\bmbddt}{{\ddot{\bmb}}}
\newcommand{\bmaddt}{{\ddot{\bma}}}
\newcommand{\bmcddt}{{\ddot{\bmc}}}
\newcommand{\bmuddt}{{\ddot{\bmu}}}
\newcommand{\bmvddt}{{\ddot{\bmv}}}
\newcommand{\bmwddt}{{\ddot{\bmw}}}
\newcommand{\bmrddt}{{\ddot{\bmr}}}
\newcommand{\bmsddt}{{\ddot{\bms}}}
\newcommand{\bmgddt}{{\ddot{\bmg}}}
\newcommand{\bmeddt}{{\ddot{\bme}}}

\newcommand{\bmbdt}{{\dot{\bmb}}}
\newcommand{\bmadt}{{\dot{\bma}}}
\newcommand{\bmcdt}{{\dot{\bmc}}}
\newcommand{\bmudt}{{\dot{\bmu}}}
\newcommand{\bmvdt}{{\dot{\bmv}}}
\newcommand{\bmwdt}{{\dot{\bmw}}}
\newcommand{\bmrdt}{{\dot{\bmr}}}
\newcommand{\bmsdt}{{\dot{\bms}}}
\newcommand{\bmgdt}{{\dot{\bmg}}}
\newcommand{\bmxdt}{{\dot{\bmx}}}
\newcommand{\bmydt}{{\dot{\bmy}}}
\newcommand{\bmedt}{{\dot{\bme}}}

\newcommand{\bmaac}{{\acute{\bma}}}
\newcommand{\bmbac}{{\acute{\bmb}}}
\newcommand{\bmcac}{{\acute{\bmc}}}
\newcommand{\bmsac}{{\acute{\bms}}}
\newcommand{\bmuac}{{\acute{\bmu}}}
\newcommand{\bmvac}{{\acute{\bmv}}}
\newcommand{\bmrac}{{\acute{\bmr}}}
\newcommand{\bmgac}{{\acute{\bmg}}}

\newcommand{\bmagr}{{\grave{\bma}}}
\newcommand{\bmbgr}{{\grave{\bmb}}}
\newcommand{\bmcgr}{{\grave{\bmc}}}
\newcommand{\bmsgr}{{\grave{\bms}}}
\newcommand{\bmugr}{{\grave{\bmu}}}
\newcommand{\bmvgr}{{\grave{\bmv}}}
\newcommand{\bmrgr}{{\grave{\bmr}}}
\newcommand{\bmggr}{{\grave{\bmg}}}

\newcommand{\aht}{{\hat{a}}}
\newcommand{\aac}{{\acute{a}}}
\newcommand{\agr}{{\grave{a}}}
\newcommand{\addt}{{\ddot{a}}}
\newcommand{\adt}{{\dot{a}}}
\newcommand{\abr}{{\bar{a}}}

\newcommand{\bht}{{\hat{b}}}
\newcommand{\bac}{{\acute{b}}}
\newcommand{\bgr}{{\grave{b}}}
\newcommand{\bddt}{{\ddot{b}}}
\newcommand{\bdt}{{\dot{b}}}
\newcommand{\bbr}{{\bar{b}}}

\newcommand{\cht}{{\hat{c}}}
\newcommand{\cac}{{\acute{c}}}
\newcommand{\cgr}{{\grave{c}}}
\newcommand{\cddt}{{\ddot{c}}}
\newcommand{\cdt}{{\dot{c}}}
\newcommand{\cbr}{{\bar{c}}}

\newcommand{\rht}{{\hat{r}}}
\newcommand{\rac}{{\acute{r}}}
\newcommand{\rgr}{{\grave{r}}}
\newcommand{\rddt}{{\ddot{r}}}
\newcommand{\rdt}{{\dot{r}}}
\newcommand{\rck}{{\check{r}}}
\newcommand{\rbr}{{\bar{r}}}

\newcommand{\vbr}{{\bar{v}}}
\newcommand{\ebr}{{\bar{e}}}

\newcommand{\fbr}{{\bar{f}}}
\newcommand{\gdt}{{\dot{g}}}
\newcommand{\ght}{{\hat{g}}}
\newcommand{\gddt}{{\ddot{g}}}
\newcommand{\gbr}{{\bar{g}}}

\newcommand{\xdt}{{\dot{x}}}
\newcommand{\ydt}{{\dot{y}}}
\newcommand{\fdt}{{\dot{f}}}
\newcommand{\kdt}{{\dot{k}}}
\newcommand{\kddt}{{\ddot{k}}}

\newcommand{\odotht}{{\hat{\odot}}}
\newcommand{\cdotht}{{\hat{\cdot}}}
\newcommand{\bmuht}{{\hat{\bmu}}}
\newcommand{\bmudtht}{{\hat{\bmudt}}}
\newcommand{\bmuddtht}{{\hat{\bmuddt}}}
\newcommand{\bmvht}{{\hat{\bmv}}}
\newcommand{\bmvdtht}{{\hat{\bmvdt}}}
\newcommand{\bmvddtht}{{\hat{\bmvddt}}}
\newcommand{\caleht}{{\hat{\cale}}}
\newcommand{\calvht}{{\hat{\calv}}}
\newcommand{\callht}{{\hat{\call}}}

\newcommand{\bmghat}{{\hat{\mathbf{g}}}}
\newcommand{\bmgbar}{{\bar{\mathbf{g}}}}
\newcommand{\bmehat}{{\hat{\mathbf{e}}}}

\newcommand{\ssr}{Schreier series}
\newcommand{\fssr}{forward Schreier series}
\newcommand{\bssr}{backward Schreier series}
\newcommand{\gm}{generator matrix}
\newcommand{\gms}{generator matrices}
\newcommand{\atmx}{alphabet matrix}
\newcommand{\atmxs}{alphabet matrices}
\newcommand{\sgen}{shift generator}
\newcommand{\sgens}{shift generators}
\newcommand{\glab}{generator label}
\newcommand{\glabs}{generator labels}
\newcommand{\mchn}{matrix chain}

\newcommand{\ellctl}{$\ell$-controllable}
\newcommand{\ellpctl}{$\ell'$-controllable}
\newcommand{\cdc}{coset decomposition chain}
\newcommand{\creps}{coset representatives}
\newcommand{\crep}{coset representative}
\newcommand{\crepc}{coset representative chain}
\newcommand{\compset}{complete set of coset representatives}
\newcommand{\fss}{full symmetry system}
\newcommand{\consub}{consistent subsystem}
\newcommand{\nss}{natural symmetry system}
\newcommand{\nsss}{natural symmetry systems}
\newcommand{\scgs}{strongly controllable group system}
\newcommand{\sctigs}{strongly controllable time invariant group system}
\newcommand{\sctigss}{strongly controllable time invariant group systems}
\newcommand{\sctigc}{strongly controllable time invariant group code}
\newcommand{\fhgs}{first homomorphism theorem for group systems}

\newcommand{\xj}{{\{X_j\}}}
\newcommand{\yi}{{\{Y_i\}}}
\newcommand{\xjt}{{\{X_j^t\}}}
\newcommand{\yit}{{\{Y_i^t\}}}
\newcommand{\yj}{{\{Y_j\}}}
\newcommand{\yk}{{\{Y_k\}}}
\newcommand{\bi}{{\{B^{(i)}\}}}

\newcommand{\calf}{{\mathcal{F}}}
\newcommand{\calb}{{\mathcal{B}}}
\newcommand{\calh}{{\mathcal{H}}}
\newcommand{\call}{{\mathcal{L}}}
\newcommand{\calk}{{\mathcal{K}}}
\newcommand{\calp}{{\mathcal{P}}}
\newcommand{\calt}{{\mathcal{T}}}

\newcommand{\calq}{{\mathcal{Q}}}
\newcommand{\calr}{{\mathcal{R}}}
\newcommand{\calu}{{\mathcal{U}}}
\newcommand{\cale}{{\mathcal{E}}}
\newcommand{\cals}{{\mathcal{S}}}
\newcommand{\calv}{{\mathcal{V}}}
\newcommand{\calw}{{\mathcal{W}}}
\newcommand{\cali}{{\mathcal{I}}}

\newcommand{\mfrkcpu}{{\mathfrak{U}}}
\newcommand{\mfrkcpr}{{\mathfrak{R}}}
\newcommand{\mfrkr}{{\mathfrak{r}}}
\newcommand{\mfrkrdt}{{\dot{\mfrkr}}}
\newcommand{\mfrkrddt}{{\ddot{\mfrkr}}}
\newcommand{\mfrku}{{\mathfrak{u}}}
\newcommand{\mfrkudt}{{\dot{\mfrku}}}
\newcommand{\mfrkuddt}{{\ddot{\mfrku}}}
\newcommand{\mfrkubr}{{\bar{\mfrku}}}

\newcommand{\cpaht}{{\hat{A}}}
\newcommand{\cpvht}{{\hat{V}}}
\newcommand{\cpuht}{{\hat{U}}}
\newcommand{\cprht}{{\hat{R}}}
\newcommand{\calbht}{{\hat{\calb}}}
\newcommand{\caluht}{{\hat{\calu}}}
\newcommand{\circht}{{\hat{\circ}}}
\newcommand{\ccircht}{{\hat{\ccirc}}}
\newcommand{\starht}{{\hat{\star}}}

\newcommand{\rmdef}{\stackrel{\rm def}{=}}
\newcommand{\imp}{{\rm im}\,p}
\newcommand{\imtheta}{{\rm im}\,\theta}
\newcommand{\imfr}{{\rm im}\,f_r}
\newcommand{\imf}{{\rm im}\,f}
\newcommand{\imfu}{{\rm im}\,f_u}
\newcommand{\imhu}{{\rm im}\,h_u}
\newcommand{\imhv}{{\rm im}\,h_v}
\newcommand{\plus}{{+}}
\newcommand{\dimnd}{{\diamond}}
\newcommand{\dggr}{{\dagger}}
\newcommand{\ddggr}{{\ddagger}}
\newcommand{\vtri}{{\vartriangle}}

%%%%%%%%%%%%%%%%%%%%%%%%%%%%%%%%%%%%%%%%%%
\newcommand{\fracfjk}[2]
{{
\frac{\calf^{#1}(\Delta_{#2}^t)}{\calf^{#1}(\Delta_{{#2}-1}^t)}
}}

\newcommand{\ssl}{{/ \!\! /}} %\sslash

\newcommand{\argu}{{\,\cdot\,}}

\newcommand{\ovast}{{\overline{*}}}
\newcommand{\ovtimes}{{\overline{\times}}}
\newcommand{\ovcirc}{{\overline{\circ}}}
\newcommand{\ovstar}{{\overline{\star}}}
\newcommand{\ovcdot}{{\overline{\cdot}}}
\newcommand{\ovphi}{{\overline{\phi}}}
\newcommand{\ovtheta}{{\overline{\theta}}}
\newcommand{\ovomega}{{\overline{\omega}}}
\newcommand{\ovalpha}{{\overline{\alpha}}}

\newcommand{\eid}{{{\stackrel{e}{=}}}}

%MOSTLY USED IN SECTION 6 AND FOLLOWING%sss
\newcommand{\triupjk}[2]{{\triangledown_{#1,#2}}}
\newcommand{\triupjkt}[3]{{\triangledown_{#1,#2}^{#3}}}
\newcommand{\triupjktarg}[4]{{\triupjkt{#1}{#2}{#3}(#4)}}
\newcommand{\prodjktarg}[4]{{{#4}_{#1,#2}^{#3}}}
\newcommand{\grpupjktarg}[5]{{(\triupjktarg{#1}{#2}{#3}{#4},\prodjktarg{#1}{#2}{#3}{#5})}}

\newcommand{\triupjktargnop}[4]{{\triupjkt{#1}{#2}{#3}\,{#4}}}
\newcommand{\grpupjktargnop}[4]{{(\triupjktargnop{#1}{#2}{#3}{#4},\circ)}}

\newcommand{\proditarg}[3]{{{#3}_{#1}^{#2}}}

\newcommand{\boxkt}[2]{{\Box_{#1}^{#2}}}
\newcommand{\boxktarg}[3]{{\boxkt{#1}{#2}(#3)}}
\newcommand{\boxktargnop}[3]{{\boxkt{#1}{#2}\,{#3}}}
%\newcommand{\boxktexpargnop}[4]{{\boxktargnop{#1^{#4}}{#2^{#4}}{#3}}}
%\newcommand{\grpboxktargnop}[4]{{(\boxktargnop{#1}{#2}{#3},\proditarg{#1}{#2}{#4})}}

%\capkcalu{}=\triupjktexpargnop{0}{\bmk}{\bmt}{\calk}{}
\newcommand{\capkcalu}[1]{{K_{\bmk^{#1}}^{\bmt^{#1}}(\calu)}}
\newcommand{\capkcalux}{{K_\bmk^\bmt(\calu)}}

\newcommand{\trilwjk}[2]{{\vartriangle_{#1,#2}}}
\newcommand{\trilwjkt}[3]{{\vartriangle_{#1,#2}^{#3}}}
\newcommand{\trilwjktarg}[4]{{\trilwjkt{#1}{#2}{#3}(#4)}}
\newcommand{\grplwjktarg}[5]{{(\trilwjktarg{#1}{#2}{#3}{#4},\prodjktarg{#1}{#2}{#3}{#5})}}

%lower elementary groups
\newcommand{\trilwjktargnop}[4]{{\trilwjkt{#1}{#2}{#3}\,{#4}}}
\newcommand{\grplwjktargnop}[4]{{(\trilwjktargnop{#1}{#2}{#3}{#4},\circ)}}

\newcommand{\rhdjkt}[3]{{\rhd_{#1,#2}^{#3}}}
\newcommand{\rhdjktarg}[4]{{\rhdjkt{#1}{#2}{#3}(#4)}}
\newcommand{\rhdgrpjktarg}[5]{{(\rhdjktarg{#1}{#2}{#3}{#4},\prodjktarg{#1}{#2}{#3}{#5})}}

%\blacktriangle "lower"
%\blacktriangledown "upper"
\newcommand{\blktridn}{{\blacktriangledown}}
\newcommand{\blktrilwktargnop}[3]{{\blacktriangle_{#1}^{#2}\,{#3}}}
\newcommand{\blktriupktarg}[3]{{\blacktriangledown_{#1}^{#2}(#3)}}
\newcommand{\blktrilwktarg}[3]{{\blacktriangle_{#1}^{#2}(#3)}}
\newcommand{\blktrilwktexpargnop}[4]{{\blacktriangle_{#1^{#4}}^{#2^{#4}}\,{#3}}}
\newcommand{\grpblktriupktarg}[4]{{(\blktriupktarg{#1}{#2}{#3},\proditarg{#1}{#2}{#4})}}

%%%%%%%%%%%%%%%%%%%%%%%%%%%%%%%%%%%%%%%%%%

\title{Any strongly controllable group system or group shift or any linear block code
is isomorphic to a generator group}

\author{Kenneth M. Mackenthun Jr. (email:  {\tt ken1212576@gmail.com})}
\maketitle

%can also use \begin{abstract},\end{abstract}
\vspace{3mm}
{\bf ABSTRACT}
\vspace{3mm}

Consider any sequence of finite groups $A^t$, where $t$ takes values in an integer index 
set $\mathbf{Z}$.  A group system $A$ is a set of sequences with components in
$A^t$ that forms a group under componentwise addition in $A^t$, for each $t\in\mathbf{Z}$.
As shown previously, any strongly controllable complete group system $A$ can be
decomposed into generators.  We study permutations of the generators when sequences
in the group system are multiplied.
We show that any strongly controllable complete group system $A$ is isomorphic to
a generator group $({\mathcal{U}},\circ)$.
The set ${\mathcal{U}}$ is a set of tensors, a double 
Cartesian product space of sets $G_k^t$, with indices $k$, for $0\le k\le\ell$, and time $t$, for
$t\in\mathbf{Z}$.  $G_k^t$ is a set of unique generator labels for the
generators in $A$ with nontrivial span for the time interval $[t,t+k]$.  We show
the generator group contains a unique elementary system, 
an infinite collection of elementary groups, one for 
each $k$ and $t$, defined on small subsets of ${\mathcal{U}}$, in the shape of triangles,
which form a tile like structure over ${\mathcal{U}}$.  There is a homomorphism from 
each elementary group to any elementary group defined on smaller tiles of the former group.
The group system $A$ may be constructed from either the generator group or elementary system.
These results have application to linear block codes, any algebraic system
that contains a linear block code, group shifts, and harmonic theory in mathematics, and
systems theory, coding theory, control theory, and related fields in engineering.

\newpage

%page style
\topmargin      =-0.25in
\evensidemargin =0.10in %1inch less than margin
\oddsidemargin  =-0.90in
\textheight     = 9.5in
\headheight     = 0.0in
\headsep        = 0.0in

\newpage
\vspace{3mm}
{\bf 1.  INTRODUCTION}
\vspace{3mm}

Linear systems theory is a well established field in engineering.
Willems reenvisaged linear system theory by starting with a set of sequences
which forms a group, called here a group system, and derived many properties of the
system from this set \cite{JW}.  Following the work of Willems, Forney and Trott \cite{FT} 
described the state group and state code of a group system, which they called a group code.
They showed that any group code that can be characterized by its local behavior (complete \cite{FT})
can be wholely specified by a sequence of connected labeled group trellis sections, 
a trellis, which may vary in time.
They explained the important idea of ``shortest length code sequences''
or generators.  In a controllable group code, a generator is a code sequence which
is nontrivial over a finite time interval $[t,t+k]$, for some time $t\in\bmcpz$ and
some finite integer $k$, and which is not a combination of shorter code sequences.
In a strongly controllable group code, all the generators have a finite length
which is uniformly bounded by a finite integer $\ell$.  Any sequence in the code
can be found by a composition of generators.  At any time $t$, any letter in a
code sequence is obtained by a combination of generators wich have nontrivial
letters at time $t$.
Loeliger and Mittelholzer \cite{LM} obtain a complementary approach to the work
of Forney and Trott by starting with a trellis having group properties instead 
of a set of sequences.

A time invariant group system is essentially a group shift \cite{LMr}.  Therefore the results
here have application to group shifts.
The study of group shifts originated with the work of Kitchens \cite{KIT} in symbolic
dynamics theory, and preceded the work of Willems.  Kitchens \cite{KIT} 
showed that any group shift has finite memory, i.e., it is a shift of finite type \cite{LMr}.

We consider the general time varying group system.  A general time varying group system 
can have nontrivial values over a finite time interval, in which case it is often called
a block code over a group \cite{FT} or linear block code \cite{MS}.
Therefore the results here also apply to block codes.

In the setting of group systems, a natural definition of a linear system is a homomorphism from a group
of inputs to the group system, an output.
Both Willems and Forney and Trott have taken the approach to systems theory that
all properties of the system can be derived starting from the sequences in the system themselves.
Willems has shown for linear systems that this approach gives a notion of an input sequence.
Similarly Forney and Trott have shown that there is a notion of input for a group system,
the sequence of first components of the generators.  However as discussed in \cite{FT},
they do not show there is a homomorphic map from the 
input sequence space to the output sequence space.  Therefore, they do not show
a group system is a linear system.   Likewise, as also discussed in \cite{FT}, the work of
Brockett and Willsky \cite{BW} on ``finite group homomorphic sequential systems''
does not suffice to realize general linear systems over groups.
One result of this paper is to show that for a strongly controllable 
complete group system, the generators themselves 
form a group, the generator group, and there is an isomorphism from the generator
group, an input group, to the group system,
the output group.  Therefore a group system is a linear system.  This 
fills a gap in the theory of Willems and Forney and Trott.
Moreover, since there is an isomorphism from the generator group to the
group system, the generator group can be used to study the group system.
The generator group is very revealing of the structure of group systems
and seems to give new insights into related studies
such as linear systems, group codes, group shifts, linear block codes, and
any algebraic structure which contains a linear block code.

Consider any sequence of finite groups $A^t$, where $t$ takes values in an integer index 
set $\mathbf{Z}$.  A group system $A$ is a set of sequences with components in $A^t$,
for each $t\in\bmcpz$, that forms a group under componentwise addition in $A^t$ \cite{FT}.
We only consider strongly controllable group systems $A$, in which there is
a fixed least integer $\ell$ such that for any time $t$, for any pair of sequences
$\bma,\bmadt\in A$, there is a sequence $\bmaddt\in A$ that agrees with $\bma$ on
$(-\infty,t]$ and agrees with $\bmadt$ on $[t+\ell,\infty)$ \cite{FT}.
We only study group systems $A$ that are complete \cite{FT}.  These are group systems which can
be completely characterized by local behavior, i.e., no global constraints are
needed.

As shown in \cite{FT}, any strongly controllable complete group system $A$ can be
decomposed into generators.  A generator $\bmg^{[t,t+k]}$ is a sequence in $A$ which 
is the identity except for a nontrivial span on the time interval $[t,t+k]$ \cite{FT}. 
A generator is a \crep\ in a \cdc\ of $A$.  We can form any sequence $\bma\in A$ by 
selecting one \crep\ from each coset in the \cdc, for each time $t$, $t\in\bmcpz$,
for $0\le k\le\ell$.  

The list of generators for sequence $\bma$ forms a tensor $\bmr$.  The
set of tensors $\bmr$ formed by $\bma\in A$ is $\calr$.  There is a
1-1 correspondence between sequences $\bma\in A$ and tensors $\bmr\in\calr$
We study permutations of the generators in $\bmr$ when sequences $\bma$
in the group system $A$ are multiplied.  
If $\bmadt,\bmaddt$ are two sequences in $A$, then their product $\bmadt\bmaddt$ 
is another sequence in $A$.  Let $\bmrdt\ra\bmadt$ and $\bmrddt\ra\bmaddt$ under the
bijection $\calr\ra A$.  We define an operation $*$ on $\calr$ by 
$\bmrdt*\bmrddt\rmdef\bmrbr$ if $\bmrbr\mapsto\bmadt\bmaddt$ under the
bijection $\calr\ra A$.  The set of tensors $\calr$ with operation $*$ forms a 
group $(\calr,*)$ called a decomposition group of $A$, and $A\simeq (\calr,*)$.

An element $\bmr\in\calr$ is equivalent to a sequence of generators $\bmg^{[t,t+k]}$ for 
$0\le k\le\ell$, for each $t\in\bmcpz$.  We replace each
generator $\bmg^{[t,t+k]}$ in $\bmr\in\calr$ with a \glab\ $g_k^t$.
Under the assignment $\bmg^{[t,t+k]}\mapsto g_k^t$,
for $0\le k\le\ell$, for each $t\in\bmcpz$, a tensor $\bmr\in\calr$
becomes a tensor $\bmu$, $\bmr\mapsto\bmu$.  Let $\calu$ be the set
of tensors $\bmu$ obtained from $\calr$ this way.
The set ${\mathcal{U}}$ is a double 
Cartesian product space of sets $G_k^t$, with indices $k$, for $0\le k\le\ell$, and time $t$, for
$t\in\mathbf{Z}$, where $G_k^t$ is a set of unique generator labels for the
generators in $A$ with nontrivial span for the time interval $[t,t+k]$.  
The operation $*$ in $(\calr,*)$ determines an operation $\circ$ on
$\calu$, and this gives a group $(\calu,\circ)$ isomorphic to $(\calr,*)$, called a 
generator group; we have $A\simeq (\calr,*)\simeq (\calu,\circ)$.
Therefore any strongly controllable complete group system $A$ is isomorphic to
a generator group $({\mathcal{U}},\circ)$.

We regard the \cdc\ of $A$ used in \cite{FT} as a spectral domain decomposition of $A$.
We use a different \cdc\ of $A$ which we think of as a time domain decomposition.
We show the time and spectral domain decompositions of $A$ can be more easily obtained
as the \crepc s of two different \cdc s of the generator group of $A$.  We show the generator
group has many other \cdc s which give other decompositions.  This gives a
harmonic theory of group systems.

We show the multiplication of two sequences in $A$ 
can be broken into local groups in $(\calu,\circ)$.
We show there is an infinite collection of local elementary groups $\grpupjktarg{0}{k}{t}{\calu}{\ccirc}$
of $(\calu,\circ)$, one for each $k$ such that $0\le k\le\ell$, and each $t\in\bmcpz$,
defined on small subsets of $\calu$, in the shape of triangles,
which form a nested tile like structure over $\calu$.  There is a homomorphism from 
each elementary group to any elementary group nested in the former group.
We define the infinite collection of elementary groups, together with a homomorphism
condition, to be an $(\ell+1)$-depth elementary system $\cale_A$.  
The homomorphism condition is that for each $k$ such that $0\le k<\ell$, for each
$t\in\bmcpz$, there is a homomorphism from 
elementary group $\grpupjktarg{0}{k}{t}{\calu}{\ccirc}$ to
the next two largest elementary groups nested in $\grpupjktarg{0}{k}{t}{\calu}{\ccirc}$. 
Then we have shown that the generator group of any \ellctl\ complete group system $A$ contains
an elementary system $\cale_A$.

Given an $(\ell+1)$-depth elementary system $\cale_A$ of a group system $A$, 
we can always recover the $(\ell+1)$-depth generator group $(\calu,\circ)$ of $A$,
up to an isomorphic and essentially identical group,
using a generalization of the first homomorphism theorem to group systems.
Therefore any \ellctl\ complete group system $A$ 
can be constructed from either the generator group $(\calu,\circ)$ of $A$ or
the $(\ell+1)$-depth elementary system $\cale_A$ of $A$.
Moreover, starting from any constructed $(\ell+1)$-depth elementary system $\cale$, 
the \fhgs\ can always construct an \ellctl\ complete group system.
Therefore the study and construction of \ellctl\ complete group systems
is also essentially the study and construction of elementary systems.
Since the $(\ell+1)$-depth elementary system has finite depth and is nested by depth, 
the construction of any $(\ell+1)$-depth elementary system is very simple.

In this paper, we study group systems as a permutation of the generators, 
without any explicit use of the concept of state as in conventional systems theory
and group codes \cite{FT} or of the future cover as in symbolic dynamics and group
shifts \cite{LMr}.  Nevertheless, we can replicate many of the results in these fields.
For example, the local elementary group $\grpupjktarg{0}{0}{t}{\calu}{\ccirc}$
is an analog of the branch group in \cite{FT} and future cover in \cite{LMr}.

In Section 2, we review some of the definitions in group systems and previous work.
In Section 3, we find a normal chain of $A$, the generators of $A$ for this normal chain,
the set $\calr$, and a formula to calculate a component $a^t$ of $A$ using a matrix in
$\calr$.  In Section 4, we find the decomposition group $(\calr,*)$, the generator
group $(\calu,\circ)$, the largest elementary group of $(\calu,\circ)$, and the 
component group of $(\calr,*)$.  We also show how to recover $A$ from the
generator group using a \fhgs.  In Section 5, we find the nested elementary groups
of $(\calu,\circ)$ and their homomorphism relation.  We show that finite or infinite
sets of elementary groups of $(\calu,\circ)$ form a group.  Then we find normal chains
of the generator group.  In Section 6, we discuss the elementary system, the global
group of the elementary system, how to find all \ellctl\ complete group systems $A$
from the elementary system, and how to construct elementary systems.

\newpage
\vspace{3mm}
{\bf 2.  GROUP SYSTEMS}
\vspace{3mm}

This section gives a very brief review of some fundamental concepts in 
\cite{FT}, and introduces some definitions used here.
We follow the notation of Forney and Trott \cite{FT} as closely as possible.
One significant difference is that subscript $k$ in \cite{FT} denotes
time; we use $t$ (an integer) in place of $k$.  Further, time is always 
indicated with a superscript.  We always
associate notation $\bone$ with the identity of any group or group of sequences; 
for example $g_\bone$ will be the identity of group $G$.

Forney and Trott study a collection of sequences with time axis defined on the
set of integers $\bmcpz$, whose components $a^t$ are taken from 
an {\it alphabet group} or {\it alphabet} $A^t$ at each time $t$, $t\in\bmcpz$.  
The set of sequences is a group under componentwise addition in $A^t$ \cite{FT}.  
We call this a {\it group system} $A$.
A sequence $\bma$ in $A$ is given by
\be
\label{c9}
\bma=\ldots,a^{t-1},a^t,a^{t+1},\ldots,
\ee
where $a^t\in A^t$ is the component at time $t$.
We assume that all elements of $A^t$ are represented in $A$ at time $t$.
The identity of $A^t$ is $a_\bone^t$, and the identity of $A$ is $\bma_\bone$.

We apply the standard definition of group isomorphism for finite groups to group systems.
Let $G$ be a group or group system.  Then $A\simeq G$ \ifof\ there is a bijection
$A\ra G$ such that if $\bmadt,\bmaddt\in A$ and $\bmadt\mapsto\gdt$,
$\bmaddt\mapsto\gddt$, then $\bmadt\bmaddt\mapsto\gdt\gddt$.  There may be
other definitions of isomorphism for group systems that are more suitable \cite{KM6} v11-v12.

In this paper, we study {\it complete} group systems  \cite{JW,FT}.  A complete
group system can be characterized by its local behavior; in particular complete
systems can be generated by their trellis diagrams \cite{FT}.
Forney and Trott construct their canonical encoder for a complete strongly 
controllable group system.  Completeness is called closure in symbolic 
dynamics \cite{FT}.  Therefore a time invariant complete group system 
$A$ is the same thing as a group shift in symbolic dynamics \cite{FT}.  
For an incomplete group system, a global constraint is 
required to fully specify the group system.  Some examples of group systems 
that require a global constraint are given in \cite{FT}.

As in \cite{FT}, we use conventional notation for time intervals.
If $m\le n$, the time interval $[m,n]$ starts at time $m$, ends at time $n$, and
has {\it length} length $n-m+1$.  We also write time interval $[m,n]$ as
$[m,n+1)$.  The time interval $[m,m]$ or $[m,m+1)$ has length 1 and is also written just $m$.

Let $A$ be a group system, and let $\bma$ be a sequence in $A$.
Using (\ref{c9}), define the projection map at time $t$, 
$\chi^t:  A\ra A^t$, by the assignment $\bma\mapsto a^t$.
Define the projection map $\chi^{[t_1,t_2]}:  A\ra A^{t_1}\times\cdots\times A^{t_2}$ 
by the assignment $\bma\mapsto (a^{t_1},\ldots,a^{t_2})$.  In general,
we say that sequence $\bma$ has {\it span} $t_2-t_1+1$ if $\bma$ is the same as the identity
sequence except for a finite segment $(a^{t_1},\ldots,a^{t_2})$ of length $t_2-t_1+1$,
where $a^{t_1}\ne a_\bone^{t_1}$ and $a^{t_2}\ne a_\bone^{t_2}$.  
We define $A^{[t_1,t_2]}$ to be the sequences in $A$ which are the identity outside time 
interval $[t_1,t_2]$.

A group system $A$ is {\it $[m,n)$-controllable} if for any
$\bmadt,\bmaddt\in A$, there exists a sequence $\bma\in A$ with
$\chi^{(-\infty,m)}(\bma)=\chi^{(-\infty,m)}(\bmadt)$ and 
$\chi^{[n,+\infty)}(\bma)=\chi^{[n,+\infty)}(\bmaddt)$.  
Then the finite segment $\chi^{[m,n)}(\bma)$ of length $n-m$ in $\bma$ 
connects the past $\chi^{(-\infty,m)}(\bmadt)$ of $\bmadt$ to
the future $\chi^{[n,+\infty)}(\bmaddt)$ of $\bmaddt$ \cite{FT}.  
A group system $A$ is {\it $l$-controllable} if there is an integer $l>0$ 
such that $A$ is $[t,t+l)$-controllable for all $t\in\bmcpz$.  
A group system $A$ is 
{\it strongly controllable} if it is $l$-controllable for some integer $l$.  
The least integer $l$ for which a group system
$A$ is strongly controllable is denoted as $\ell$.
In this paper we study strongly controllable group systems.

For each $t\in\bmcpz$, define $X^t$ to be the set of all sequences $\bma$ in $A$ for which 
$a^n=a_\bone^n$ for $n<t$, where $a_\bone^n$ is the identity of $A^n$ at time $n$ (see Figure
\ref{statedefn}).  For each $t\in\bmcpz$, define $Y^t$ to be the set of all sequences $\bma$ 
in $A$ for which $a^n=a_\bone^n$ for $n>t$.
The {\it canonic state space} $\Sigma^t$ of $A$ at time $t$ is defined to be 
$$
\Sigma^t\rmdef\frac{A}{Y^{t-1}X^t}.
$$
(Note that $A$ is the same as $C$ in \cite{FT}, $Y^{t-1}$ is the same as $C^{t^-}$, 
and $X^t$ is the same as $C^{t^+}$.)
From Figure \ref{statedefn}, it is evident the definition of the state at time $t$
involves a split between time $t-1$ and time $t$ \cite{FT}.  The canonic state space 
is unique.  The group system satisfies the {\it axiom of state}:  whenever
two sequences pass through the same state at a given time,
the concatenation of the past of either with the future of the other
is a valid sequence \cite{FT}.  In this paper, we just use the identity state
or zero state of $\Sigma^t$ at each time $t\in\bmcpz$.

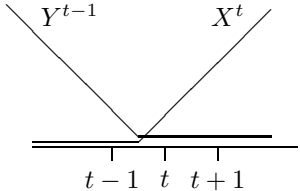
\begin{figure}[h]
\centering

\begin{picture}(100,80)(0,-10)

%lines
\put(0,0){\line(1,0){100}}
\put(0,2){\line(1,0){40}}
\put(40,4){\line(1,0){50}}
\put(40,2){\line(1,1){50}}
\put(40,4){\line(-1,1){50}}

%labels
\put(30,-9){\makebox(0,0)[t]{$t-1$}}
\put(50,-9){\makebox(0,0)[t]{$t$}}
\put(70,-9){\makebox(0,0)[t]{$t+1$}}
\put(30,-5){\line(0,1){5}}
\put(50,-5){\line(0,1){5}}
\put(70,-5){\line(0,1){5}}
\put(3,50){\makebox(0,0)[l]{$Y^{t-1}$}}
\put(80,50){\makebox(0,0)[r]{$X^t$}}

\end{picture}

\caption{Definition of $Y^{t-1}$ and $X^t$.}
\label{statedefn}

\end{figure}

We now review some results from \cite{FT} on the construction of $A$ from its
fundamental components, the generators.  Assume a group system $A$ is
\ellctl\ and complete.  Forney and Trott \cite{FT} 
define the $k$-controllable subcode $A_k$ of an \ellctl\ group code $A$, for $0\le k\le\ell$.  
The $k$-{\it controllable subcode} $A_k$ of a group code $A$ is defined as the set of 
combinations of sequences of span $k+1$ or less:
$$
A_k=\prod_tA^{[t,t+k]}.
$$
They show
\be
\label{norm1}
A_0\subset A_1\subset\ldots A_{k-1}\subset A_k\subset\ldots A_\ell=A
\ee
is a normal series.  A chain coset decomposition yields a one-to-one correspondence
\be
\label{norm2}
A\lra A_0\times (A_1/A_0)\times\cdots\times (A_k/A_{k-1})\times\cdots\times (A_\ell/A_{\ell-1}).
\ee

For $1\le k\le\ell$, the quotient groups $(A_k/A_{k-1})$ may be further evaluated as follows.
In their Code Granule Theorem \cite{FT}, they show
$A_k/A_{k-1}$ is isomorphic to a direct product of quotient groups $\Gamma^{[t,t+k]}$,
\be
\label{norm3}
A_k/A_{k-1}\simeq\prod_t\Gamma^{[t,t+k]},
\ee
where $\Gamma^{[t,t+k]}$ is defined by
$$  
\Gamma^{[t,t+k]}\rmdef\frac{A^{[t,t+k]}}{A^{[t,t+k)}A^{(t,t+k]}}.
$$
$\Gamma^{[t,t+k]}$ is called a {\it granule}.
A \crep\ of $\Gamma^{[t,t+k]}$ is called a {\it Forney-Trott generator} $\bmg_{FT}^{[t,t+k]}$
of $A$.  The \crep\ of $A^{[t,t+k)}A^{(t,t+k]}$ is always taken
to be the identity sequence.  In case
$\Gamma^{[t,t+k]}$ is isomorphic to the identity group, the identity
sequence is the only coset representative.  
A nonidentity generator is an element of
$A^{[t,t+k]}$ but not of $A^{[t,t+k)}$ or of $A^{(t,t+k]}$, so its span
is exactly $k+1$.  Thus every nonidentity generator
is a codeword that cannot be expressed as a combination of shorter
codewords \cite{FT}.

If $Q$ is any quotient group, let $[Q]$ denote a set of \creps\
of $Q$, or a transversal of $Q$.  Let $[\Gamma^{[t,t+k]}]$ be a transversal of $\Gamma^{[t,t+k]}$.
It follows from (\ref{norm3}) that the set $\prod_t \left[\Gamma^{[t,t+k]}\right]$
is a set of coset representatives for the cosets of $A_{k-1}$ in $A_k$.
We know that the set of \creps\ of the granule $\Gamma^{[t,t+k]}$, or of transversal 
$[\Gamma^{[t,t+k]}]$, is a set of generators $\{\bmg_{FT}^{[t,t+k]}\}$.
Then from (\ref{norm2}), any sequence $\bma$ 
can be uniquely evaluated as a product
$$
\bma=A_0\prod_{k=1}^{\ell} \prod_{t=-\infty}^{+\infty} \bmg_{FT}^{[t,t+k]}.
$$
Any element of $A^{[t,t]}$ is a \crep\ $\bmg_{FT}^{[t,t]}$.
Then $A_0=\prod_t \{\bmg_{FT}^{[t,t]}\}$.  It follows that 
(Generator Theorem \cite{FT}) every sequence $\bma$ 
can be uniquely expressed as a product
\be
\label{encftc}
\bma=\prod_{k=0}^{\ell} \prod_{t=-\infty}^{+\infty} \bmg_{FT}^{[t,t+k]}
\ee
of generators $\bmg_{FT}^{[t,t+k]}$.  Thus every sequence $\bma$
is a product of some sequence of generators, and conversely, every
sequence of generators corresponds to some sequence $\bma$.
Then a component $a^t$ of $\bma$ is given by
\be
\label{encftd}
a^t=\prod_{k=0}^\ell \left(\prod_{j=0}^k \chi^t(\bmg_{FT}^{[t-j,t-j+k]})\right).
\ee

A {\it basis} of $A$ is a smallest set of shortest length generators 
that is sufficient to generate the group system $A$ \cite{FY1}.
It follows from the encoder in \cite{FT} that a basis of $A$ is a set of \creps\ 
of $\Gamma^{[t,t+k]}$, for $0\le k\le\ell$, for each $t\in\bmcpz$.
The encoder in (\ref{encftc}) forms output $\bma$ from a sequence of generators selected from
the basis.  For each time $t\in\bmcpz$, for each $k$ such that $0\le k\le\ell$, 
a single generator $\bmg_{FT}^{[t,t+k]}$ is selected from set $[\Gamma^{[t,t+k]}]$.
In general, the set of sequences that can be selected from the basis to form $A$ is a
proper subset of the double Cartesian product \cite{FT}
\be
\label{cprod}
\bigotimes_{t=+\infty}^{t=-\infty} \bigotimes_{0\le k\le\ell} [\Gamma^{[t,t+k]}].
\ee
But if $A$ is complete, then the set of sequences that forms $A$ is exactly
the double Cartesian product (\ref{cprod}) \cite{FT}.

\newpage
\vspace{3mm}
{\bf 3.  THE NORMAL CHAIN OF $A$}%hhh
\vspace{3mm}

Recall that for each $t\in\bmcpz$, we have defined $X^t$ to be the set of all sequences 
$\bma$ in $A$ for which $a^n=a_\bone^n$ for $n<t$, where $a_\bone^n$ is the identity of $A^n$ 
at time $n$.  And for each $t\in\bmcpz$, we have defined $Y^t$ to be the set of all sequences $\bma$ 
in $A$ for which $a^n=a_\bone^n$ for $n>t$.

It is clear that $X^t\lhd A$ and $Y^t\lhd A$ for each $t\in\bmcpz$. 
Then the group $A$ has two normal series (and chief series)
\be
\label{chnx}
\bma_\bone\cdots\subset X^{t+1}\subset X^t\subset X^{t-1}\subset\cdots\subset X^{t-j+1}\subset X^{t-j}\subset\cdots\subset X^{t-\ell+1}\subset X^{t-\ell}\subset\cdots A,
\ee
and
\be
\label{chny}
\bma_\bone\cdots\subset Y^{t-1}\subset Y^t\subset Y^{t+1}\subset\cdots\subset Y^{t+k-1}\subset Y^{t+k}\subset\cdots\subset Y^{t+\ell-1}\subset Y^{t+\ell}\subset\cdots A.
\ee
In (\ref{chnx}) we have chosen index $j$ such that $0\le j\le\ell$ and in (\ref{chny})
we have chosen index $k$ such that $0\le k\le\ell$.
The Schreier refinement theorem used to prove the Jordan-H\"{o}lder 
theorem \cite{ROT} shows how to obtain a refinement of two normal series
by inserting one into the other.  We can obtain a refinement of (\ref{chnx})
by inserting (\ref{chny}) between each two successive terms of (\ref{chnx}).
This gives the normal series shown in (\ref{sminf}).  The normal series
is given by the $\ell+3$ rows in the middle of (\ref{sminf}) and the rows denoted by the two
vertical ellipses near the top and bottom.  The bottom row and top row in (\ref{sminf}) are
limiting groups, explained further below.  The normal series is
an infinite series of columns, with each column an infinite series of groups.
We have only shown $\ell+1$ columns of the infinite series in (\ref{sminf}). 
Since (\ref{chnx}) and (\ref{chny}) are chief series, the normal chain 
(\ref{sminf}) is a chief series.

%%%%%%%%%%%%%%%%%%%%%%%%%%%%%%%%%%%%%%%%%
\begin{figure}

\be
\label{sminf}
\begin{array}{cccccccc}
& \shrtpll & \shrtpll && \shrtpll && \shrtpll & \\

& X^{t+1}(X^t) & X^t(X^{t-1}) & \cdots & X^{t-j+1}(X^{t-j}) & \cdots & X^{t-\ell+1}(X^{t-\ell}) & \\

& \cup & \cup && \cup && \cup & \\

& \vdots & \vdots & \vdots & \vdots & \vdots & \vdots & \\

& \vdots & \vdots & \vdots & \vdots & \vdots & \vdots & \\

& \cup & \cup && \cup && \cup & \\

& X^{t+1}(X^t\cap Y^{t+\ell+1}) & X^t(X^{t-1}\cap Y^{t+\ell}) & \cdots & X^{t-j+1}(X^{t-j}\cap Y^{t+\ell-j+1}) & \cdots & X^{t-\ell+1}(X^{t-\ell}\cap Y^{t+1}) & \\

& \cup & \cup && \cup && \cup & \\

& X^{t+1}(X^t\cap Y^{t+\ell}) & X^t(X^{t-1}\cap Y^{t+\ell-1}) & \cdots & X^{t-j+1}(X^{t-j}\cap Y^{t+\ell-j}) & \cdots & X^{t-\ell+1}(X^{t-\ell}\cap Y^{t}) & \\

& \cup & \cup && \cup && \cup & \\

& X^{t+1}(X^t\cap Y^{t+\ell-1}) & X^t(X^{t-1}\cap Y^{t+\ell-2}) & \cdots & X^{t-j+1}(X^{t-j}\cap Y^{t+\ell-j-1}) & \cdots & X^{t-\ell+1}(X^{t-\ell}\cap Y^{t-1}) & \\

& \cup & \cup && \cup && \cup & \\

& \cdots & \cdots & \cdots & \cdots & \cdots & \cdots & \\

& \cup & \cup && \cup && \cup & \\

& X^{t+1}(X^t\cap Y^{t+k}) & X^t(X^{t-1}\cap Y^{t+k-1}) & \cdots & X^{t-j+1}(X^{t-j}\cap Y^{t+k-j}) & \cdots & X^{t-\ell+1}(X^{t-\ell}\cap Y^{t+k-\ell}) & \\

& \cup & \cup && \cup && \cup & \\

& X^{t+1}(X^t\cap Y^{t+k-1}) & X^t(X^{t-1}\cap Y^{t+k-2}) & \cdots & X^{t-j+1}(X^{t-j}\cap Y^{t+k-j-1}) & \cdots & X^{t-\ell+1}(X^{t-\ell}\cap Y^{t+k-\ell-1}) & \\

& \cup & \cup && \cup && \cup & \\

\cdots & \cdots & \cdots & \cdots & \cdots & \cdots & \cdots & \cdots \\

& \cup & \cup && \cup && \cup & \\

& X^{t+1}(X^t\cap Y^{t+j}) & X^t(X^{t-1}\cap Y^{t+j-1}) & \cdots & X^{t-j+1}(X^{t-j}\cap Y^{t}) & \cdots & X^{t-\ell+1}(X^{t-\ell}\cap Y^{t+j-\ell}) & \\

& \cup & \cup && \cup && \cup & \\

& X^{t+1}(X^t\cap Y^{t+j-1}) & X^t(X^{t-1}\cap Y^{t+j-2}) & \cdots & X^{t-j+1}(X^{t-j}\cap Y^{t-1}) & \cdots & X^{t-\ell+1}(X^{t-\ell}\cap Y^{t+j-\ell-1}) & \\

& \cup & \cup && \cup && \cup & \\

& \cdots & \cdots & \cdots & \cdots & \cdots & \cdots & \\

& \cup & \cup && \cup && \cup & \\

& X^{t+1}(X^t\cap Y^{t+1}) & X^t(X^{t-1}\cap Y^{t}) & \cdots & X^{t-j+1}(X^{t-j}\cap Y^{t-j+1}) & \cdots & X^{t-\ell+1}(X^{t-\ell}\cap Y^{t-\ell+1}) & \\

& \cup & \cup && \cup && \cup & \\

& X^{t+1}(X^t\cap Y^t) & X^t(X^{t-1}\cap Y^{t-1}) & \cdots & X^{t-j+1}(X^{t-j}\cap Y^{t-j}) & \cdots & X^{t-\ell+1}(X^{t-\ell}\cap Y^{t-\ell}) & \\

& \cup & \cup && \cup && \cup & \\

& X^{t+1}(X^t\cap Y^{t-1}) & X^t(X^{t-1}\cap Y^{t-2}) & \cdots & X^{t-j+1}(X^{t-j}\cap Y^{t-j-1}) & \cdots & X^{t-\ell+1}(X^{t-\ell}\cap Y^{t-\ell-1}) & \\

& \cup & \cup && \cup && \cup & \\

& \vdots & \vdots & \vdots & \vdots & \vdots & \vdots & \\

& \vdots & \vdots & \vdots & \vdots & \vdots & \vdots & \\

& \cup & \cup && \cup && \cup & \\

& X^{t+1} & X^t & \cdots & X^{t-j+1} & \cdots & X^{t-\ell+1} & \\

& \shrtpll & \shrtpll && \shrtpll && \shrtpll &
\end{array}
\ee
\end{figure}
%%%%%%%%%%%%%%%%%%%%%%%%%%%%%%%%%%%%%%%%%%%

We now show that (\ref{sminf}) is indeed a refinement of (\ref{chnx}).
First we show that in each column the group in the bottom row is contained
in all groups in the infinite column of groups, and the group in the top row
contains all groups in the infinite column of groups.
Any term in (\ref{sminf}) is of the form
$$
X^{i+1}(X^i\cap Y^{i+m})
$$
for some integer pair $i,m\in\bmcpz$.  For example, term $X^{t-j+1}(X^{t-j}\cap Y^{t+k-j})$
is of the form $X^{i+1}(X^i\cap Y^{i+m})$ for $i=t-j$ and $m=k$.   Fix $i\in\bmcpz$.  
For any $m\in\bmcpz$, we have $X^{i+1}\subset X^{i+1}(X^i\cap Y^{i+m})$
Therefore the group in the bottom row is contained in all groups 
in the infinite column of groups.  For any $m\in\bmcpz$,
it is clear that $X^{i+1}(X^i\cap Y^{i+m})\subset X^{i+1}(X^i)$.
Therefore the group in the top row contains all groups in the 
infinite column of groups.  Now note that $X^{i+1}(X^i)=X^i$.
This means that a group in the top row is the same as the group
in the bottom row in the next column.  But now note the groups in
the bottom row form the same sequence as (\ref{chnx}).  Therefore (\ref{sminf}) is 
indeed a refinement of (\ref{chnx}).

The factor $X^{t-j+1}$ in term $X^{t-j+1}(X^{t-j}\cap Y^{t+k-j})$ in (\ref{sminf})
is called an {\it integration factor}, and the factor $(X^{t-j}\cap Y^{t+k-j})$
in term $X^{t-j+1}(X^{t-j}\cap Y^{t+k-j})$ is called a
{\it derivative factor}.  The derivative factors in each row of a finite band of rows
in (\ref{sminf}), absent the normal chain given here, are essentially what is used in 
\cite{FT} to find a \cdc\ for the spectral domain (see also \cite{LM}).  
The product of the derivative factors in each row
of the finite band corresponds to a $k$-controllable subcode of \cite{FT}.

We now show the normal chain (\ref{sminf}) contains all the sequences $\bma\in A$.  Since $A$
is complete, any $\bma\in A$ can be specified by its projection over finite time intervals
\cite{FT}.  Then we just have to show that the normal chain contains the projection of $\bma$
over any finite time interval $[t,t+n]$, $\chi^{[t,t+n]}(\bma)$.  But since $A$ is 
\ellctl, the group $(X^i\cap Y^{i+m})$ contains $\chi^{[t,t+n]}(\bma)$
for $i\le t-\ell$ and $i+m\ge t+n+\ell$.  Therefore the group
$X^{i+1}(X^i\cap Y^{i+m})$ in (\ref{sminf}) contains $\chi^{[t,t+n]}(\bma)$
for $i\le t-\ell$ and $i+m\ge t+n+\ell$.  This gives the following.

\begin{thm}
Any sequence $\bma\in A$ is completely specified by the normal chain (\ref{sminf}).
\end{thm}

Since any sequence $\bma\in A$ is specified by the normal chain (\ref{sminf}), we can use the
\cdc\ of (\ref{sminf}) to find any sequence $\bma$ in $A$.
In any normal chain, we may form the quotient group of two successive groups in the chain.
Then a normal chain of groups, as in (\ref{sminf}), gives a series of quotient groups.
A general quotient group obtained from (\ref{sminf}) is of the form
\be
\label{qgx}
\Lambda^{[i,i+m]}\rmdef\frac{X^{i+1}(X^i\cap Y^{i+m})}{X^{i+1}(X^i\cap Y^{i+m-1})}
\ee
for any integer pair $i,m\in\bmcpz$.  We call $\Lambda^{[i,i+m]}$ the {\it time domain granule}.
Note that the time domain granule has half infinite extent while the granule
$\Gamma^{[t,t+k]}$ of \cite{FT} has finite extent.
We wish to find a transversal of the time domain granule.  
The \creps\ of the time domain granule are called {\it generators}.
Since $(X^{i+1}\cap Y^{i+m})\subset X^{i+1}$, we have
$X^{i+1}(X^i\cap Y^{i+m-1})=X^{i+1}(X^i\cap Y^{i+m-1})(X^{i+1}\cap Y^{i+m})$,
and we may rewrite (\ref{qgx}) as
\be
\label{qgx0}
\Lambda^{[i,i+m]}=\frac{X^{i+1}(X^i\cap Y^{i+m})}{X^{i+1}(X^i\cap Y^{i+m-1})(X^{i+1}\cap Y^{i+m})}.
\ee

We now find a transversal of (\ref{qgx0}).  An element of the numerator group is of 
the form $\bmx\bmy$, where $\bmx\in X^{i+1}$ and $\bmy\in (X^i\cap Y^{i+m})$.
Then a coset of normal subgroup $X^{i+1}(X^i\cap Y^{i+m-1})(X^{i+1}\cap Y^{i+m})$ is 
\begin{align*}
\bmx\bmy X^{i+1}(X^i\cap Y^{i+m-1})(X^{i+1}\cap Y^{i+m}) 
&=\bmx X^{i+1}\bmy(X^i\cap Y^{i+m-1})(X^{i+1}\cap Y^{i+m}) \\
&=X^{i+1}\bmy(X^i\cap Y^{i+m-1})(X^{i+1}\cap Y^{i+m}).
\end{align*}

\begin{thm}
A coset of normal subgroup $X^{i+1}(X^i\cap Y^{i+m-1})(X^{i+1}\cap Y^{i+m})$ is 
$X^{i+1}\bmy(X^i\cap Y^{i+m-1})(X^{i+1}\cap Y^{i+m})$ where $\bmy\in (X^i\cap Y^{i+m})$.
A \crep\ of coset $X^{i+1}\bmy(X^i\cap Y^{i+m-1})(X^{i+1}\cap Y^{i+m})$ is
$\bmxdt\bmydt$ where $\bmxdt$ is in $X^{i+1}$ and $\bmydt$ is in
$\bmy(X^i\cap Y^{i+m-1})(X^{i+1}\cap Y^{i+m})$.
A transversal of the quotient group (\ref{qgx0}) is a selection of one \crep\
$\bmxdt\bmydt$ from each coset $X^{i+1}\bmy(X^i\cap Y^{i+m-1})(X^{i+1}\cap Y^{i+m})$.
\end{thm}
We may always select a \crep\ $\bmxdt\bmydt$ of coset $X^{i+1}\bmy(X^i\cap Y^{i+m-1})(X^{i+1}\cap Y^{i+m})$
to be $\bma_\bone\bmydt$ where $\bma_\bone$ is the identity of $X^{i+1}$.
But note that $\bmy(X^i\cap Y^{i+m-1})(X^{i+1}\cap Y^{i+m})$ is a coset of
normal subgroup $(X^i\cap Y^{i+m-1})(X^{i+1}\cap Y^{i+m})$ in quotient group
\be
\label{qgx1}
\frac{(X^i\cap Y^{i+m})}{(X^i\cap Y^{i+m-1})(X^{i+1}\cap Y^{i+m})}.
\ee
Then $\bmydt$ is a \crep\ of (\ref{qgx1}).  This gives the following.

\begin{cor}
A transversal of the quotient group (\ref{qgx1}) is a transversal
of the quotient group (\ref{qgx0}), but the reverse is only true if $\bmxdt\bmydt=\bma_\bone\bmydt$ 
or $\bmxdt=\bma_\bone$ for each \crep\ $\bmxdt\bmydt$ of (\ref{qgx0}).
\end{cor}

We have just shown that a transversal of (\ref{qgx1}) is a transversal of (\ref{qgx0}).
It is not surprising then that there is a homomorphism from (\ref{qgx0}) to
(\ref{qgx1}).  In fact this result is an application of the Zassenhaus lemma
used in the proof of the Schreier refinement theorem \cite{ROT}.  We 
first restate the following lemma, excised from the proof of the Zassenhaus 
lemma (see p.\ 100 of \cite{ROT}).

\begin{lem}
\label{lem26xx}
{\bf (from proof of Zassenhaus lemma)}
Let $U\lhd U^*$ and $V\lhd V^*$ be four subgroups of a group $G$.
Then $D=(U^*\cap V)(U\cap V^*)$ is a normal subgroup of $U^*\cap V^*$.
If $g\in U(U^*\cap V^*)$, then $g=uu^*$ for $u\in U$ and $u^*\in U^*\cap V^*$.
Define function $f:  U(U^*\cap V^*)\ra (U^*\cap V^*)/D$ by $f(g)=f(uu^*)=Du^*$.
Then $f$ is a well defined homomorphism with kernel $U(U^*\cap V)$ and
$$
\frac{U(U^*\cap V^*)}{U(U^*\cap V)}\simeq\frac{U^*\cap V^*}{D}.
$$
\end{lem}

Note that (\ref{qgx0}) is equivalent to (\ref{qgx}).
We now use Lemma \ref{lem26xx} to show there is a homomorphism from (\ref{qgx}) to
(\ref{qgx1}).  Let $U=X^{i+1}$ and $U^*=X^i$.  Let $V=Y^{i+m-1}$ and $V^*=Y^{i+m}$.
Note that $U\lhd U^*$ and $V\lhd V^*$.  Then
\begin{align*}
\frac{U(U^*\cap V^*)}{U(U^*\cap V)} &=
\frac{X^{i+1}(X^i\cap Y^{i+m})}{X^{i+1}(X^i\cap Y^{i+m-1})},
\end{align*}
and
$$
\frac{U^*\cap V^*}{D}=\frac{U^*\cap V^*}{(U^*\cap V)(U\cap V^*)}
=\frac{(X^i\cap Y^{i+m})}{(X^i\cap Y^{i+m-1})(X^{i+1}\cap Y^{i+m})}.
$$
Now the function $f$ of Lemma \ref{lem26xx} is a homomorphism from (\ref{qgx}) to
(\ref{qgx1}).

\begin{thm}
\label{dmnh}
There is a homomorphism from the quotient group (\ref{qgx}), or equivalently (\ref{qgx0}),
to the quotient group (\ref{qgx1}), given by the function $f$ of the Zassenhaus lemma.
\end{thm}

We now choose a \crep\ or generator of the time domain granule (\ref{qgx}) formed 
by the normal chain (\ref{sminf}) for any integer pair $i,m\in\bmcpz$.  Fix any $i\in\bmcpz$.
There are three cases to consider for (\ref{qgx}), depending on the value of $m$.
For $m<0$, we have $(X^i\cap Y^{i+m})=\bma_\bone$ and so (\ref{qgx}) 
reduces to $X^{i+1}(\bma_\bone)/X^{i+1}(\bma_\bone)=
X^{i+1}/X^{i+1}$.  Then \crep\ $\bmxdt\bmydt$ is $\bmxdt\bma_\bone$.  We may always select $\bmxdt$
to be $\bma_\bone$.  Then for $m<0$, we can choose the \crep\ or generator of (\ref{qgx}) to be $\bma_\bone$.

We next consider the case $0\le m\le\ell$.  If we select a \crep\ $\bmxdt\bmydt$ of 
coset $X^{i+1}\bmy(X^i\cap Y^{i+m-1})(X^{i+1}\cap Y^{i+m})$
to be $\bma_\bone\bmydt$ where $\bma_\bone$ is the identity of $X^{i+1}$, then a transversal 
of the quotient group (\ref{qgx1}) is a transversal of the quotient group (\ref{qgx0}).
But for $0\le m\le\ell$ note that (\ref{qgx1}) is the same as the Forney-Trott granule 
$\Gamma^{[i,i+m]}$, or {\it spectral domain granule}, for $A$,
\be
\label{spctrl}
\Gamma^{[i,i+m]}=\frac{A^{[i,i+m]}}{A^{[i,i+m-1]}A^{[i+1,i+m]}}.
\ee
This gives the following.

\begin{cor}
\label{cor10}
For $0\le m\le\ell$, a transversal of the spectral domain granule (\ref{spctrl}) is a transversal of 
the time domain granule (\ref{qgx}), but the reverse is only true if $\bmxdt\bmydt=\bma_\bone\bmydt$ 
or $\bmxdt=\bma_\bone$ for each \crep\ $\bmxdt\bmydt$ of the time domain granule.
\end{cor}
Then for $0\le m\le\ell$, we can choose the \crep\ or generator of (\ref{qgx}) to be $\bma_\bone\bmydt$.
In this case, the \crep\ or generator of (\ref{qgx}) is the same as the \crep\ $\bmg^{[i,i+m]}$
of the spectral domain granule (\ref{spctrl}) of $A$.  Therefore the \crep\ of the time domain
granule can be chosen to be the same as the Forney-Trott generator of $A$.  Note that for $0\le m\le\ell$, 
Theorem \ref{dmnh} shows there is a homomorphism from the time domain granule to the spectral 
domain granule given by the function $f$ of the Zassenhaus lemma.

Lastly we consider the case $m>\ell$.  Since $A$ is \ellctl, in (\ref{qgx1}) there can be no
elements of $(X^i\cap Y^{i+m})$ that are not elements of $(X^i\cap Y^{i+m-1})(X^{i+1}\cap Y^{i+m})$.
Then
$$
X^i\cap Y^{i+m}=(X^i\cap Y^{i+m-1})(X^{i+1}\cap Y^{i+m}),
$$
and quotient group (\ref{qgx1}) is trivial.  Then  quotient group (\ref{qgx0}) and 
(\ref{qgx}) is trivial.  Then we can choose the \crep\ $\bmxdt\bmydt$ or generator of (\ref{qgx}) 
to be $\bmxdt\bmydt=\bma_\bone\bma_\bone$ or $\bma_\bone$.

%%%%%%%%%%%%%%%%%%%%%%%%%%%%%%%%%%%%%%%%%
\begin{figure}

\be
\label{smgen}
\begin{array}{llllllll}
& \vdots & \vdots & \vdots & \vdots & \vdots & \vdots & \\
& \vdots & \vdots & \vdots & \vdots & \vdots & \vdots & \\
& \bma_\bone          & \bma_\bone            & \cdots & \bma_\bone              & \cdots & \bma_\bone & \\
& \bmg^{[t,t+\ell]}   & \bmg^{[t-1,t-1+\ell]}   & \cdots & \bmg^{[t-j,t-j+\ell]}   & \cdots & \bmg^{[t-\ell,t]} & \\
& \bmg^{[t,t+\ell-1]} & \bmg^{[t-1,t-1+\ell-1]} & \cdots & \bmg^{[t-j,t-j+\ell-1]} & \cdots & \bmg^{[t-\ell,t-1]} & \\
& \cdots              & \cdots                & \cdots & \cdots                  & \cdots & \cdots & \\
& \bmg^{[t,t+k]}      & \bmg^{[t-1,t-1+k]}    & \cdots & \bmg^{[t-j,t-j+k]}      & \cdots & \bmg^{[t-\ell,t-\ell+k]} & \\
& \bmg^{[t,t+k-1]}    & \bmg^{[t-1,t-1+k-1]}  & \cdots & \bmg^{[t-j,t-j+k-1]}    & \cdots & \bmg^{[t-\ell,t-\ell+k-1]} & \\
\cdots & \cdots       & \cdots                & \cdots & \cdots                  & \cdots & \cdots & \cdots \\
& \bmg^{[t,t+j]}      & \bmg^{[t-1,t-1+j]}    & \cdots & \bmg^{[t-j,t]}          & \cdots & \bmg^{[t-\ell,t-\ell+j]} & \\
& \bmg^{[t,t+j-1]}    & \bmg^{[t-1,t-1+j-1]}  & \cdots & \bmg^{[t-j,t-1]}        & \cdots & \bmg^{[t-\ell,t-\ell+j-1]} & \\
& \cdots              & \cdots                & \cdots & \cdots                  & \cdots & \cdots & \\
& \bmg^{[t,t+1]}      & \bmg^{[t-1,t]}        & \cdots & \bmg^{[t-j,t-j+1]}      & \cdots & \bmg^{[t-\ell,t-\ell+1]} & \\
& \bmg^{[t,t]}        & \bmg^{[t-1,t-1]}      & \cdots & \bmg^{[t-j,t-j]}        & \cdots & \bmg^{[t-\ell,t-\ell]} & \\
& \bma_\bone          & \bma_\bone            & \cdots & \bma_\bone              & \cdots & \bma_\bone & \\
& \vdots & \vdots & \vdots & \vdots & \vdots & \vdots & \\
& \vdots & \vdots & \vdots & \vdots & \vdots & \vdots &
\end{array}
\ee
\end{figure}
%%%%%%%%%%%%%%%%%%%%%%%%%%%%%%%%%%%%%%%%%%%

We have found generators of the time domain granule for each $i\in\bmcpz$ and
ranges $m<0$, $0\le m\le\ell$, and $m>\ell$.  
We can arrange the selected set of generators in the same order as (\ref{sminf}), as
shown in (\ref{smgen}).  For $m<0$, there are an infinite number of rows at the bottom 
of (\ref{smgen}) which are filled with the identity generator; for $m>\ell$ there are 
an infinite number of rows at the top of (\ref{smgen}) which are filled with the identity generator.
There are a finite number of rows, $\ell+1$, which can have nontrivial generators $\bmg^{[i,i+m]}$
for $0\le m\le\ell$.  Let $\bmr$ be the infinite series of finite columns given by generators in
the finite band in (\ref{smgen}), i.e., the generators $\bmg^{[i,i+m]}$ for $0\le m\le\ell$,
for each $i\in\bmcpz$.
Note that $\bmr$ is a tensor.  Let $\calr$ be the set of all possible tensors $\bmr$.

We can arrange the sequence of generators in the finite band in (\ref{smgen}) into the product
\be
\label{prod}
\cdots\bmg^{[i+1,i+1+\ell]}\bmg^{[i,i]}\bmg^{[i,i+1]}\cdots\bmg^{[i,i+\ell]}
\bmg^{[i-1,i-1]}\bmg^{[i-1,i]}\cdots\bmg^{[i-1,i-1+\ell]}\bmg^{[i-2,i-2]}\cdots.
\ee
Each generator $\bmg^{[i,i+m]}$ is a sequence in $A$.  Then the product (\ref{prod}) 
is understood to mean a multiplication of sequences in $A$.

\begin{thm}
The sequence (\ref{prod}) is a well defined product in $A$.  Therefore (\ref{prod}) 
is a well defined sequence $\bma$ in $A$, where $\bma$ is given by
\be
\label{enctdx}
\bma=\prod_{i=+\infty}^{-\infty} \left(\prod_{m=0}^\ell \bmg^{[i,i+m]})\right).
\ee
\end{thm}

\begin{prf}
Each term in (\ref{prod}) is a sequence in $A$.  The sequences in product (\ref{prod})
are multiplied according to the definition of product in $A$, which means
component by component multiplication for $-\infty<i<+\infty$.  The infinite
product (\ref{prod}) is well defined if for each $i\in\bmcpz$, there are only 
finitely many terms $\bmg^{[i,i+m]}$ having a nontrivial component \cite{FT}.
Therefore the final result is a well defined sequence in $A$.
\end{prf}

Choosing \creps\ or generators $\bmg^{[i,i+m]}$ of the time domain granules $\Lambda^{[i,i+m]}$
formed by the normal chain (\ref{sminf}) forms a {\it basis} 
$\calb$ of $A$.  If a basis $\calb$ of $A$ is chosen, then $\calr$ is chosen.

\begin{thm}
\label{thm9}
Find a basis $\calb$ of $A$.  Then $\calr$ is chosen, and there is a 
bijection $\alpha:  \calr\ra A$ given by assignment $\alpha:  \bmr\mapsto\bma$, 
where $\bma$ is an encoding of $\bmr$ using product (\ref{enctdx}) on the 
generators in $\bmr$.
\end{thm}

\begin{prf}
Since (\ref{sminf}) forms a \cdc\ of $A$, and since (\ref{enctdx}) uses a selection of
one representative from each nontrivial quotient group in the chain, then any $\bma\in A$ can
be obtained by the composition (\ref{enctdx}).  The assignment $\alpha:  \bmr\mapsto\bma$
is a bijection since any unique selection of \creps\ gives a unique $\bma\in A$.
\end{prf}

A component $a^t$ of $\bma$ at any time $t$ can be obtained by the 
composition of the time $t$ component $\chi^t$ of generators in (\ref{smgen}).
There are only a finite number of generators in (\ref{smgen}) with nontrivial components
at time $t$.  These are generators $\bmg^{[t-j,t-j+k]}$, for $0\le j\le\ell$, for $j\le k\le\ell$,
shown in the triangular matrix (\ref{smgent}).
The components at time $t$ of all other generators in (\ref{smgen}) are 
the identity element $a_\bone^t$ of $A^t$.  This gives the following.

\begin{lem}
\label{lem9}
The only quotient groups, or time domain granules, formed from (\ref{sminf}) that have transversals 
with nontrivial components at time $t$ are of the form
\be
\label{qgqg}
\Lambda^{[t-j,t-j+k]}=\frac{X^{t-j+1}(X^{t-j}\cap Y^{t+k-j})}{X^{t-j+1}(X^{t-j}\cap Y^{t+k-j-1})}
\ee
for $0\le j\le\ell$, for $j\le k\le\ell$.
\end{lem}
Then a component $a^t$ of $\bma$ can be obtained by the composition of components
$\chi^t(\bmg^{[t-j,t-j+k]})$ for $0\le j\le\ell$, for $j\le k\le\ell$, as
\be
\label{enctdc}
a^t=\prod_{j=0}^\ell \left(\prod_{k=j}^\ell \chi^t(\bmg^{[t-j,t-j+k]})\right).
\ee

%%%%%%%%%%%%%%%%%%%%%%%%%%%%%%%%%%%%%%%%%
\begin{figure}

\be
\label{smgent}
\begin{array}{llllll}
\bmg^{[t,t+\ell]}   & \bmg^{[t-1,t-1+\ell]}   & \cdots & \bmg^{[t-j,t-j+\ell]}   & \cdots & \bmg^{[t-\ell,t]} \\
\bmg^{[t,t+\ell-1]} & \bmg^{[t-1,t-1+\ell-1]} & \cdots & \bmg^{[t-j,t-j+\ell-1]} & \cdots & \\      
\cdots              & \cdots                  & \cdots & \cdots                  & \cdots & \\      
\bmg^{[t,t+k]}      & \bmg^{[t-1,t-1+k]}      & \cdots & \bmg^{[t-j,t-j+k]}      &        & \\      
\bmg^{[t,t+k-1]}    & \bmg^{[t-1,t-1+k-1]}    & \cdots & \bmg^{[t-j,t-j+k-1]}    &        & \\      
\cdots              & \cdots                  & \cdots & \cdots                  &        & \\      
\bmg^{[t,t+j]}      & \bmg^{[t-1,t-1+j]}      & \cdots & \bmg^{[t-j,t]}          &        & \\      
\bmg^{[t,t+j-1]}    & \bmg^{[t-1,t-1+j-1]}    & \cdots &                         &        & \\      
\cdots              & \cdots                  & \cdots &                         &        & \\      
\bmg^{[t,t+1]}      & \bmg^{[t-1,t]}          &        &                         &        & \\      
\bmg^{[t,t]}        &                         &        &                         &        &         
\end{array}
\ee
\end{figure}
%%%%%%%%%%%%%%%%%%%%%%%%%%%%%%%%%%%%%%%%%%%

The encoder in (\ref{enctdc}) can be written as $a^t=\prod_{j=0}^\ell h_j^{t-j}$, where
the inner term in parentheses is some function of time, say $h_j^{t-j}$.
Thus the encoder has the form of a time convolution, reminiscent of a linear system.  
The encoder (\ref{encftd}) in \cite{FT} does not have a convolution
property.  For this reason, we say the encoder (\ref{enctdx}) or (\ref{enctdc}) 
is a {\it time domain encoder},
and the encoder (\ref{encftc}) or (\ref{encftd}) in \cite{FT} a {\it spectral domain encoder}.
The time domain encoder has a granule based construction of input granules and 
state granules in the same way as the spectral domain encoder in \cite{FT}.
Developed in another way \cite{KM6}, the encoder (\ref{enctdc}) is an estimator 
which corrects its sequence with a new estimate at each time $t$.

We have seen from Corollary \ref{cor10} (see also \cite{KM5}) that the \crep\ of the time domain
granule of $A$ can be chosen to be the same as the Forney-Trott generator of the spectral domain granule
of $A$.  In this case, the generator $\bmg^{[t-j,t-j+k]}$ in (\ref{enctdc}) can be the same
as the generator $\bmg_{FT}^{[t-j,t-j+k]}$ in (\ref{encftd}).
Then comparing (\ref{encftd}) and (\ref{enctdc}), we see that the time domain encoder
is just an interchange of the double product in the spectral domain encoder.  
If both (\ref{encftd}) and (\ref{enctdc}) use the same
generators and the alphabet group $A^t$ is abelian, then the interchange
does not matter and both encoders give the same output at time $t$.
If $A^t$ is nonabelian, this may not be the case.

The spectral domain encoder is a {\it minimal} encoder because there is
a bijection between the state granules at any time $t$ and the states of the canonical
realization \cite{FT}.  Since the time domain encoder can use the same generators as the
spectral domain encoder, the state granules of the time domain and spectral domain
encoder can be the same.  Therefore the time domain encoder is also a minimal encoder.

Fix time $t\in\bmcpz$.  For $0\le j\le\ell$, for $j\le k\le\ell$, let generator $\bmg^{[t-j,t-j+k]}$ 
be a representative in quotient group $\Lambda^{[t-j,t-j+k]}$.  We assume $\bmg^{[t-j,t-j+k]}$ has the
sequence of components
\be
\label{tps0}
\bmg^{[t-j,t-j+k]}=
\ldots,a_\bone^{t-j-2},a_\bone^{t-j-1},r_{0,k}^{t-j},r_{1,k}^{t-j},\ldots,r_{m,k}^{t-j},\ldots,r_{k,k}^{t-j},a_\bone^{t-j+k+1},a_\bone^{t-j+k+2},\ldots,
\ee
where $0\le m\le k$.  For $0\le m\le k$, the superscript $t-j$ in $r_{m,k}^{t-j}$ indicates 
that $r_{m,k}^{t-j}$ is in a generator $\bmg^{[t-j,t-j+k]}$ whose nontrivial components 
start at time $t-j$.  The component at time $t-j$ in $\bmg^{[t-j,t-j+k]}$ is $r_{0,k}^{t-j}$.
In general, the component $r_{m,k}^{t-j}$ is at time $t-j+m$.  Note that the component at time $t$ in 
generator $\bmg^{[t-j,t-j+k]}$ is $r_{j,k}^{t-j}$, which is some letter $a^t\in A^t$.

A transversal $[\Lambda^{[t-j,t-j+k]}]$ of $\Lambda^{[t-j,t-j+k]}$ is a selection of one
representative from each coset of (\ref{qgqg}).  We always take the \crep\ $\bmg^{[t-j,t-j+k]}$
of the denominator in (\ref{qgqg}) to be the identity $\bma_\bone$ of $A$.
Since $\chi^t(\bmg^{[t-j,t-j+k]})=r_{j,k}^{t-j}$, we let $\{r_{j,k}^{t-j}\}$
be the set of representatives at time $t$ in generators $\bmg^{[t-j,t-j+k]}$
in $[\Lambda^{[t-j,t-j+k]}]$.

\begin{lem}
\label{distinct}
Fix time $t\in\bmcpz$.  Fix $j$ such that $0\le j\le\ell$ and fix $k$ such that $j\le k\le\ell$.
There is a bijection $[\Lambda^{[t-j,t-j+k]}]\ra \{r_{j,k}^{t-j}\}$.
\end{lem}

\begin{prf}
Using (\ref{qgx}) and (\ref{qgx0}), (\ref{qgqg}) can be rewritten as
\be
\label{distinct1}
\Lambda^{[t-j,t-j+k]}
=\frac{X^{t-j+1}(X^{t-j}\cap Y^{t+k-j})}{X^{t-j+1}(X^{t-j}\cap Y^{t+k-j-1})(X^{t-j+1}\cap Y^{t+k-j})}.
\ee
We use proof by contradiction.
Suppose two generators $\bmg_1^{[t-j,t-j+k]}$ and $\bmg_2^{[t-j,t-j+k]}$ in set 
$[\Lambda^{[t-j,t-j+k]}]$ share the same representative in set $\{r_{j,k}^{t-j}\}$.  Then
there is a sequence $\bma^{[t-j,t-1]}$ in $A$ which is trivial outside the interval
$[t-j,t-1]$, and a sequence $\bma^{[t+1,t-j+k]}$ in $A$ which is trivial outside the interval
$[t+1,t-j+k]$, such that $\bmg_1^{[t-j,t-j+k]}=(\bma^{[t-j,t-1]}\bma^{[t+1,t-j+k]})\bmg_2^{[t-j,t-j+k]}$.
But $\bma^{[t-j,t-1]}$ is an element of denominator term $(X^{t-j}\cap Y^{t+k-j-1})$
in (\ref{distinct1}), and $\bma^{[t+1,t-j+k]}$
is an element of denominator term $(X^{t-j+1}\cap Y^{t+k-j})$ in (\ref{distinct1}).
Then the two generators differ by a sequence $(\bma^{[t-j,t-1]}\bma^{[t+1,t-j+k]})$ which is an
element in the denominator of (\ref{distinct1}).  Then both generators
must be in the same coset of (\ref{distinct1}), a contradiction.
\end{prf}

Since $\chi^t(\bmg^{[t-j,t-j+k]})=r_{j,k}^{t-j}$, we can write the time $t$
components of the generators in (\ref{smgent}) as the triangular matrix (\ref{rttf}).
%%%%%%%%%%%%%%%%%%%%%%%%%%%%%%%%%%%%%%%%%
\be
\label{rttf}
\begin{array}{llllllllll}
  r_{0,\ell}^t   & r_{1,\ell}^{t-1}   & \cdots & \cdots & r_{j,\ell}^{t-j}   & \cdots & \cdots & \cdots & r_{\ell-1,\ell}^{t-\ell+1}   & r_{\ell,\ell}^{t-\ell} \\
  r_{0,\ell-1}^t & r_{1,\ell-1}^{t-1} & \cdots & \cdots & r_{j,\ell-1}^{t-j} & \cdots & \cdots & \cdots & r_{\ell-1,\ell-1}^{t-\ell+1} & \\
  \vdots & \vdots & \vdots & \vdots & \vdots & \vdots & \vdots & \vdots && \\
  r_{0,k}^t & r_{1,k}^{t-1} & \cdots & \cdots & r_{j,k}^{t-j} & \cdots & r_{k,k}^{t-k} &&& \\
  \vdots & \vdots & \vdots & \vdots & \vdots & \vdots &&&& \\
  \cdots & \cdots & \cdots & \cdots & r_{j,j}^{t-j} &&&&& \\
  \vdots & \vdots & \vdots &&&&&&& \\
  r_{0,2}^t   & r_{1,2}^{t-1} & r_{2,2}^{t-2} &&&&&&& \\
  r_{0,1}^t   & r_{1,1}^{t-1} &&&&&&&& \\
  r_{0,0}^t   &&&&&&&&&
\end{array}
\ee
%%%%%%%%%%%%%%%%%%%%%%%%%%%%%%%%%%%%%%%%%
Since all the entries in $\triupjktarg{0}{0}{t}{\bmr}$ are alphabet letters in alphabet $A^t$, 
we call (\ref{rttf}) the {\it alphabet matrix}.
Then we see that each sequence $\bmr\in\calr$ gives a sequence of \atmxs\
\be
\label{seqstms}
\ldots,\triupjktarg{0}{0}{t}{\bmr},\triupjktarg{0}{0}{t-1}{\bmr},\ldots.
\ee
Let $\triupjktarg{0}{0}{t}{\calr}$ be the set of all \atmxs\ $\triupjktarg{0}{0}{t}{\bmr}$.
In other words, $\triupjktarg{0}{0}{t}{\calr}\rmdef\{\triupjktarg{0}{0}{t}{\bmr}:  \bmr\in\calr\}$.

We have written a sequence $\bma\in A$ as $\ldots,a^{t-1},a^t,\ldots$.
We have chosen to write the sequence of \atmxs\ (\ref{seqstms}) in reverse time order
since it reflects the order of indices in (\ref{smgen}), 
which reflects the order of indices in (\ref{sminf}).  It is a natural ordering
for this problem.  But writing the sequence in
reverse time order on a piece of paper does not change the physics of time:
the component at time epoch $t-1$ still occurs before the component at time epoch $t$,
as it does for $\bma$.

Using the \atmx\ $\triupjktarg{0}{0}{t}{\bmr}$, we can rewrite (\ref{enctdc}) in the equivalent form as
\begin{align}
a^t
\label{enctda}
&=\prod_{j=0}^\ell \left(\prod_{k=j}^\ell \chi^t(\bmg^{[t-j,t-j+k]})\right) \\
\label{enctd}
&=\prod_{j=0}^\ell \left(\prod_{k=j}^\ell r_{j,k}^{t-j}\right),
\end{align} 
where the inner product in parentheses in (\ref{enctd})  
is just the product of terms in the $j$-th column of $\triupjktarg{0}{0}{t}{\bmr}$.
By the convention used here, equation (\ref{enctd})
is evaluated as
\be
\label{enctd1}
a^t=r_{0,0}^tr_{0,1}^tr_{0,2}^t\cdots r_{0,\ell}^tr_{1,1}^{t-1}\cdots r_{1,\ell}^{t-1}r_{2,2}^{t-2}
\cdots r_{j,j}^{t-j}\cdots r_{j,k}^{t-j}\cdots r_{j,\ell}^{t-j}\cdots 
r_{\ell-1,\ell-1}^{t-\ell+1} r_{\ell-1,\ell}^{t-\ell+1} r_{\ell,\ell}^{t-\ell}.
\ee

Using (\ref{enctd}) we can give the following enhancement of Theorem \ref{thm9}.

\begin{thm}
\label{thm11}
Find a basis $\calb$ of $A$.  Then $\calr$ is chosen, and there is a 
bijection $\alpha:  \calr\ra A$ given by assignment $\alpha:  \bmr\mapsto\bma$, 
where component $a^t$ of $\bma$ is an encoding of the representatives in
$\triupjktarg{0}{0}{t}{\bmr}$ of $\bmr$ using (\ref{enctd}), for each $t\in\bmcpz$.
\end{thm}

For each $t\in\bmcpz$,
define a map $\alpha^t:  \triupjktarg{0}{0}{t}{\calr}\ra A^t$ by the assignment
$\alpha^t:  \triupjktarg{0}{0}{t}{\bmr}\mapsto a^t$ if $a^t$ is given by 
the map in encoders (\ref{enctda})-(\ref{enctd}).
In general we have $|\triupjktarg{0}{0}{t}{\calr}|>|A^t|$ so this is a many to one map.
This gives the following corollary to Theorem \ref{thm11}.

\begin{cor}
\label{cor11}
Find a basis $\calb$ of $A$.  Then $\calr$ is chosen, and there is a 
bijection $\alpha:  \calr\ra A$ given by assignment $\alpha:  \bmr\mapsto\bma$, 
where component $a^t$ of $\bma$ is given by the assignment
$\alpha^t:  \triupjktarg{0}{0}{t}{\bmr}\mapsto a^t$, for each $t\in\bmcpz$.
\end{cor}

The encoder in (\ref{enctda})-(\ref{enctd}) forms output $\bma$ from a sequence of 
generators selected from basis $\calb$, which forms a tensor $\bmr\in\calr$.  
For each time $t\in\bmcpz$, for each $k$ such that $0\le k\le\ell$, 
a single generator $\bmg^{[t,t+k]}$ is selected from set $[\Lambda^{[t,t+k]}]$.
Since $A$ is complete \cite{FT}, the set of tensors $\bmr\in\calr$ that can be selected in this way 
to form $A$ is the double Cartesian product
\be
\label{cprod1}
\bigotimes_{t=+\infty}^{t=-\infty} \bigotimes_{0\le k\le\ell} [\Lambda^{[t,t+k]}].
\ee

We have let $\bmr\in\calr$ be the infinite series of finite columns given by generators in
(\ref{smgen}).  For each $t\in\bmcpz$, for $0\le k\le\ell$, let $\bmr_g^{[t,t+k]}$ be the 
tensor $\bmr\in\calr$ when all the generators in (\ref{smgen}) other than $\bmg^{[t,t+k]}$ 
are trivial.  Then $\alpha:  \bmr_g^{[t,t+k]}\mapsto\bmg^{[t,t+k]}$.

\newpage
{\bf 4.  THE DECOMPOSITION GROUP AND GENERATOR GROUP}
\vspace{3mm}

\vspace{3mm}
{\bf 4.1  The decomposition group}%hhh
\vspace{3mm}

Let $\calb$ be a basis of $A$.  Let $\calr$ be the set of tensors
determined by $\calb$.  From Theorem \ref{thm9}, there is a 
bijection $\alpha:  \calr\ra A$ with assignment $\alpha:  \bmr\mapsto\bma$
if $\bma$ is an encoding of $\bmr$ using product (\ref{enctdx}) on the 
generators in $\bmr$.
We define an operation $*$ on $\calr$ by using the operation in $A$.
Let $\bmrdt,\bmrddt\in\calr$.  Let $\alpha:  \bmrdt\mapsto\bmadt$ and 
$\alpha:  \bmrddt\mapsto\bmaddt$.  Define an operation $*$ on $\calr$ by
\be
\label{eqno13}
\bmrdt*\bmrddt\rmdef\bmrbr
\ee
if $\alpha:  \bmrbr\mapsto\bmadt\bmaddt$.

\begin{lem}
\label{lem37}
The operation $*$ is well defined.
\end{lem}

\begin{prf}
Let $\bmrgr,\bmrac\in\calr$ such that $\bmrgr=\bmrdt$ and $\bmrac=\bmrddt$.
We have to show that $\bmrdt*\bmrddt=\bmrgr*\bmrac$.  But if $\bmrgr=\bmrdt$,
then $\alpha:  \bmrgr\mapsto\bmadt$, and similarly $\alpha:  \bmrac\mapsto\bmaddt$.
Then both $\bmrdt*\bmrddt$ and $\bmrgr*\bmrac$ are determined by $\bmadt\bmaddt$.
\end{prf}

\begin{thm}
\label{thm31}
The set $\calr$ with operation $*$ forms a group $(\calr,*)$,
and $A\simeq(\calr,*)$ under the bijection $\alpha:  \calr\ra A$.
\end{thm}

\begin{prf}
We first show the operation $*$ is associative.  Let 
$\bmr,\bmrdt,\bmrddt\in\calr$.  We need to show
\be
\label{show0}
(\bmr*\bmrdt)*\bmrddt=\bmr*(\bmrdt*\bmrddt).
\ee
Let $\alpha:  \bmr\mapsto\bma$, $\alpha:  \bmrdt\mapsto\bmadt$, and
$\alpha:  \bmrddt\mapsto\bmaddt$.  Then (\ref{show0}) is the same as showing
$$
(\bma\bmadt)\bmaddt=\bma(\bmadt\bmaddt).
$$
But this follows since operation in $A$ is associative.

Let $\bma_\bone$ be the identity of $A$.  Let $\alpha:  \bmr_\bone\mapsto\bma_\bone$.
We show $\bmr_\bone$ is the identity of $(\calr,*)$.  Let $\bmr\in\calr$.
We need to show $\bmr_\bone*\bmr=\bmr$ and $\bmr*\bmr_\bone=\bmr$.
But this is the same as showing $\bma_\bone\bma=\bma$ and $\bma\bma_\bone=\bma$, 
where $\alpha:  \bmr\mapsto\bma$.  But this follows since $\bma_\bone$ is the identity 
of $A$.

Let $\bmr\in\calr$.  We show $\bmr$ has an inverse in $(\calr,*)$.
Let $\alpha:  \bmr\mapsto\bma$.  The group element $\bma$ has an inverse
$\bmabr$ in $A$ such that $\bmabr\bma=\bma_\bone$
and $\bma\bmabr=\bma_\bone$.  Let $\alpha:  \bmrbr\mapsto\bmabr$.
It follows that $\bmrbr*\bmr=\bmr_\bone$ and $\bmr*\bmrbr=\bmr_\bone$.

Together these results show that $(\calr,*)$ is a group.
By the definition of operation $*$ given in (\ref{eqno13}),
$(\calr,*)$ is just an isomorphic copy of
$A$ under bijection $\alpha$.
\end{prf}
If $\alpha:  \bmr\mapsto\bma$, then $\bmr$ is the decomposition of $\bma$ into its generators.
For this reason, we call $(\calr,*)$ the {\it decomposition group} of $A$.

In a sense,
the decomposition group $(\calr,*)$ shows the permutation of the generators when
sequences in $A$ are multiplied.  However the group $(\calr,*)$ is a global group defined on
sequences $\bmr$ in $\calr$.  It is not clear that the permutation of the generators in the 
infinite sequences $\bmr\in\calr$ under multiplication determines a component group 
on set $\triupjktarg{0}{0}{t}{\calr}$ for each $t\in\bmcpz$.  We show that
$(\calr,*)$ has a component group in Subsection 4.4.

\vspace{3mm}
{\bf 4.2  The generator group}%hhh
\vspace{3mm}

%We are using tnsr129 in the following; also see tnsr117, they are basically the same.

First we discuss how to label each generator in basis $\calb$ uniquely.
The time index $t$ and length $k$ specify each generator $\bmg^{[t,t+k]}$ 
in basis $\calb$ uniquely up to the set of generators of the same length $k$ 
and time $t$.  The generators are \creps\ in the normal chain (\ref{sminf}) of $A$,
and therefore they must be unique.  The set of generators of the same length $k$ 
and time $t$ are the representatives in $[\Lambda^{[t,t+k]}]$.
For generators of the same length $k$ and time $t$,
we define a set of unique identifiers $G_k^t\rmdef\{g_{k}^t\}$.
We call each $g_{k}^t$ in set $G_k^t$ a {\it generator label}.
Then there is a bijection $[\Lambda^{[t,t+k]}]\ra G_k^t$ with assignment
$\bmg^{[t,t+k]}\mapsto g_k^t$.

Previously we constructed a tensor set $\calr$.  A tensor $\bmr\in\calr$
is equivalent to a collection of generators $\bmg^{[t,t+k]}$ for 
$0\le k\le\ell$, for each $t\in\bmcpz$, as shown by the nontrivial generators in
the middle rows of (\ref{smgen}).  For $0\le k\le\ell$, for each $t\in\bmcpz$, 
we now replace each generator $\bmg^{[t,t+k]}$ in $\bmr\in\calr$ 
with a single generator label $g_{k}^t$, where
$\bmg^{[t,t+k]}\mapsto g_k^t$ in the bijection $[\Lambda^{[t,t+k]}]\ra G_k^t$. 
Under the assignment $\bmg^{[t,t+k]}\mapsto g_{k}^t$
for $0\le k\le\ell$, for each $t\in\bmcpz$, a tensor $\bmr\in\calr$, as shown by the
middle rows of (\ref{smgen}), becomes a tensor $\bmu$ as shown in (\ref{ugttf}).
The tensor $\bmu$ has the same time reverse ordering as $\bmr$.
Let $\calu$ be the set of tensors $\bmu$ obtained from $\calr$ this way.
This gives the following.

\begin{lem}
\label{lem57}
There is a bijection $\beta:  \calr\ra\calu$ given by the
assignment $\beta:  \bmr\mapsto\bmu$ where each generator
$\bmg^{[t,t+k]}$ in $\bmr$ is replaced by a single \glab\ $g_{k}^t$ in $\bmu$, for 
$0\le k\le\ell$, for each $t\in\bmcpz$, where
$\bmg^{[t,t+k]}\mapsto g_k^t$ in the bijection $[\Lambda^{[t,t+k]}]\ra G_k^t$.
\end{lem}
Using (\ref{cprod1}) with bijections $\beta: \calr\ra\calu$ and $[\Lambda^{[t,t+k]}]\ra G_k^t$, 
set $\calu$ is just the double Cartesian product
\be
\label{input4}
\bigotimes_{t=+\infty}^{t=-\infty} \bigotimes_{0\le k\le\ell} G_k^t.
\ee

%%%%%%%%%%%%%%%%%%%%%%%%%%%%%%%%%%%%%%%%%
\be
\label{ugttf}
\begin{array}{llllllll}
        &  g_{\ell}^t   & g_{\ell}^{t-1}   & \cdots & g_{\ell}^{t-j}   & \cdots & g_{\ell}^{t-\ell}   & \\
        &  g_{\ell-1}^t & g_{\ell-1}^{t-1} & \cdots & g_{\ell-1}^{t-j} & \cdots & g_{\ell-1}^{t-\ell} & \\
        &  \cdots       & \cdots           &        & \cdots           &        & \cdots              & \\
        &  g_{k}^t      & g_{k}^{t-1}      & \cdots & g_{k}^{t-j}      & \cdots & g_{k}^{t-\ell}      & \\
        &  g_{k-1}^t    & g_{k-1}^{t-1}    & \cdots & g_{k-1}^{t-j}    & \cdots & g_{k-1}^{t-\ell}    & \\
\cdots  &  \cdots       & \cdots           & \cdots & \cdots           & \cdots & \cdots & \cdots       \\
        &  g_{j}^t      & g_{j}^{t-1}      & \cdots & g_{j}^{t-j}      & \cdots & g_{j}^{t-\ell}      & \\
        &  g_{j-1}^t    & g_{j-1}^{t-1}    & \cdots & g_{j-1}^{t-j}    & \cdots & g_{j-1}^{t-\ell}    & \\
        &  \cdots       & \cdots           &        & \cdots           &        & \cdots              & \\
        &  g_{1}^t      & g_{1}^{t-1}      & \cdots & g_{1}^{t-j}      & \cdots & g_{1}^{t-\ell}      & \\
        &  g_{0}^t      & g_{0}^{t-1}      & \cdots & g_{0}^{t-j}      & \cdots & g_{0}^{t-\ell}      &
\end{array}
\ee   
%%%%%%%%%%%%%%%%%%%%%%%%%%%%%%%%%%%%%%%%%

We say an element $\bmu\in\calu$ is a nontrivial {\it generator} 
$\bmu_{g,k}^t$ of $(\calu,\circ)$ if $\bmu_{g,k}^t$
contains one and only one nontrivial \glab\ $g_k^t$ for some $k$ such that $0\le k\le\ell$ 
and some time $t\in\bmcpz$.  For each $k$ such that $0\le k\le\ell$ 
and each $t\in\bmcpz$, we always assume there is a trivial generator $\bmu_{g,k}^t$
of $(\calu,\circ)$ which is the identity $\bmu_\bone$ of $(\calu,\circ)$.
Under the bijection $\beta:  \calr\ra\calu$, a generator 
$\bmr_g^{[t,t+k]}$ of $(\calr,*)$ is mapped to a generator $\bmu_{g,k}^t$ of $(\calu,\circ)$, or 
$\beta:  \bmr_g^{[t,t+k]}\mapsto\bmu_{g,k}^t$.

We now develop a compressed version of $(\calr,*)$ called $(\calu,\circ)$.
The operation $*$ in $(\calr,*)$ determines an operation $\circ$ on
$\calu$.  Let $\bmudt,\bmuddt\in\calu$.  Let $\beta:  \bmrdt\mapsto\bmudt$ and
$\beta:  \bmrddt\mapsto\bmuddt$.  Define an operation $\circ$ on $\calu$ by
\be
\label{eqno23}
\bmudt\circ\bmuddt\rmdef\bmubr
\ee
if $\beta:  \bmrdt*\bmrddt\mapsto\bmubr$.

\begin{lem}
The operation $\circ$ is well defined.
\end{lem}

\begin{prf}
The proof is similar to the proof of Lemma \ref{lem37}.
\end{prf}

\begin{thm}
\label{thm88a}
The set $\calu$ with operation $\circ$ forms a group $(\calu,\circ)$,
and $(\calr,*)\simeq(\calu,\circ)$ under the bijection 
$\beta:  \calr\ra\calu$.
\end{thm}

\begin{prf}
The proof is similar to the proof of Theorem \ref{thm31}.
\end{prf}

We know that each $\bmr\in\calr$ in group $(\calr,*)$ corresponds to a sequence of generators.
We see that $\bmu\in\calu$ in group $(\calu,\circ)$ demonstrates this sequence 
using a single generator label in place of each generator.  For this reason we call 
$(\calu,\circ)$ the {\it generator group} of $A$.

\vspace{3mm}
{\bf 4.3  The elementary groups of $(\calu,\circ)$}%hhh
\vspace{3mm}

In this subsection we define a component group on $(\calu,\circ)$.  We first define a
matrix $\triupjktarg{0}{0}{t}{\bmu}$ in $\bmu$ which is congruent
in shape to \atmx\ $\triupjktarg{0}{0}{t}{\bmr}$ in $\bmr$.  Given $\bmu$ in (\ref{ugttf}),
we define $\triupjktarg{0}{0}{t}{\bmu}$ to be the triangle in $\bmu$ with lower
vertex $g_0^t$ and upper vertices $g_\ell^t$ and $g_\ell^{t-\ell}$,
as shown in (\ref{rcttf}).  Note that all the entries in \atmx\ $\triupjktarg{0}{0}{t}{\bmr}$
are at the same time $t$, but in matrix $\triupjktarg{0}{0}{t}{\bmu}$, 
all entries in the same column are at the same time, but entries in different columns are
at different times.
%%%%%%%%%%%%%%%%%%%%%%%%%%%%%%%%%%%%%%%%%
\be
\label{rcttf}
\begin{array}{llllllllll}
  g_\ell^t       & g_\ell^{t-1}       & \cdots & \cdots & g_\ell^{t-j}     & \cdots & \cdots        & \cdots & g_\ell^{t-\ell+1}     & g_\ell^{t-\ell} \\
  g_{\ell-1}^t   & g_{\ell-1}^{t-1}   & \cdots & \cdots & g_{\ell-1}^{t-j} & \cdots & \cdots        & \cdots & g_{\ell-1}^{t-\ell+1} & \\
  \vdots         & \vdots             & \vdots & \vdots & \vdots           & \vdots & \vdots        & \vdots && \\
  g_k^t          & g_k^{t-1}          & \cdots & \cdots & g_k^{t-j}        & \cdots & g_k^{t-k}     &&& \\
  \vdots         & \vdots             & \vdots & \vdots & \vdots           & \vdots &&&& \\
  \cdots         & \cdots             & \cdots & \cdots & g_j^{t-j}        &&&&& \\
  \vdots         & \vdots             & \vdots &&&&&&& \\
  g_2^t          & g_2^{t-1}          & g_2^{t-2} &&&&&&& \\
  g_1^t          & g_1^{t-1}          &&&&&&&& \\
  g_0^t                               &&&&&&&&&
\end{array}
\ee  
%%%%%%%%%%%%%%%%%%%%%%%%%%%%%%%%%%%%%%%%%
Note that we can think of tensor $\bmu$ in (\ref{ugttf}) as a sequence of
matrices $\ldots,\triupjktarg{0}{0}{t}{\bmu},\triupjktarg{0}{0}{t-1}{\bmu},\ldots$ which overlap
to form $\bmu$.  Define $\triupjktarg{0}{0}{t}{\calu}$ to be the set
$\{\triupjktarg{0}{0}{t}{\bmu}:  \bmu\in\calu\}$.

Define a mapping $\beta^t:  \triupjktarg{0}{0}{t}{\calr}\ra\triupjktarg{0}{0}{t}{\calu}$
given by assignment $\beta^t:  \triupjktarg{0}{0}{t}{\bmr}\mapsto\triupjktarg{0}{0}{t}{\bmu}$
if $\beta:  \bmr\mapsto\bmu$ in the bijection $\beta:  \calr\ra\calu$.

\begin{thm}
\label{thm66}
For each $t\in\bmcpz$, the map $\beta^t:  \triupjktarg{0}{0}{t}{\calr}\ra\triupjktarg{0}{0}{t}{\calu}$
is a bijection.
\end{thm}

\begin{prf}
We first show the map $\beta^t$ is well defined.  Let $\bmr,\bmrdt\in\calr$ such that
$\triupjktarg{0}{0}{t}{\bmr}=\triupjktarg{0}{0}{t}{\bmrdt}$.  Let $\beta:  \bmr\mapsto\bmu$ 
and $\beta:  \bmrdt\mapsto\bmudt$.  Then 
$\beta^t: \triupjktarg{0}{0}{t}{\bmr}\mapsto\triupjktarg{0}{0}{t}{\bmu}$
and $\beta^t: \triupjktarg{0}{0}{t}{\bmrdt}\mapsto\triupjktarg{0}{0}{t}{\bmudt}$.
We need to show $\triupjktarg{0}{0}{t}{\bmu}=\triupjktarg{0}{0}{t}{\bmudt}$.

If $\triupjktarg{0}{0}{t}{\bmr}=\triupjktarg{0}{0}{t}{\bmrdt}$, then 
for each $j$, $0\le j\le\ell$, and each $k$, $j\le k\le\ell$, the representative 
$r_{j,k}^{t-j}$ in $\triupjktarg{0}{0}{t}{\bmr}$ in (\ref{rttf}) is the same as the representative 
$\rdt_{j,k}^{t-j}$ in $\triupjktarg{0}{0}{t}{\bmrdt}$.  Since from 
Lemma \ref{distinct}, for each $j$, $0\le j\le\ell$, and each $k$, $j\le k\le\ell$,
there is a bijection $[\Lambda^{[t-j,t-j+k]}]\ra \{r_{j,k}^{t-j}\}$, then for these
same indices, generators
$\bmg^{[t-j,t-j+k]}$ in $\bmr$ and generators $\bmgdt^{[t-j,t-j+k]}$ in $\bmrdt$
must be the same.  Therefore the mapping $\beta:  \bmr\mapsto\bmu$ and $\beta:  \bmrdt\mapsto\bmudt$
must give $\triupjktarg{0}{0}{t}{\bmu}=\triupjktarg{0}{0}{t}{\bmudt}$.

We now show the map $\beta^t$ is 1-1.
Suppose $\triupjktarg{0}{0}{t}{\bmu}=\triupjktarg{0}{0}{t}{\bmudt}$.  
Let $\beta:  \bmr\mapsto\bmu$ and $\beta:  \bmrdt\mapsto\bmudt$.  We need to show 
$\triupjktarg{0}{0}{t}{\bmr}=\triupjktarg{0}{0}{t}{\bmrdt}$.  If
$\triupjktarg{0}{0}{t}{\bmu}=\triupjktarg{0}{0}{t}{\bmudt}$, then
for each $j$, $0\le j\le\ell$, and each $k$, $j\le k\le\ell$, the \glab\
$g_k^{t-j}$ in $\triupjktarg{0}{0}{t}{\bmu}$ in (\ref{rcttf}) is the same as the \glab\
$\gdt_k^{t-j}$ in $\triupjktarg{0}{0}{t}{\bmudt}$.  Since there is a bijection
$[\Lambda^{[t,t+k]}]\ra G_k^t$ for each $t\in\bmcpz$ and each $k$, $0\le k\le\ell$, then
for each $j$, $0\le j\le\ell$, and each $k$, $j\le k\le\ell$, generators
$\bmg^{[t-j,t-j+k]}$ in $\bmr$ and generators $\bmgdt^{[t-j,t-j+k]}$ in $\bmrdt$
must be the same.  But this means $\triupjktarg{0}{0}{t}{\bmr}=\triupjktarg{0}{0}{t}{\bmrdt}$.

We now show the map $\beta^t$ is onto $\triupjktarg{0}{0}{t}{\calu}$.  Let
$\triupjktarg{0}{0}{t}{\bmu}$ be any element of $\triupjktarg{0}{0}{t}{\calu}$.  
It is clear there is an $\bmr\in\calr$ such that $\beta:  \bmr\ra\bmu$.  But if
$\beta:  \bmr\ra\bmu$, then we have
$\beta^t:  \triupjktarg{0}{0}{t}{\bmr}\mapsto\triupjktarg{0}{0}{t}{\bmu}$
by definition.  Therefore $\beta^t$ is onto.  Therefore $\beta^t$ is a bijection 
$\triupjktarg{0}{0}{t}{\calr}\ra\triupjktarg{0}{0}{t}{\calu}$.
\end{prf}
We use this result in Subsection 4.4 to define a component group of $(\calr,*)$.

We now define a local group of $(\calu,\circ)$.
Recall that for each $t\in\bmcpz$, we have defined $X^t$ to be the set of all sequences 
$\bma$ in $A$ for which $a^n=a_\bone^n$ for $n<t$, where $a_\bone^n$ is the identity of $A^n$ 
at time $n$.  And for each $t\in\bmcpz$, we have defined $Y^t$ to be the set of all sequences $\bma$ 
in $A$ for which $a^n=a_\bone^n$ for $n>t$.  It is clear that $X^t\lhd A$ and $Y^t\lhd A$ for 
each $t\in\bmcpz$. 

Fix $t\in\bmcpz$.
In $\calr$, $X^t$ is the set of all $\bmr\in\calr$ with trivial generators $\bmg^{[t',t'+k]}$,
$0\le k\le\ell$, for all $t'\in\bmcpz$ except for $t'\ge t$.  Call this subset
of $\calr$ as $\calr^{t+}$.  Since $X^t$ is a normal subgroup of $A$,
then $(\calr^{t+},*)$ is a normal subgroup of $(\calr,*)$, under the bijection
$\alpha^{-1}$ of the isomorphism from $A$ to $(\calr,*)$.
In $\calu$, this is the set of all $\bmu$ with trivial \glabs\
$g_k^{t'}$, $0\le k\le\ell$, except for $t'\ge t$.  Call this subset
of $\calu$ as $\calu^{t+}$.  Since $(\calr^{t+},*)$ is a normal subgroup of $(\calr,*)$,
then $(\calu^{t+},\circ)$ is a normal subgroup of $(\calu,\circ)$, under the bijection
$\beta$ of the isomorphism from $(\calr,*)$ to $(\calu,\circ)$.
This shows that $(\calu^{t+},\circ)$ is the analog in $(\calu,\circ)$
of $X^t$ in $A$.

In $\calr$, $Y^t$ is the set of all $\bmr\in\calr$ with trivial generators $\bmg^{[t',t'+k]}$,
$0\le k\le\ell$, for all $t'\in\bmcpz$ except for $t'\le t-k$.  Call this subset
of $\calr$ as $\calr^{t-}$.  Since $Y^t$ is a normal subgroup of $A$,
then $(\calr^{t-},*)$ is a normal subgroup of $(\calr,*)$, under the bijection
$\alpha^{-1}$ of the isomorphism from $A$ to $(\calr,*)$.
In $\calu$, this is the set of all $\bmu$ with trivial \glabs\
$g_k^{t'}$, $0\le k\le\ell$, except for $t'\le t-k$.  Call this subset
of $\calu$ as $\calu^{t-}$.  Since $(\calr^{t-},*)$ is a normal subgroup of $(\calr,*)$,
then $(\calu^{t-},\circ)$ is a normal subgroup of $(\calu,\circ)$, under the bijection
$\beta$ of the isomorphism from $(\calr,*)$ to $(\calu,\circ)$.
This shows that $(\calu^{t-},\circ)$ is the analog in $(\calu,\circ)$
of $Y^t$ in $A$.

Note that $\calu^{(t+1)+}$ and $\calu^{(t-1)-}$ cover $\calu$ except for $\triupjktarg{0}{0}{t}{\bmu}$.
In other words the union of set $\calu^{(t+1)+}$ and set $\calu^{(t-1)-}$
is the set of all $\bmu\in\calu$ which are trivial 
in $\triupjktarg{0}{0}{t}{\bmu}$; call this subset of $\calu$ as $\triupjktargnop{0}{0}{t}{\calk}$.
Then the product of $(\calu^{(t+1)+},\circ)$ and $(\calu^{(t-1)-},\circ)$ is a normal subgroup
$\grpupjktargnop{0}{0}{t}{\calk}$ of $(\calu,\circ)$.
Then we have a quotient group $(\calu,\circ)/\grpupjktargnop{0}{0}{t}{\calk}$.  
The projection of the cosets in $(\calu,\circ)/\grpupjktargnop{0}{0}{t}{\calk}$ on set
$\triupjktarg{0}{0}{t}{\calu}$ is $\triupjktarg{0}{0}{t}{\calu}$.  
We can use $\triupjktarg{0}{0}{t}{\calu}$ to define a local group on $(\calu,\circ)$.

Fix time $t$.  Let
$\triupjktarg{0}{0}{t}{\bmudt},\triupjktarg{0}{0}{t}{\bmuddt}\in\triupjktarg{0}{0}{t}{\calu}$.
Define an operation $\prodjktarg{0}{0}{t}{\ccirc}$ on set $\triupjktarg{0}{0}{t}{\calu}$ by
\be
\label{defy}
\triupjktarg{0}{0}{t}{\bmudt}\prodjktarg{0}{0}{t}{\ccirc}\triupjktarg{0}{0}{t}{\bmuddt}\rmdef\triupjktarg{0}{0}{t}{\bmudt\circ\bmuddt}.
\ee

\begin{lem}
\label{lem62}
Fix time $t$.  The operation $\prodjktarg{0}{0}{t}{\ccirc}$ on set $\triupjktarg{0}{0}{t}{\calu}$ 
is well defined.
\end{lem}

\begin{prf}
Choose any $\bmugr,\bmuac\in\calu$ 
such that $\triupjktarg{0}{0}{t}{\bmugr}=\triupjktarg{0}{0}{t}{\bmudt}$ and 
$\triupjktarg{0}{0}{t}{\bmuac}=\triupjktarg{0}{0}{t}{\bmuddt}$.
To show the operation is well defined, we need to show
$$
\triupjktarg{0}{0}{t}{\bmugr}\prodjktarg{0}{0}{t}{\ccirc}\triupjktarg{0}{0}{t}{\bmuac}=
\triupjktarg{0}{0}{t}{\bmudt}\prodjktarg{0}{0}{t}{\ccirc}\triupjktarg{0}{0}{t}{\bmuddt},
$$
or what is the same, $\triupjktarg{0}{0}{t}{\bmugr\circ\bmuac}=\triupjktarg{0}{0}{t}{\bmudt\circ\bmuddt}$.
But if $\triupjktarg{0}{0}{t}{\bmugr}=\triupjktarg{0}{0}{t}{\bmudt}$, then
$\bmugr$ and $\bmudt$ are in the same coset of $(\calu,\circ)/\grpupjktargnop{0}{0}{t}{\calk}$.
And if $\triupjktarg{0}{0}{t}{\bmuac}=\triupjktarg{0}{0}{t}{\bmuddt}$, then
$\bmuac$ and $\bmuddt$ are in the same coset of $(\calu,\circ)/\grpupjktargnop{0}{0}{t}{\calk}$.
But then $\bmugr\circ\bmuac$ and $\bmudt\circ\bmuddt$ are in the same coset of
$(\calu,\circ)/\grpupjktargnop{0}{0}{t}{\calk}$.
Then $\triupjktarg{0}{0}{t}{\bmugr\circ\bmuac}=\triupjktarg{0}{0}{t}{\bmudt\circ\bmuddt}$.
\end{prf}

\begin{thm}
\label{thm63}
Fix time $t$.  The set $\triupjktarg{0}{0}{t}{\calu}$ with operation $\prodjktarg{0}{0}{t}{\ccirc}$ 
forms a group $\grpupjktarg{0}{0}{t}{\calu}{\ccirc}$.
\end{thm}

\begin{prf}
First we show the operation $\prodjktarg{0}{0}{t}{\ccirc}$ is associative.
Let $\triupjktarg{0}{0}{t}{\bmu},\triupjktarg{0}{0}{t}{\bmudt},\triupjktarg{0}{0}{t}{\bmuddt}\in\triupjktarg{0}{0}{t}{\calu}$.
We need to show
$$
(\triupjktarg{0}{0}{t}{\bmu}\prodjktarg{0}{0}{t}{\ccirc}\triupjktarg{0}{0}{t}{\bmudt})\prodjktarg{0}{0}{t}{\ccirc}\triupjktarg{0}{0}{t}{\bmuddt}
=\triupjktarg{0}{0}{t}{\bmu}\prodjktarg{0}{0}{t}{\ccirc}(\triupjktarg{0}{0}{t}{\bmudt}\prodjktarg{0}{0}{t}{\ccirc}\triupjktarg{0}{0}{t}{\bmuddt}).
$$
But using (\ref{defy}) we have
\begin{align*}
(\triupjktarg{0}{0}{t}{\bmu}\prodjktarg{0}{0}{t}{\ccirc}\triupjktarg{0}{0}{t}{\bmudt})\prodjktarg{0}{0}{t}{\ccirc}\triupjktarg{0}{0}{t}{\bmuddt}
&=\triupjktarg{0}{0}{t}{\bmu\circ\bmudt}\prodjktarg{0}{0}{t}{\ccirc}\triupjktarg{0}{0}{t}{\bmuddt} \\
&=\triupjktarg{0}{0}{t}{(\bmu\circ\bmudt)\circ\,\bmuddt},
\end{align*}
and
\begin{align*}
\triupjktarg{0}{0}{t}{\bmu}\prodjktarg{0}{0}{t}{\ccirc}(\triupjktarg{0}{0}{t}{\bmudt}\prodjktarg{0}{0}{t}{\ccirc}\triupjktarg{0}{0}{t}{\bmuddt})
&=\triupjktarg{0}{0}{t}{\bmu}\prodjktarg{0}{0}{t}{\ccirc}\triupjktarg{0}{0}{t}{\bmudt\circ\bmuddt} \\
&=\triupjktarg{0}{0}{t}{\bmu\circ\,(\bmudt\circ\bmuddt)}.
\end{align*}
Therefore the operation $\prodjktarg{0}{0}{t}{\ccirc}$ is associative since the operation $\circ$ in 
group $(\calu,\circ)$ is associative.

Let $\bmu_\bone$ be the identity of $(\calu,\circ)$.  We show
$\triupjktarg{0}{0}{t}{\bmu_\bone}$ is the identity of $\grpupjktarg{0}{0}{t}{\calu}{\ccirc}$.
Let $\bmu\in\calu$ and $\triupjktarg{0}{0}{t}{\bmu}\in\triupjktarg{0}{0}{t}{\calu}$.  
But using (\ref{defy}) we have
\begin{align*}
\triupjktarg{0}{0}{t}{\bmu_\bone}\prodjktarg{0}{0}{t}{\ccirc}\triupjktarg{0}{0}{t}{\bmu} &=\triupjktarg{0}{0}{t}{\bmu_\bone\circ\bmu} \\
&=\triupjktarg{0}{0}{t}{\bmu}
\end{align*}
and
\begin{align*}
\triupjktarg{0}{0}{t}{\bmu}\prodjktarg{0}{0}{t}{\ccirc}\triupjktarg{0}{0}{t}{\bmu_\bone} &=\triupjktarg{0}{0}{t}{\bmu\circ\bmu_\bone} \\
&=\triupjktarg{0}{0}{t}{\bmu}.
\end{align*}

Fix $\bmu\in\calu$.  Let $\bmubr$ be the inverse of $\bmu$ in $(\calu,\circ)$.
We show $\triupjktarg{0}{0}{t}{\bmubr}$ is the inverse of $\triupjktarg{0}{0}{t}{\bmu}$
in $\grpupjktarg{0}{0}{t}{\calu}{\ccirc}$.  But using (\ref{defy}) we have
\begin{align*}
\triupjktarg{0}{0}{t}{\bmubr}\prodjktarg{0}{0}{t}{\ccirc}\triupjktarg{0}{0}{t}{\bmu} &=\triupjktarg{0}{0}{t}{\bmubr\circ\bmu} \\
&=\triupjktarg{0}{0}{t}{\bmu_\bone}
\end{align*}
and
\begin{align*}
\triupjktarg{0}{0}{t}{\bmu}\prodjktarg{0}{0}{t}{\ccirc}\triupjktarg{0}{0}{t}{\bmubr} &=\triupjktarg{0}{0}{t}{\bmu\circ\bmubr} \\
&=\triupjktarg{0}{0}{t}{\bmu_\bone}.
\end{align*}
Together these results show $\grpupjktarg{0}{0}{t}{\calu}{\ccirc}$ is a group.
\end{prf}

\begin{cor}
\label{cor63a}
For each $t\in\bmcpz$, the group $\grpupjktarg{0}{0}{t}{\calu}{\ccirc}$ determined by $(\calu,\circ)$
is unique.
\end{cor}

\begin{prf}
From the proof of Lemma \ref{lem62}, for any
$\triupjktarg{0}{0}{t}{\bmu},\triupjktarg{0}{0}{t}{\bmudt}\in\triupjktarg{0}{0}{t}{\calu}$,
the operation $\triupjktarg{0}{0}{t}{\bmu}\prodjktarg{0}{0}{t}{\ccirc}\triupjktarg{0}{0}{t}{\bmudt}$
is unique.
\end{prf}
We say the group $\grpupjktarg{0}{0}{t}{\calu}{\ccirc}$, for each $t\in\bmcpz$, 
is an {\it upper elementary group} of $(\calu,\circ)$.

For each $t\in\bmcpz$, define a map $\theta^t:  \calu\ra\triupjktarg{0}{0}{t}{\calu}$ 
by the assignment $\theta^t:  \bmu\mapsto\triupjktarg{0}{0}{t}{\bmu}$.
Note that $\theta^t$ is {\bf not} a time projection since the \glabs\ in $\triupjktarg{0}{0}{t}{\bmu}$ 
occur at different times.  Therefore the map $\theta^t$ includes memory of the
\glabs\ in $\calu$.  In this sense, it is different from the projection
$\chi^t$ of $A$.

\begin{thm}
\label{homou}
For each $t\in\bmcpz$,
there is a homomorphism from $(\calu,\circ)$ to $\grpupjktarg{0}{0}{t}{\calu}{\ccirc}$ given by the map
$\theta^t:  \calu\ra\triupjktarg{0}{0}{t}{\calu}$ with assignment
$\theta^t:  \bmu\ra\triupjktarg{0}{0}{t}{\bmu}$.  Also there is an isomorphism
$(\calu,\circ)/\grpupjktargnop{0}{0}{t}{\calk}\simeq\grpupjktarg{0}{0}{t}{\calu}{\ccirc}$.
\end{thm}

\begin{prf}
The homomorphism follows immediately from the definition of operation $\prodjktarg{0}{0}{t}{\ccirc}$
in (\ref{defy}).  We have 
$(\calu,\circ)/\grpupjktargnop{0}{0}{t}{\calk}\simeq\grpupjktarg{0}{0}{t}{\calu}{\ccirc}$
since $\grpupjktargnop{0}{0}{t}{\calk}$ is the kernel of the homomorphism.
\end{prf}

We now show that the group $(\calu,\circ)$ and the collection of upper elementary groups 
$\grpupjktarg{0}{0}{t}{\calu}{\ccirc}$ contain redundant information in that the global operation 
in $(\calu,\circ)$ can be recovered from the collection of upper elementary groups.

\begin{lem}
\label{lem95a}
The product operation $\bmudt\circ\bmuddt$ in $(\calu,\circ)$ is uniquely determined by the evaluation of
\be
\label{eq100a}
\triupjktarg{0}{0}{t}{\bmudt}\prodjktarg{0}{0}{t}{\ccirc}\triupjktarg{0}{0}{t}{\bmuddt}
\ee
in group $\grpupjktarg{0}{0}{t}{\calu}{\ccirc}$, for $t=+\infty$ to $-\infty$.
\end{lem}

\begin{prf}
The product $\bmudt\circ\bmuddt$ is uniquely determined by the evaluation of 
$\triupjktarg{0}{0}{t}{\bmudt\circ\bmuddt}$ for $t=+\infty$ to $-\infty$.  
But by definition we have 
$$
\triupjktarg{0}{0}{t}{\bmudt\circ\bmuddt}=
\triupjktarg{0}{0}{t}{\bmudt}\prodjktarg{0}{0}{t}{\ccirc}\triupjktarg{0}{0}{t}{\bmuddt}
$$
for each $t\in\bmcpz$.
\end{prf}

Evidently the local information in the upper elementary group $\grpupjktarg{0}{0}{t}{\calu}{\ccirc}$,
for each $t\in\bmcpz$, is enough to determine the multiplication of the infinite sequences 
$\bmu\in\calu$.  This means $(\calu,\circ)$ is completely specified by local operations 
on local sets.

\vspace{3mm}
{\bf 4.4  The component group of $(\calr,*)$}%hhh
\vspace{3mm}

Fix time $t$.  Let
$\triupjktarg{0}{0}{t}{\bmrdt},\triupjktarg{0}{0}{t}{\bmrddt}\in\triupjktarg{0}{0}{t}{\calr}$.
Let $\beta:  \bmrdt\mapsto\bmudt$ and $\beta:  \bmrddt\mapsto\bmuddt$.
Define an operation $\prodjktarg{0}{0}{t}{\otimes}$ on set $\triupjktarg{0}{0}{t}{\calr}$ by
\be
\label{defxx}
\triupjktarg{0}{0}{t}{\bmrdt}\prodjktarg{0}{0}{t}{\otimes}\triupjktarg{0}{0}{t}{\bmrddt}\rmdef(\beta^t)^{-1}(\triupjktarg{0}{0}{t}{\bmudt\circ\bmuddt}).
\ee

\begin{lem}
%\label{lem62}
Fix time $t$.  The operation $\prodjktarg{0}{0}{t}{\otimes}$ on set $\triupjktarg{0}{0}{t}{\calr}$ 
is well defined.
\end{lem}

\begin{prf}
Choose any $\bmrgr,\bmrac\in\calr$ 
such that $\triupjktarg{0}{0}{t}{\bmrgr}=\triupjktarg{0}{0}{t}{\bmrdt}$ and 
$\triupjktarg{0}{0}{t}{\bmrac}=\triupjktarg{0}{0}{t}{\bmrddt}$.
Let $\beta:  \bmrgr\mapsto\bmugr$ and $\beta:  \bmrac\mapsto\bmuac$.
To show the operation is well defined, we need to show
$$
\triupjktarg{0}{0}{t}{\bmrgr}\prodjktarg{0}{0}{t}{\otimes}\triupjktarg{0}{0}{t}{\bmrac}=
\triupjktarg{0}{0}{t}{\bmrdt}\prodjktarg{0}{0}{t}{\otimes}\triupjktarg{0}{0}{t}{\bmrddt},
$$
or what is the same,
\be
\label{eq47}
(\beta^t)^{-1}(\triupjktarg{0}{0}{t}{\bmugr\circ\bmuac})=(\beta^t)^{-1}(\triupjktarg{0}{0}{t}{\bmudt\circ\bmuddt}).
\ee
But if $\triupjktarg{0}{0}{t}{\bmrgr}=\triupjktarg{0}{0}{t}{\bmrdt}$ and 
$\triupjktarg{0}{0}{t}{\bmrac}=\triupjktarg{0}{0}{t}{\bmrddt}$, then
$\triupjktarg{0}{0}{t}{\bmugr}=\triupjktarg{0}{0}{t}{\bmudt}$ and 
$\triupjktarg{0}{0}{t}{\bmuac}=\triupjktarg{0}{0}{t}{\bmuddt}$.
But the operation $\ccirc$ in group $\grpupjktarg{0}{0}{t}{\calu}{\ccirc}$
is well defined, so 
$$
\triupjktarg{0}{0}{t}{\bmugr}\prodjktarg{0}{0}{t}{\ccirc}\triupjktarg{0}{0}{t}{\bmuac}=
\triupjktarg{0}{0}{t}{\bmudt}\prodjktarg{0}{0}{t}{\ccirc}\triupjktarg{0}{0}{t}{\bmuddt}.
$$
This gives $\triupjktarg{0}{0}{t}{\bmugr\circ\bmuac}=\triupjktarg{0}{0}{t}{\bmudt\circ\bmuddt}$,
which proves (\ref{eq47}).
\end{prf}

Fix time $t$.  Let
$\triupjktarg{0}{0}{t}{\bmrdt},\triupjktarg{0}{0}{t}{\bmrddt}\in\triupjktarg{0}{0}{t}{\calr}$.
Define an operation $\prodjktarg{0}{0}{t}{\cast}$ on set $\triupjktarg{0}{0}{t}{\calr}$ by
\be
\label{defx}
\triupjktarg{0}{0}{t}{\bmrdt}\prodjktarg{0}{0}{t}{\cast}\triupjktarg{0}{0}{t}{\bmrddt}\rmdef\triupjktarg{0}{0}{t}{\bmrdt*\bmrddt}.
\ee

\begin{lem}
\label{lem32}
Fix time $t$.  The operation $\prodjktarg{0}{0}{t}{\cast}$ on set $\triupjktarg{0}{0}{t}{\calr}$ 
is well defined.
\end{lem}

\begin{prf}
Since $\beta^t:  \triupjktarg{0}{0}{t}{\calr}\mapsto\triupjktarg{0}{0}{t}{\calu}$
is a bijection, definition (\ref{defxx}) is equivalent to definition (\ref{defx}).
\end{prf}

\begin{thm}
\label{thm41}
Fix time $t$.  The set $\triupjktarg{0}{0}{t}{\calr}$ with operation $\prodjktarg{0}{0}{t}{\cast}$ 
forms a group $\grpupjktarg{0}{0}{t}{\calr}{\cast}$.
\end{thm}

\begin{prf}
The proof is exactly the same as for Theorem \ref{thm63} with a change of notation.
\end{prf}

\begin{cor}
\label{cor63aa}
For each $t\in\bmcpz$, the group $\grpupjktarg{0}{0}{t}{\calr}{\cast}$ determined by $(\calr,*)$
is unique.
\end{cor}

\begin{prf}
Let $\bmrdt,\bmrddt\in\calr$.  Fix time $t$.  Let
$\triupjktarg{0}{0}{t}{\bmrdt},\triupjktarg{0}{0}{t}{\bmrddt}\in\triupjktarg{0}{0}{t}{\calr}$.
Choose any $\bmrgr,\bmrac\in\calr$ 
such that $\triupjktarg{0}{0}{t}{\bmrgr}=\triupjktarg{0}{0}{t}{\bmrdt}$ and 
$\triupjktarg{0}{0}{t}{\bmrac}=\triupjktarg{0}{0}{t}{\bmrddt}$.
From Lemma \ref{lem32}, since the operation $\prodjktarg{0}{0}{t}{\cast}$ is well defined, we have
$$
\triupjktarg{0}{0}{t}{\bmrgr}\prodjktarg{0}{0}{t}{\cast}\triupjktarg{0}{0}{t}{\bmrac}=
\triupjktarg{0}{0}{t}{\bmrdt}\prodjktarg{0}{0}{t}{\cast}\triupjktarg{0}{0}{t}{\bmrddt},
$$
This means for any 
$\triupjktarg{0}{0}{t}{\bmrdt},\triupjktarg{0}{0}{t}{\bmrddt}\in\triupjktarg{0}{0}{t}{\calr}$,
the operation $\triupjktarg{0}{0}{t}{\bmrdt}\prodjktarg{0}{0}{t}{\cast}\triupjktarg{0}{0}{t}{\bmrddt}$
is unique.  Therefore the group $\grpupjktarg{0}{0}{t}{\calr}{\cast}$ is unique.
\end{prf}
We say the group $\grpupjktarg{0}{0}{t}{\calr}{\cast}$, for each $t\in\bmcpz$, 
is a {\it component group} of $(\calr,*)$.

\begin{cor}
\label{cor65}
For each $t\in\bmcpz$,
there is an isomorphism $\grpupjktarg{0}{0}{t}{\calr}{\cast}\simeq\grpupjktarg{0}{0}{t}{\calu}{\ccirc}$ 
under the bijection $\beta^t:  \triupjktarg{0}{0}{t}{\calr}\ra\triupjktarg{0}{0}{t}{\calu}$ with
assignment $\beta^t:  \triupjktarg{0}{0}{t}{\bmr}\ra\triupjktarg{0}{0}{t}{\bmu}$
if $\beta:  \bmr\mapsto\bmu$ under the bijection $\calr\ra\calu$.
\end{cor}

\begin{prf}
Let $\bmrdt,\bmrddt\in\calr$.
Let $\beta:  \bmrdt\mapsto\bmudt$ and $\beta:  \bmrddt\mapsto\bmuddt$ under the
bijection $\calr\ra\calu$.  Then 
$\triupjktarg{0}{0}{t}{\bmrdt}\mapsto\triupjktarg{0}{0}{t}{\bmudt}$ and
$\triupjktarg{0}{0}{t}{\bmrddt}\mapsto\triupjktarg{0}{0}{t}{\bmuddt}$.
We have to show that
$$
\triupjktarg{0}{0}{t}{\bmrdt}\prodjktarg{0}{0}{t}{\cast}\triupjktarg{0}{0}{t}{\bmrddt}\mapsto
\triupjktarg{0}{0}{t}{\bmudt}\prodjktarg{0}{0}{t}{\ccirc}\triupjktarg{0}{0}{t}{\bmuddt}.
$$
But this is the same as showing
\be
\label{assign3a}
\triupjktarg{0}{0}{t}{\bmrdt*\bmrddt}\mapsto\triupjktarg{0}{0}{t}{\bmudt\circ\bmuddt}.
\ee
But if $\beta:  \bmrdt\mapsto\bmudt$ and $\beta:  \bmrddt\mapsto\bmuddt$, then
$\beta:  \bmrdt*\bmrddt\mapsto\bmudt\circ\bmuddt$, and so (\ref{assign3a}) holds.
\end{prf}

For each $t\in\bmcpz$,
define a map $\phi^t:  \calr\ra\triupjktarg{0}{0}{t}{\calr}$ by the assignment 
$\phi^t:  \bmr\mapsto\triupjktarg{0}{0}{t}{\bmr}$.  The map $\phi^t$ is the
projection of $(\calr,*)$ at time $t$.  Note that $\phi^t(\calr)=\triupjktarg{0}{0}{t}{\calr}$.

\begin{thm}
\label{homor}
There is a homomorphism from $(\calr,*)$ to $\grpupjktarg{0}{0}{t}{\calr}{\cast}$ given 
by the map $\phi^t:  \calr\ra\triupjktarg{0}{0}{t}{\calr}$, for each $t\in\bmcpz$.
\end{thm}

\begin{prf}
This follows immediately from the definition of operation $\prodjktarg{0}{0}{t}{\cast}$ in (\ref{defx}).
\end{prf}

\begin{thm}
\label{homor1}
There is a homomorphism from $\grpupjktarg{0}{0}{t}{\calr}{\cast}$ to $A^t$ given 
by the map $\alpha^t$, for each $t\in\bmcpz$.
\end{thm}

\begin{prf}
Let $\alpha:  \bmrdt\mapsto\bmadt$ and $\alpha:  \bmrddt\mapsto\bmaddt$.  Then
$\alpha:  \bmrdt*\bmrddt\mapsto\bmadt\bmaddt$.
From Corollary \ref{cor11}, we have
$\alpha^t:  \triupjktarg{0}{0}{t}{\bmrdt}\mapsto\adt^t$ and
$\alpha^t:  \triupjktarg{0}{0}{t}{\bmrddt}\mapsto\addt^t$.  Let
$\bmabr\rmdef\bmadt\bmaddt$.  Then $\abr^t=\adt^t\addt^t$.  Then from Corollary \ref{cor11}, we have
$\alpha^t:  \triupjktarg{0}{0}{t}{\bmrdt*\bmrddt}\mapsto\adt^t\addt^t$.
But by definition $\triupjktarg{0}{0}{t}{\bmrdt*\bmrddt}=
\triupjktarg{0}{0}{t}{\bmrdt}\prodjktarg{0}{0}{t}{\cast}\triupjktarg{0}{0}{t}{\bmrddt}$.  Then
$\alpha^t:  \triupjktarg{0}{0}{t}{\bmrdt}\prodjktarg{0}{0}{t}{\cast}\triupjktarg{0}{0}{t}{\bmrddt}\mapsto\adt^t\addt^t$,
where $\alpha^t:  \triupjktarg{0}{0}{t}{\bmrdt}\mapsto\adt^t$ and
$\alpha^t:  \triupjktarg{0}{0}{t}{\bmrddt}\mapsto\addt^t$.
\end{prf}

The following lemma is proven in the same way as Lemma \ref{lem95a}.

\begin{lem}
\label{lem95aa}
The product operation $\bmrdt*\bmrddt$ in $(\calr,*)$ is uniquely determined by the evaluation of
\be
\label{eq100aa}
\triupjktarg{0}{0}{t}{\bmrdt}\prodjktarg{0}{0}{t}{\cast}\triupjktarg{0}{0}{t}{\bmrddt}
\ee
for $t=+\infty$ to $-\infty$.
\end{lem}

We have found a component group $\grpupjktarg{0}{0}{t}{\calr}{\cast}$ of $(\calr,*)$
from an elementary group of $(\calu,\circ)$.
Evidently the local information in the component group,
for each $t\in\bmcpz$, is enough to determine the multiplication of the infinite sequences 
$\bmr\in\calr$.  The component group gives
the local permutation of the generators in $\calr$ when sequences
in $A$ with these generators are multiplied.
The component group $\grpupjktarg{0}{0}{t}{\calr}{\cast}$ resembles the branch
group of a group trellis discussed in \cite{FT}, but is not the same.  The branch
group is obtained by a brute force determination of the state group of the system \cite{FT},
but the group $\grpupjktarg{0}{0}{t}{\calr}{\cast}$ is obtained from permutations 
of the local generators active at time $t$ when sequences in $A$ are multiplied.

Collecting Theorems \ref{thm31} and \ref{thm88a}, we have the following result.

\begin{thm}
\label{homo7}
There is an isomorphism from $A$ to $(\calu,\circ)$ given by the composition map
$\xi\rmdef\beta\bullet\alpha^{-1}:  A \ra\calu$,
where isomorphism $A\simeq (\calr,*)$ is given by the bijection $\alpha:  \calr\ra A$
in Theorem \ref{thm31}, and isomorphism $(\calr,*)\simeq(\calu,\circ)$ is given by the bijection 
$\beta:  \calr\ra\calu$ in Theorem \ref{thm88a}.
\end{thm}
Then we can summarize the results of this paper so far by the chain
%%%%%%%%%%%%%%%%%%%%%%%%%%%%%%%%%%%%%%%%%
\be
\label{chain7a}
\begin{array}{lllll}
A   & {\stackrel{\simeq}{\ra}}   & (\calr,*) & {\stackrel{\simeq}{\ra}}   & (\calu,\circ)    
\end{array}
\ee
%%%%%%%%%%%%%%%%%%%%%%%%%%%%%%%%%%%%%%%%%
where $A\simeq (\calr,*)$ and $(\calr,*)\simeq (\calu,\circ)$ under bijections
$\alpha:  \calr\ra A$ and $\beta:  \calr\ra\calu$, respectively.

Note that $(\calr,*)$ is not a group system but it is easily made into one.
Since $(\calr,*)$ has a component group,
we can make group $(\calr,*)$ into a group system in a natural way.
All the entries in $\triupjktarg{0}{0}{t}{\calr}$ occur at time $t$.
Therefore it is natural to define a group system by the assignment
\be
\label{assignr}
\bmr\mapsto\ldots,\triupjktarg{0}{0}{t}{\bmr},\triupjktarg{0}{0}{t-1}{\bmr},\ldots.
\ee
It is easy to see the resulting set of sequences with a componentwise group addition
defined by $\grpupjktarg{0}{0}{t}{\calr}{\cast}$ is a group system.

In a similar way, note that $(\calu,\circ)$ is a group but it is not a group system.
However it has elementary groups $\grpupjktarg{0}{0}{t}{\calu}{\ccirc}$ which overlap in time.
If we separate the elementary groups in time using the map
\be
\label{assignu}
\bmu\mapsto\ldots,\triupjktarg{0}{0}{t}{\bmu},\triupjktarg{0}{0}{t-1}{\bmu},\ldots,
\ee
then the resulting set of sequences is a group system with component groups
$\grpupjktarg{0}{0}{t}{\calu}{\ccirc}$.

The assignment (\ref{assignr}) is a 1-1 correspondence between the set of $\bmr\in\calr$
and the set of sequences formed by $\bmr$.  Similarly, 
the assignment (\ref{assignu}) is a 1-1 correspondence between the set of $\bmu\in\calu$
and the set of sequences formed by $\bmu$.  Consider the bijection $\beta:  \calr\ra\calu$.
If $\beta:  \bmr\mapsto\bmu$, under the bijection 
$\beta^t:  \triupjktarg{0}{0}{t}{\calr}\ra\triupjktarg{0}{0}{t}{\calu}$ for each 
$t\in\bmcpz$, a sequence of triangles 
$\ldots,\triupjktarg{0}{0}{t}{\bmr},\triupjktarg{0}{0}{t-1}{\bmr},\ldots$
of $\bmr$ in (\ref{assignr}) becomes a sequence of triangles
$\ldots,\triupjktarg{0}{0}{t}{\bmu},\triupjktarg{0}{0}{t-1}{\bmu},\ldots$, which then
overlap to form $\bmu$ in (\ref{assignu}).  We can also go in reverse.  This gives the following.

\begin{cor}
\label{cor66}
Let $\beta:  \calr\ra\calu$ with assignment $\beta:  \bmr\mapsto\bmu$.
Given $\bmu\in\calu$, we can find $\bmr\in\calr$ such that $\beta:  \bmr\mapsto\bmu$
by first finding the sequence of triangles in (\ref{assignu}), and then
finding $(\beta^t)^{-1}:  \triupjktarg{0}{0}{t}{\bmu}\mapsto\triupjktarg{0}{0}{t}{\bmr}$
for each $t\in\bmcpz$ to find the sequence of triangles
\be
\label{seqtri}
\ldots,\triupjktarg{0}{0}{t}{\bmr},\triupjktarg{0}{0}{t-1}{\bmr},\ldots,
\ee
which then finds $\bmr$ in (\ref{assignr}).
\end{cor}

\vspace{3mm}
{\bf 4.5  Recovery of group system $A$ from generator group $(\calu,\circ)$}%hhh
\vspace{3mm}

We show in this subsection we can recover $A$ from its generator group $(\calu,\circ)$.
We prove the following simple extension of the
first homomorphism theorem to show that it is possible to use a homomorphism to construct
a group system from any input group.  We can think of the following construction theorem 
as a {\it first homomorphism theorem for group systems}.
In Subsection 6.3, we use the \fhgs\ to find all \ellctl\ complete groups systems $A$
up to isomorphism.

\begin{thm}
\label{fhgs}
Consider any group $\msfcpg$.  Suppose there is a homomorphism $p^t:  \msfcpg\ra \msfcpg^t$ from $\msfcpg$
to a group $\msfcpg^t$ for each $t\in\bmcpz$.  In general group $\msfcpg^t$ may be different for each 
$t\in\bmcpz$.  Define the direct product group $(\msfcpg_\amalg,\plus)$ 
by $(\msfcpg_\amalg,\plus)\rmdef\cdots\times \msfcpg^t\times \msfcpg^{t+1}\times\cdots$.
There is a homomorphism $p:  G\ra G_\amalg$, from $\msfcpg$ to the direct product group 
$(\msfcpg_\amalg,\plus)$, defined by
\be
\label{eqthm37}
p(\msfg)\rmdef\ldots,p^t(\msfg),p^{t+1}(\msfg),\ldots.
\ee
Define $\msfg_\amalg\rmdef\ldots,p^t(\msfg),p^{t+1}(\msfg),\ldots$.  Then 
$p:  G\ra G_\amalg$ with assignment $p:  \msfg\mapsto\msfg_\amalg$.  Then
$$
\msfcpg/\msfcpg_K\simeq\imp,
$$
where group $\imp$ is the image of the homomorphism $p$, and where 
$\msfcpg_K$ is the kernel of the homomorphism $p$.  We have $\imp$ is a group system 
defined by a componentwise operation in $\msfcpg^t$ for each $t\in\bmcpz$.  Lastly we have 
$\msfcpg\simeq\imp$ \ifof\ the kernel $\msfcpg_K$ of the homomorphism $p$ is the identity.
\end{thm}

\begin{prf}
Since there is a homomorphism $p^t:  \msfcpg\ra \msfcpg^t$ from $\msfcpg$ to a group 
$\msfcpg^t$ for each $t\in\bmcpz$, we must have $p^t(\msfgdt\msfgddt)=p^t(\msfgdt)p^t(\msfgddt)$ 
for each $t\in\bmcpz$.  Then
\begin{align*}
p(\msfgdt\msfgddt) &=\ldots,p^t(\msfgdt\msfgddt),p^{t+1}(\msfgdt\msfgddt),\ldots \\
                   &=\ldots,p^t(\msfgdt)p^t(\msfgddt),p^{t+1}(\msfgdt)p^{t+1}(\msfgddt),\ldots \\
                   &=p(\msfgdt)\plus p(\msfgddt).
\end{align*}
Then there is a homomorphism $p$ from $\msfcpg$ to the direct product group $(\msfcpg_\amalg,\plus)$.
We have $\msfcpg/\msfcpg_K\simeq\imp$ from the first homomorphism theorem.
We have $\imp$ is a group system since the global operation $\plus$ in $\imp$ is defined 
by a componentwise operation in $\msfcpg^t$ for each $t\in\bmcpz$.
\end{prf}
We refer to (\ref{eqthm37}) by the notation $p=\ldots,p^t,p^{t+1},\ldots$.

We now ask whether we can reverse the chain in (\ref{chain7a}), i.e., starting with a generator
group $(\calu,\circ)$, can we recover $A$.  We now show that we can recover $A$ directly 
from group $(\calu,\circ)$ using the \fhgs.

\begin{thm}
\label{thm71}
The \fhgs\ constructs a group system $\imfu$ with component group
$A^t$ from an $(\ell+1)$-depth generator group $(\calu,\circ)$ using a homomorphism $f_u$.
\end{thm}

\begin{prf}
The composition $\alpha^t\bullet(\beta^t)^{-1}\bullet\theta^t$ is a map 
$\calu\ra\triupjktarg{0}{0}{t}{\calu}\ra\triupjktarg{0}{0}{t}{\calr}\ra A^t$ given by the assignment 
$\alpha^t\bullet(\beta^t)^{-1}\bullet\theta^t:  \bmu\mapsto\triupjktarg{0}{0}{t}{\bmu}\mapsto\triupjktarg{0}{0}{t}{\bmr}\mapsto a^t$.
From Theorem \ref{homou}, we know there is a homomorphism from $(\calu,\circ)$ to 
$\grpupjktarg{0}{0}{t}{\calu}{\ccirc}$ given by the map $\theta^t$; from Corollary \ref{cor65}, we know there is an isomorphism
from $\grpupjktarg{0}{0}{t}{\calu}{\ccirc}$ to $\grpupjktarg{0}{0}{t}{\calr}{\cast}$ given by the map $(\beta^t)^{-1}$; 
and from Theorem \ref{homor1}, we know there is a homomorphism from 
$\grpupjktarg{0}{0}{t}{\calr}{\cast}$ to $A^t$ given by the map $\alpha^t$.  
Let $f_u^t\rmdef\alpha^t\bullet(\beta^t)^{-1}\bullet\theta^t$.  
Then $f_u^t:  \calu\ra A^t$ is a
homomorphism from $(\calu,\circ)$ to $A^t$ for each $t\in\bmcpz$.  Consider the Cartesian product 
$A_\amalg\rmdef\cdots\times A^t\times A^{t+1}\times\cdots$ (note here $A^t$ is interpreted
as a set).  Define the direct product group
$(A_\amalg,+)$ by $(A_\amalg,+)\rmdef\cdots\times A^t\times A^{t+1}\times\cdots$.
Then from Theorem \ref{fhgs}, using $(\calu,\circ)$ for information group $\msfcpg$
and alphabet group $A^t$ for $\msfcpg^t$, $t\in\bmcpz$, 
there is a homomorphism $f_u:  \calu\ra A_\amalg$, from $(\calu,\circ)$
to the direct product group $(A_\amalg,+)$, defined by
$$
f_u(\bmu)\rmdef\ldots,f_u^t(\bmu),f_u^{t+1}(\bmu),\ldots.
$$
Define
\begin{align*}
\bma_\amalg\rmdef &\ldots,f_u^t(\bmu),f_u^{t+1}(\bmu),\ldots \\
            = &\ldots,a^t,a^{t+1},\ldots.
\end{align*}
Then $f_u:  \calu\ra A_\amalg$
with assignment $f_u:  \bmu\mapsto\bma_\amalg$.
We can think of $\bma_\amalg$ as a ``sliding block" mapping of $\bmu$.  Applying the first
homomorphism theorem for groups, we have
$$
(\calu,\circ)/(\calu,\circ)_K\simeq\imfu,
$$
where group $\imfu$ is the image of the homomorphism $f_u$, and where 
$(\calu,\circ)_K$ is the kernel of the homomorphism $f_u$.  Since group $\imfu$ is a
subgroup of the direct product group $(A_\amalg,+)$, then $\imfu$ is a
group system where global operation $+$ defines the componentwise 
operation in group $A^t$ for each $t\in\bmcpz$.
\end{prf}

\begin{thm}
\label{thm72}
The homomorphism $f_u$ is a bijection $f_u:  \calu\ra A$.  Then $\imfu=A$.
Therefore the group system $\imfu$ is \ellctl\ and complete.
\end{thm}

\begin{prf}
Fix $\bmu\in\calu$.  Let
$$
\bmr_\amalg\rmdef\ldots,\triupjktarg{0}{0}{t}{\bmr},\triupjktarg{0}{0}{t-1}{\bmr},\ldots,
$$
where $\triupjktarg{0}{0}{t}{\bmr}$ is given by the assignment $(\beta^t)^{-1}\bullet\theta^t:  \bmu\mapsto\triupjktarg{0}{0}{t}{\bmu}\mapsto\triupjktarg{0}{0}{t}{\bmr}$ 
for each $t\in\bmcpz$ of Theorem \ref{thm71}.
From Corollary \ref{cor66}, the assignment 
$(\beta^t)^{-1}\bullet\theta^t:  \bmu\mapsto\triupjktarg{0}{0}{t}{\bmu}\mapsto\triupjktarg{0}{0}{t}{\bmr}$ 
for each $t\in\bmcpz$ is the same as the assignment 
$\beta^{-1}:  \bmu\mapsto\bmr_\amalg$ of bijection $\beta^{-1}:  \calu\ra\calr$.  
Then $\bmr_\amalg\in\calr$.  Let
$$
\bma_\amalg'\rmdef\ldots,a^t,a^{t+1},\ldots,
$$
where $a^t$ is given by the assignment
$\alpha^t:  \triupjktarg{0}{0}{t}{\bmr}\mapsto a^t$ for each $t\in\bmcpz$ of Theorem \ref{thm71}.  
From Corollary \ref{cor11}, the assignment 
$\alpha^t:  \triupjktarg{0}{0}{t}{\bmr}\mapsto a^t$ for each $t\in\bmcpz$
is the same as the assignment $\alpha:  \bmr_\amalg\mapsto\bma_\amalg'$ of 
bijection $\alpha:  \calr\ra A$.  Then $\bma_\amalg'\in A$.  Then the assignment 
$\alpha^t\bullet(\beta^t)^{-1}\bullet\theta^t:  \bmu\mapsto\triupjktarg{0}{0}{t}{\bmu}\mapsto\triupjktarg{0}{0}{t}{\bmr}\mapsto a^t$
for each $t\in\bmcpz$ of Theorem \ref{thm71} is the same as the assignment 
$\alpha\bullet\beta^{-1}:  \bmu\mapsto\bmr_\amalg\mapsto\bma_\amalg'$ 
of the bijection $\alpha\bullet\beta^{-1}:  \calu\ra\calr\ra A$.
But the map $\alpha^t\bullet(\beta^t)^{-1}\bullet\theta^t$ is the same as
the map $f_u^t$.  Then the bijection $\alpha\bullet\beta^{-1}:  \calu\ra A$
is the same as map $f_u$, since $f_u=\ldots,f_u^t,f_u^{t+1},\ldots$.  Then
$f_u(\bmu)=\bma_\amalg'$, and $f_u$ is a bijection $f_u:  \calu\ra A$.
Then $\imfu=A$.

Since $\imfu=A$, the group system $\imfu$ is \ellctl\ and complete.
\end{prf}
With (\ref{chain7a}), these results give the chain shown in (\ref{chain7b}).  As shown 
we can recover $A$ directly from group $(\calu,\circ)$ using the \fhgs.
%%%%%%%%%%%%%%%%%%%%%%%%%%%%%%%%%%%%%%%%%
\be
\label{chain7b}
\begin{array}{lllll}
\da & \la                       & \la       & {\stackrel{=}{\la}}        & \imfu                                  \\
\da &                           &           &                            & \ua\,f_u=\ldots,f_u^t,f_u^{t+1},\ldots \\
A   & {\stackrel{\simeq}{\ra}}  & (\calr,*) & {\stackrel{\simeq}{\ra}}   & (\calu,\circ)    
\end{array}
\ee
%%%%%%%%%%%%%%%%%%%%%%%%%%%%%%%%%%%%%%%%

\begin{cor}
\label{cor80}
We have $(\calu,\circ)\simeq\imfu=A$.
\end{cor}

\begin{prf}
The kernel of the homomorphism $f_u$ is the identity so $(\calu,\circ)\simeq\imfu=A$.
\end{prf}

\newpage
\vspace{3mm}
{\bf 5.  PROPERTIES OF THE GENERATOR GROUP}
\vspace{3mm}

\vspace{3mm}
{\bf 5.1  The nested elementary groups of $(\calu,\circ)$}%hhh
\vspace{3mm}

For $0\le k\le\ell$ and $t\in\bmcpz$, and for each $\bmu\in\calu$, 
we let $\triupjktarg{0}{k}{t}{\bmu}$ be the triangle of \glabs\
in $\bmu$ in (\ref{ugttf}) with lower vertex $r_{0,k}^t$ 
and upper vertices $r_{0,\ell}^t$ and $r_{0,\ell}^{t-(\ell-k)}$.
Note that for $0\le k\le\ell$, $\triupjktarg{0}{k}{t}{\bmu}$ is a subtriangle in
$\triupjktarg{0}{0}{t}{\bmu}$ in (\ref{rcttf}), and for $k=0$, $\triupjktarg{0}{k}{t}{\bmu}$
is the same as $\triupjktarg{0}{0}{t}{\bmu}$ in (\ref{rcttf}).
For $0\le k\le\ell$ and $t\in\bmcpz$, let $\triupjktarg{0}{k}{t}{\calu}$ 
be the set of all possible triangles $\triupjktarg{0}{k}{t}{\bmu}$,
$\triupjktarg{0}{k}{t}{\calu}\rmdef\{\triupjktarg{0}{k}{t}{\bmu} : \bmu\in\calu\}$.
Note that for $k=0$, $\triupjktarg{0}{k}{t}{\calu}$ is just the set $\triupjktarg{0}{0}{t}{\calu}$
previously defined.

We now show there are smaller elementary groups on sets $\triupjktarg{0}{k}{t}{\calu}$
nested in $\grpupjktarg{0}{0}{t}{\calu}{\ccirc}$, for $0\le k\le\ell$, for each $t\in\bmcpz$.
We can use the same approach as for defining $\grpupjktarg{0}{0}{t}{\calu}{\ccirc}$.
We restate this approach now.
Recall that we have defined $(\calu^{t+},\circ)$ and $(\calu^{t-},\circ)$ for each $t\in\bmcpz$,
and these are normal subgroups of $(\calu,\circ)$.  Fix $t\in\bmcpz$.  Fix $k$ such that $0\le k\le\ell$.
Note that $\calu^{(t+1)+}$ and $\calu^{(t+k-1)-}$ cover $\calu$ except for $\triupjktarg{0}{k}{t}{\bmu}$.
In other words the union of set $\calu^{(t+1)+}$ and set $\calu^{(t+k-1)-}$
is the set of all $\bmu\in\calu$ which are trivial 
in $\triupjktarg{0}{k}{t}{\bmu}$; call this subset of $\calu$ as $\triupjktargnop{0}{k}{t}{\calk}$.
Then the product of $(\calu^{(t+1)+},\circ)$ and $(\calu^{(t+k-1)-},\circ)$ is a normal subgroup
$\grpupjktargnop{0}{k}{t}{\calk}$ of $(\calu,\circ)$.
Then there is a quotient group $(\calu,\circ)/\grpupjktargnop{0}{k}{t}{\calk}$.  
The projection of the cosets in $(\calu,\circ)/\grpupjktargnop{0}{k}{t}{\calk}$ on set
$\triupjktarg{0}{k}{t}{\calu}$ is $\triupjktarg{0}{k}{t}{\calu}$.  
We use this to define a group on $\triupjktarg{0}{k}{t}{\calu}$.

Fix time $t$.  Fix $k$ such that $0\le k\le\ell$.
Let $\triupjktarg{0}{k}{t}{\bmudt},\triupjktarg{0}{k}{t}{\bmuddt}\in\triupjktarg{0}{k}{t}{\calu}$.
Define an operation $\prodjktarg{0}{k}{t}{\ccirc}$ on $\triupjktarg{0}{k}{t}{\calu}$ by
\be
\label{opdef8}
\triupjktarg{0}{k}{t}{\bmudt}\prodjktarg{0}{k}{t}{\ccirc}\triupjktarg{0}{k}{t}{\bmuddt}
\rmdef\triupjktarg{0}{k}{t}{\bmudt\circ\bmuddt}.
\ee

\begin{lem}
\label{lem74a}
Fix time $t$.  Fix $k$ such that $0\le k\le\ell$.
The operation $\prodjktarg{0}{k}{t}{\ccirc}$ on $\triupjktarg{0}{k}{t}{\calu}$ 
is well defined.
\end{lem}

\begin{prf}
The proof is the same as Lemma \ref{lem62}.
\end{prf}

\begin{thm}
%\label{thm55}
Fix time $t$.  Fix $k$ such that $0\le k\le\ell$.
The set $\triupjktarg{0}{k}{t}{\calu}$ with operation $\prodjktarg{0}{k}{t}{\ccirc}$ forms a 
group $\grpupjktarg{0}{k}{t}{\calu}{\ccirc}$.
\end{thm}

\begin{prf}
The proof is the same as Theorem \ref{thm63}.
\end{prf}

\begin{cor}
\label{cor55a}
For each $t\in\bmcpz$, for $0\le k\le\ell$, the group $\grpupjktarg{0}{k}{t}{\calu}{\ccirc}$ 
determined by $(\calu,\circ)$ is unique.
\end{cor}

\begin{prf}
The proof is the same as Corollary \ref{cor63a}.
\end{prf}
We call the groups $\grpupjktarg{0}{k}{t}{\calu}{\ccirc}$,
for each $t\in\bmcpz$, for $0\le k\le\ell$, the {\it upper elementary groups} of $(\calu,\circ)$.
The upper elementary group $\grpupjktarg{0}{0}{t}{\calu}{\ccirc}$, for each $t\in\bmcpz$,
is the component group of $(\calu,\circ)$.

The operation in elementary group $\grpupjktarg{0}{k}{t}{\calu}{\ccirc}$ just
depends on the elements in set $\triupjktarg{0}{k}{t}{\calu}$, and is otherwise 
independent of the remaining portion of $\calu$.  Then we can think of the collection of
elementary groups $\grpupjktarg{0}{k}{t}{\calu}{\ccirc}$, for $0\le k\le\ell$, for $t\in\bmcpz$, 
as a decomposition of $(\calu,\circ)$ into local groups since each group
$\grpupjktarg{0}{k}{t}{\calu}{\ccirc}$ is wholely defined by a local set
$\triupjktarg{0}{k}{t}{\calu}$.

Essentially $(\calu,\circ)$ is defined
to be the permutation of the \glabs\ in $\calu$ when two sequences
in $A$ with these generators are multiplied.  The group $\grpupjktarg{0}{\ell}{t}{\calu}{\ccirc}$
shows that when two sequences in $A$ are multiplied, the longest generator $g_\ell^t$ of the result
only depends on the longest generators at time $t$ of the two multiplicands.  Next, the group
$\grpupjktarg{0}{\ell-1}{t}{\calu}{\ccirc}$ contains \glabs\ of two longest generators and one
next longest generator.  This group shows that when two sequences in $A$ are multiplied, 
the next longest generator $g_{\ell-1}^t$ of the result only depends on the longest generators
at times $t$ and $t-1$ and next longest generators at time $t$ of the two multiplicands.  And so on.
This approach can be used to calculate $\bmudt\circ\bmuddt$ from all the upper elementary
groups.

We now give some homomorphism properties of upper elementary groups.
For each $t\in\bmcpz$, for $0\le k\le\ell$, 
define a map $\theta_{0,k}^t:  \calu\ra\triupjktarg{0}{k}{t}{\calu}$ 
by the assignment $\theta_{0,k}^t:  \bmu\mapsto\triupjktarg{0}{k}{t}{\bmu}$.
From the definition of operation $\prodjktarg{0}{k}{t}{\ccirc}$ in upper elementary group
$\grpupjktarg{0}{k}{t}{\calu}{\ccirc}$, the following homomorphism is clear.

\begin{thm}
\label{homo3}
Fix any $t\in\bmcpz$.  Fix $k$ such that $0\le k\le\ell$.
There is a homomorphism from $(\calu,\circ)$ to any 
upper elementary group $\grpupjktarg{0}{k}{t}{\calu}{\ccirc}$
given by the map $\theta_{0,k}^t:  \calu\ra\triupjktarg{0}{k}{t}{\calu}$ with assignment
$\theta_{0,k}^t:  \bmu\mapsto\triupjktarg{0}{k}{t}{\bmu}$.  Also there is an isomorphism
$(\calu,\circ)/\grpupjktargnop{0}{k}{t}{\calk}\simeq\grpupjktarg{0}{k}{t}{\calu}{\ccirc}$.
\end{thm}

\begin{prf}
The proof is the same as Theorem \ref{homou}.
\end{prf}

Next we show there are homomorphisms among the upper elementary groups.
Fix any $t\in\bmcpz$.  Fix $k$ such that $0\le k\le\ell$.  Consider set $\triupjktarg{0}{k}{t}{\calu}$.
For any $\bmu\in\calu$, $\triupjktarg{0}{j}{t-j}{\bmu}$ is a subtriangle of
$\triupjktarg{0}{k}{t}{\bmu}$ for $0\le j\le k$.   Then set
$\triupjktarg{0}{j}{t-j}{\calu}$ is nested inside set
$\triupjktarg{0}{k}{t}{\calu}$ for $0\le j\le k$.  Define the projection map
$\theta_{k,j}^t:  \triupjktarg{0}{k}{t}{\calu}\ra\triupjktarg{0}{j}{t-j}{\calu}$ by the assignment
$\theta_{k,j}^t:  \triupjktarg{0}{k}{t}{\bmu}\mapsto\triupjktarg{0}{j}{t-j}{\bmu}$.

\begin{lem}
The projection map $\theta_{k,j}^t:  \triupjktarg{0}{k}{t}{\calu}\ra\triupjktarg{0}{j}{t-j}{\calu}$
is well defined.
\end{lem}

\begin{prf}
Consider any other $\bmudt\in\calu$ such that $\triupjktarg{0}{k}{t}{\bmudt}=\triupjktarg{0}{k}{t}{\bmu}$.
Since $\bmu$ and $\bmudt$ agree on $\triupjktarg{0}{k}{t}{\calu}$, they must agree on
$\triupjktarg{0}{j}{t-j}{\calu}$.  Then we must have 
$\triupjktarg{0}{j}{t-j}{\bmudt}=\triupjktarg{0}{j}{t-j}{\bmu}$, and so $\theta_{k,j}^t$
is well defined.
\end{prf}

\begin{thm}
\label{thm41a}
Fix any $t\in\bmcpz$.  Fix $k$ such that $0\le k\le\ell$.  Consider
group $\grpupjktarg{0}{k}{t}{\calu}{\ccirc}$.
For any $0\le j\le k$, there is a homomorphism from group $\grpupjktarg{0}{k}{t}{\calu}{\ccirc}$ to group
$\grpupjktarg{0}{j}{t-j}{\calu}{\ccirc}$ given by the projection map
$\theta_{k,j}^t:  \triupjktarg{0}{k}{t}{\calu}\ra\triupjktarg{0}{j}{t-j}{\calu}$ with assignment
$\theta_{k,j}^t:  \triupjktarg{0}{k}{t}{\bmu}\mapsto\triupjktarg{0}{j}{t-j}{\bmu}$.
\end{thm}

\begin{prf}
Let $\triupjktarg{0}{k}{t}{\bmudt},\triupjktarg{0}{k}{t}{\bmuddt}\in\triupjktarg{0}{k}{t}{\calu}$.  
Consider the projections 
$\theta_{k,j}^t:  \triupjktarg{0}{k}{t}{\bmudt}\mapsto\triupjktarg{0}{j}{t-j}{\bmudt}$ 
and $\theta_{k,j}^t:  \triupjktarg{0}{k}{t}{\bmuddt}\mapsto\triupjktarg{0}{j}{t-j}{\bmuddt}$.  
We need to show
$$
\theta_{k,j}^t(\triupjktarg{0}{k}{t}{\bmudt}\ccirc_{0,k}^t\triupjktarg{0}{k}{t}{\bmuddt})=
\theta_{k,j}^t(\triupjktarg{0}{k}{t}{\bmudt})\ccirc_{0,j}^{t-j}\theta_{k,j}^t(\triupjktarg{0}{k}{t}{\bmuddt}).
$$
But
\begin{align*}
\theta_{k,j}^t(\triupjktarg{0}{k}{t}{\bmudt}\ccirc_{0,k}^t\triupjktarg{0}{k}{t}{\bmuddt}) 
&=\theta_{k,j}^t(\triupjktarg{0}{k}{t}{\bmudt\circ\bmuddt}) \\
&=\triupjktarg{0}{j}{t-j}{\bmudt\circ\bmuddt} \\
&=\triupjktarg{0}{j}{t-j}{\bmudt}\ccirc_{0,j}^{t-j}\triupjktarg{0}{j}{t-j}{\bmuddt} \\
&=\theta_{k,j}^t(\triupjktarg{0}{k}{t}{\bmudt})\ccirc_{0,j}^{t-j}\theta_{k,j}^t(\triupjktarg{0}{k}{t}{\bmuddt}).
\end{align*}
\end{prf}
We say the homomorphism in Theorem \ref{thm41a} from a larger upper elementary group to a 
smaller one nested inside is a {\it nested homomorphism property}.  This gives the following.

\begin{thm}
\label{thm41b}
The generator group $(\calu,\circ)$ of any \ellctl\ complete group system $A$ is a collection of 
upper elementary groups $\grpupjktarg{0}{k}{t}{\calu}{\ccirc}$ of $(\calu,\circ)$
for each $t\in\bmcpz$, for $0\le k\le\ell$, which have a nested homomorphism property.
\end{thm}

Because the upper elementary groups of $(\calu,\circ)$ are indexed by $0\le k\le\ell$,
we say generator group $(\calu,\circ)$ is an $(\ell+1)$-{\it depth generator group}.

\begin{thm}
\label{thm41c}
Any $(\ell+1)$-depth generator group $(\calu,\circ)$ determines a unique collection of upper elementary groups.
\end{thm}

\begin{prf}
This follows from Corollary \ref{cor55a}.
\end{prf}

There are upper elementary groups $\grpupjktarg{0}{k}{t}{\calr}{\cast}$
with a nested homomorphism property in $(\calr,*)$ as well.
These can be found by the same approach as used in Subsection 4.4 to find
$\grpupjktarg{0}{0}{t}{\calr}{\cast}$ from $\grpupjktarg{0}{0}{t}{\calu}{\ccirc}$.
The nested upper elementary groups $\grpupjktarg{0}{k}{t}{\calr}{\cast}$ in $(\calr,*)$ 
are isomorphic to the nested upper elementary groups $\grpupjktarg{0}{0}{t}{\calu}{\ccirc}$
in $(\calu,\circ)$; more details can be found in \cite{KM6}.  

The matrix $\triupjktarg{0}{0}{t}{\bmu}$ has an upper triangular form.  
We now define a matrix with a lower triangular form.  Given $\bmu$ in (\ref{ugttf}),
for each $t\in\bmcpz$, we define $\trilwjktarg{0}{\ell}{t}{\bmu}$
to be the triangle in $\bmu$ with lower vertices $g_0^{t+\ell}$ and $g_0^t$
and upper vertex $g_\ell^t$, as shown in (\ref{lowertri}).  
%%%%%%%%%%%%%%%%%%%%%%%%%%%%%%%%%%%%%%%%%
\be
\label{lowertri}
\begin{array}{llllllllllll}
                     &                    &                    &        &        &                &        &        &                  &  g_\ell^t     \\
                     &                    &                    &        &        &                &        &        & g_{\ell-1}^{t+1} &  g_{\ell-1}^t \\
                     &                    &                    &        &        &                &        & \vdots & \vdots           &  \vdots       \\
                     &                    &                    &        &        & g_k^{t+\ell-k} & \cdots & \cdots & g_k^{t+1}        &  g_k^t        \\
                     &                    &                    &        & \vdots & \vdots         & \vdots & \vdots & \vdots           &  \vdots       \\
                     &                    & g_2^{t+\ell-2}     & \cdots & \cdots & g_2^{t+\ell-k} & \cdots & \cdots & g_2^{t+1}        &  g_2^t        \\
                     & g_1^{t+\ell-1}     & g_1^{t+\ell-2}     & \cdots & \cdots & g_1^{t+\ell-k} & \cdots & \cdots & g_1^{t+1}        &  g_1^t        \\
 g_0^{t+\ell}        & g_0^{t+\ell-1}     & g_0^{t+\ell-2}     & \cdots & \cdots & g_0^{t+\ell-k} & \cdots & \cdots & g_0^{t+1}        &  g_0^t      
\end{array}
\ee
%%%%%%%%%%%%%%%%%%%%%%%%%%%%%%%%%%%%%%%%%

In general, for $0\le k\le\ell$ and $t\in\bmcpz$, and for each $\bmu\in\calu$, 
we let $\trilwjktarg{0}{k}{t}{\bmu}$ be the triangle of \glabs\
in $\bmu$ in (\ref{ugttf}) with lower vertices $g_0^{t+k}$ and $g_0^t$
and upper vertex $g_k^t$.
Note that for $0\le k\le\ell$, $\trilwjktarg{0}{k}{t}{\bmu}$ is a subtriangle in
$\trilwjktarg{0}{\ell}{t}{\bmu}$ in (\ref{lowertri}), and for $k=0$, $\trilwjktarg{0}{k}{t}{\bmu}$
is the same as $\trilwjktarg{0}{\ell}{t}{\bmu}$ in (\ref{lowertri}).
For $0\le k\le\ell$ and $t\in\bmcpz$, let $\trilwjktarg{0}{k}{t}{\calu}$ 
be the set of all possible triangles $\trilwjktarg{0}{k}{t}{\bmu}$,
$\trilwjktarg{0}{k}{t}{\calu}\rmdef\{\trilwjktarg{0}{k}{t}{\bmu} : \bmu\in\calu\}$.

Fix $t\in\bmcpz$ and consider $A^{[t,t+k]}$; this is a normal subgroup of $A$ \cite{FT}.
In $\calr$, this is the set
of all $\bmr\in\calr$ with trivial generators for all $t'\in\bmcpz$ except 
for generators $\bmg^{[m,n]}$ where $t\le m\le n\le t+k$.  Call this subset
of $\calr$ as $\calr_{0,k}^t$.  Since $A^{[t,t+k]}$ is a normal subgroup of $A$,
then $(\calr_{0,k}^t,*)$ is a normal subgroup of $(\calr,*)$, under the bijection
$\alpha^{-1}$ of the isomorphism from $A$ to $(\calr,*)$.
In $\calu$, this is the set of all $\bmu$ with trivial 
entries except for those in $\trilwjktarg{0}{k}{t}{\bmu}$; define this subset of $\calu$
to be $\trilwjktargnop{0}{k}{t}{\call}$.  Since $(\calr_{0,k}^t,*)$ is a normal subgroup of $(\calr,*)$,
then $\grplwjktargnop{0}{k}{t}{\call}$ is a normal subgroup of $(\calu,\circ)$, under the bijection
$\beta$ of the isomorphism from $(\calr,*)$ to $(\calu,\circ)$.
This shows that $\grplwjktargnop{0}{k}{t}{\call}$ is the analog in $(\calu,\circ)$
of $A^{[t,t+k]}$ in $A$.  We say the group $\grplwjktargnop{0}{k}{t}{\call}$, for each $t\in\bmcpz$,
is a {\it lower elementary group} of $(\calu,\circ)$.  Note that a lower elementary 
group $\grplwjktargnop{0}{k}{t}{\call}$ is defined on a subset of $\calu$,
but an upper elementary group $\grpupjktarg{0}{0}{t}{\calu}{\ccirc}$
is defined on a local set $\triupjktarg{0}{0}{t}{\calu}$ of $\calu$.

The groups $\grplwjktargnop{0}{k}{t+1}{\call}$ and $\grplwjktargnop{0}{k}{t}{\call}$
are normal subgroups of $(\calu,\circ)$.  Then the product
$\grplwjktargnop{0}{k}{t+1}{\call}\grplwjktargnop{0}{k}{t}{\call}$ is a 
normal subgroup of $(\calu,\circ)$.  Then the product is a normal subgroup
of $\grplwjktargnop{0}{k+1}{t}{\call}$ of $(\calu,\circ)$.
The quotient group
$\grplwjktargnop{0}{k+1}{t}{\call}/\grplwjktargnop{0}{k}{t+1}{\call}\grplwjktargnop{0}{k}{t}{\call}$
is the analog in the generator group of the granule
$A^{[t,t+k+1]}/(A^{[t,t+k+1)}A^{(t,t+k+1]})$ in $A$. Unlike the upper elementary groups,
the lower elementary groups are defined on $\calu$, and so there is no nested homomorphism property
from $\grplwjktargnop{0}{k+1}{t}{\call}$ to $\grplwjktargnop{0}{k}{t+1}{\call}$
and $\grplwjktargnop{0}{k}{t}{\call}$.

\vspace{3mm}
{\bf 5.2  Set of elementary groups}%hhh
\vspace{3mm}

The tensor $\bmu$ in (\ref{ugttf}) is indexed by ordered pairs of the form $(k,t)$,
where subscript $k$ satisfies $0\le k\le\ell$ and superscript $t\in\bmcpz$.  Then tensor $\bmu$ gives
an {\it index tensor} $\bmn$ of ordered pairs $(k,t)$, as shown in (\ref{top}).
%%%%%%%%%%%%%%%%%%%%%%%%%%%%%%%%%%%%%%%%%
\be
\label{top}
\begin{array}{llllllll}
        &  (\ell,t)   & (\ell,t-1)   & \cdots & (\ell,t-j)   & \cdots & (\ell,t-\ell)   & \\
        &  (\ell-1,t) & (\ell-1,t-1) & \cdots & (\ell-1,t-j) & \cdots & (\ell-1,t-\ell) & \\
        &  \cdots     & \cdots       &        & \cdots       &        & \cdots          & \\
        &  (k,t)      & (k,t-1)      & \cdots & (k,t-j)      & \cdots & (k,t-\ell)      & \\
        &  (k-1,t)    & (k-1,t-1)    & \cdots & (k-1,t-j)    & \cdots & (k-1,t-\ell)    & \\
\cdots  &  \cdots     & \cdots       & \cdots & \cdots       & \cdots & \cdots          & \cdots  \\
        &  (j,t)      & (j,t-1)      & \cdots & (j,t-j)      & \cdots & (j,t-\ell)      & \\
        &  (j-1,t)    & (j-1,t-1)    & \cdots & (j-1,t-j)    & \cdots & (j-1,t-\ell)    & \\
        &  \cdots     & \cdots       &        & \cdots       &        & \cdots          & \\
        &  (1,t)      & (1,t-1)      & \cdots & (1,t-j)      & \cdots & (1,t-\ell)      & \\
        &  (0,t)      & (0,t-1)      & \cdots & (0,t-j)      & \cdots & (0,t-\ell)      &
\end{array}
\ee   
%%%%%%%%%%%%%%%%%%%%%%%%%%%%%%%%%%%%%%%%%
We use a similar notation to describe triangular subsets in index tensor $\bmn$ that was
previously used for tensor $\bmr\in\calr$ and $\bmu\in\calu$.  For $0\le k\le\ell$ and $t\in\bmcpz$,
let {\it upper index triangle} $\triupjktarg{0}{k}{t}{\bmn}$ be the ordered pairs $(k,t)$ in (\ref{top})
specified by the triangle with lower vertex $(k,t)$ and upper vertices 
$(\ell,t)$ and $(\ell,t-(\ell-k))$.  As an example, $\triupjktarg{0}{0}{t}{\bmu}$
in (\ref{rcttf}) has the index triangle $\triupjktarg{0}{0}{t}{\bmn}$ shown in (\ref{rttfind}).
%%%%%%%%%%%%%%%%%%%%%%%%%%%%%%%%%%%%%%%%%
\be
\label{rttfind}
\begin{array}{llllllllll}
  (0,\ell)   & (1,\ell)   & \cdots & \cdots & (j,\ell)   & \cdots & \cdots & \cdots & (\ell-1,\ell)   & (\ell,\ell) \\
  (0,\ell-1) & (1,\ell-1) & \cdots & \cdots & (j,\ell-1) & \cdots & \cdots & \cdots & (\ell-1,\ell-1) & \\
  \vdots     & \vdots     & \vdots & \vdots & \vdots     & \vdots & \vdots & \vdots && \\
  (0,k)      & (1,k)      & \cdots & \cdots & (j,k)      & \cdots & (k,k) &&& \\
  \vdots     & \vdots     & \vdots & \vdots & \vdots     & \vdots &&&& \\
  \cdots     & \cdots     & \cdots & \cdots & (j,j) &&&&& \\
  \vdots     & \vdots     & \vdots &&&&&&& \\
  (0,2)      & (1,2)      & (2,2) &&&&&&& \\
  (0,1)      & (1,1)      &&&&&&&& \\
  (0,0)      &&&&&&&&&
\end{array}
\ee
%%%%%%%%%%%%%%%%%%%%%%%%%%%%%%%%%%%%%%%%%
Similarly, for $0\le k\le\ell$ and $t\in\bmcpz$, we let
{\it lower index triangle} $\trilwjktarg{0}{k}{t}{\bmn}$ be the ordered pairs $(k,t)$ in (\ref{top})
specified by the triangle with lower vertices $(0,t+k)$ and $(0,t)$
and upper vertex $(k,t)$.

We define a sequence of ordered pairs $(k,t)$ in index tensor $\bmn$.
Let $\bmt$ be a sequence of times $\bmt=\ldots,t,t',\ldots$, where times $t,t'\in\bmcpz$.
The sequence $\bmt=\ldots,t,t',\ldots$ is written in reverse time order,
so that $t'<t$ as in (\ref{top}).  The sequence $\bmt$ may be the integers $\bmcpz$
or a finite or infinite subset of $\bmcpz$, in reverse time order.
Let $\bmk$ be a sequence of integers $\ldots,k,k',\ldots$ with each integer $k$
satisfying $0\le k\le\ell$.  The {\it paired sequence} $(\bmk,\bmt)$ is a sequence
of ordered pairs $(k,t)$ in $\bmn$, defined by
$$
(\bmk,\bmt)\rmdef\ldots,(k,t),(k',t'),\ldots,
$$
such that each $k$ in sequence $\bmk$ is paired with a $t$ in sequence $\bmt$ and vice versa.

Fix index tensor $\bmn$.  We next consider a sequence of lower elementary triangles 
in $\bmn$ indexed by the paired sequence $(\bmk,\bmt)$.  Define
$\blktrilwktarg{\bmk}{\bmt}{\bmn}$ to be the sequence of lower elementary triangles
\be
\label{seq1}
\blktrilwktarg{\bmk}{\bmt}{\bmn}\rmdef
\ldots,\trilwjktarg{0}{k}{t}{\bmn},\trilwjktarg{0}{k'}{t'}{\bmn},\ldots
\ee
indexed by paired sequence
$$
(\bmk,\bmt)=\ldots,(k,t),\ldots,(k',t'),\ldots.
$$
Note that triangles in $\blktrilwktarg{\bmk}{\bmt}{\bmn}$ may overlap.  We assume
the sequence (\ref{seq1}) is purged so
that any two triangles in sequence $\blktrilwktarg{\bmk}{\bmt}{\bmn}$ are disjoint, i.e.,
there are entries in one triangle that are not in the other.  In other words,
there is no triangle in the sequence that is wholely contained in another triangle.

Next consider a sequence of upper elementary triangles in $\bmn$ indexed by the
paired sequence $(\bmk_u,\bmt_u)$.  Define
$\blktriupktarg{\bmk_u}{\bmt_u}{\bmn}$ to be the sequence of upper elementary triangles
\be
\label{seq2}
\blktriupktarg{\bmk_u}{\bmt_u}{\bmn}\rmdef
\ldots,\triupjktarg{0}{k_u}{t_u}{\bmn},\triupjktarg{0}{k_u'}{t_u'}{\bmn},\ldots
\ee
indexed by paired sequence
$$
(\bmk_u,\bmt_u)=\ldots,(k_u,t_u),\ldots,(k_u',t_u'),\ldots.
$$
Note that triangles in $\blktriupktarg{\bmk_u}{\bmt_u}{\bmn}$ may overlap.  We assume
the sequence (\ref{seq2}) is purged so
that any two triangles in sequence $\blktrilwktarg{\bmk_u}{\bmt_u}{\bmn}$ are disjoint.

We now show that given a sequence of lower elementary triangles (\ref{seq1}) with paired 
sequence $(\bmk,\bmt)$, there is a sequence of upper elementary triangles (\ref{seq2}) with paired 
sequence $(\bmk_u,\bmt_u)$, and the two sequences of triangles partition $\bmn$.  In this case, we say
$(\bmk_u,\bmt_u)$ is the {\it complementary paired sequence of} $(\bmk,\bmt)$.
The reverse is also true, and we say
$(\bmk,\bmt)$ is the complementary paired sequence of $(\bmk_u,\bmt_u)$. 

\begin{lem}
\label{lem47}
For any paired sequence $(\bmk,\bmt)$, there is a complementary paired sequence $(\bmk_u,\bmt_u)$
of $(\bmk,\bmt)$, and $\blktrilwktarg{\bmk}{\bmt}{\bmn}$ and 
$\blktriupktarg{\bmk_u}{\bmt_u}{\bmn}$ partition $\bmn$ into two sawtooth patterns.  
For any paired sequence $(\bmk_u',\bmt_u')$, there is a complementary paired sequence $(\bmk',\bmt')$
of $(\bmk_u',\bmt_u')$, and $\blktrilwktarg{\bmk'}{\bmt'}{\bmn}$ and 
$\blktriupktarg{\bmk_u'}{\bmt_u'}{\bmn}$ partition $\bmn$ into two sawtooth patterns. 
\end{lem}

\begin{prf}
We prove the first assertion.
Assume we are given a sequence of triangles $\blktrilwktarg{\bmk}{\bmt}{\bmn}$
indexed by paired sequence $(\bmk,\bmt)$.  We need to show 
there is a complementary paired sequence $(\bmk_u,\bmt_u)$
of $(\bmk,\bmt)$, and $\blktrilwktarg{\bmk}{\bmt}{\bmn}$ and 
$\blktriupktarg{\bmk_u}{\bmt_u}{\bmn}$ partition $\bmn$ into two sawtooth patterns.
Let $(k^*,t^*)$ be any ordered pair in $\bmn$ not contained in $\blktrilwktarg{\bmk}{\bmt}{\bmn}$.
If $(k^*,t^*)$ is not in $\blktrilwktarg{\bmk}{\bmt}{\bmn}$, then it is readily seen that triangle
$\triupjktarg{0}{k^*}{t^*}{\bmn}$ cannot intersect with any triangle $\trilwjktarg{0}{k}{t}{\bmn}$
in $\blktrilwktarg{\bmk}{\bmt}{\bmn}$, otherwise $(k^*,t^*)$ would
be in $\blktrilwktarg{\bmk}{\bmt}{\bmn}$.  Consider the union of all
triangles $\triupjktarg{0}{k^*}{t^*}{\bmn}$ over all $(k^*,t^*)$ not in 
$\blktrilwktarg{\bmk}{\bmt}{\bmn}$.  From this union, purge triangles that are wholely
contained in larger triangles.  Then we obtain a sequence of disjoint triangles
$\blktriupktarg{\bmk_u}{\bmt_u}{\bmn}$ which is indexed by a paired sequence $(\bmk_u,\bmt_u)$,
none of which intersect with any triangle $\trilwjktarg{0}{k}{t}{\bmn}$
in $\blktrilwktarg{\bmk}{\bmt}{\bmn}$.
Clearly all $(k^*,t^*)$ not in $\blktrilwktarg{\bmk}{\bmt}{\bmn}$ are in 
$\blktriupktarg{\bmk_u}{\bmt_u}{\bmn}$.  Then $\blktrilwktarg{\bmk}{\bmt}{\bmn}$
and $\blktriupktarg{\bmk_u}{\bmt_u}{\bmn}$ partition $\bmn$, and $(\bmk_u,\bmt_u)$ 
is a complementary paired sequence of $(\bmk,\bmt)$.

The set $\blktrilwktarg{\bmk}{\bmt}{\bmn}$ forms a sawtooth pattern in $\bmn$ 
with lower triangular teeth indexed by paired sequence $(\bmk,\bmt)$.  
The set $\blktriupktarg{\bmk_u}{\bmt_u}{\bmn}$ forms a sawtooth pattern in $\bmn$ 
with upper triangular teeth indexed by complementary paired sequence 
$(\bmk_u,\bmt_u)$ of $(\bmk,\bmt)$.  Since an ordered pair must be in either set, the two sets
partition $\bmn$ into two sawtooth patterns.

The second assertion is proven in an analogous way.
\end{prf}

Fix tensor $u\in\calu$.  We next consider a sequence of triangles in $\bmu$ indexed by
$\blktrilwktarg{\bmk}{\bmt}{\bmn}$.  Define
$\blktrilwktarg{\bmk}{\bmt}{\bmu}$ to be the sequence of lower elementary triangles
\be
\label{seq3}
\blktrilwktarg{\bmk}{\bmt}{\bmu}\rmdef
\ldots,\triupjktarg{0}{k}{t}{\bmu},\triupjktarg{0}{k'}{t'}{\bmu},\ldots
\ee
indexed by paired sequence $(\bmk,\bmt)=\ldots,(k,t),\ldots,(k',t'),\ldots$.

Next consider a sequence of triangles in $\bmu$ indexed by
$\blktriupktarg{\bmk_u}{\bmt_u}{\bmn}$.  Define
$\blktriupktarg{\bmk_u}{\bmt_u}{\bmu}$ to be the sequence of upper elementary triangles
\be
\label{seq4}
\blktriupktarg{\bmk_u}{\bmt_u}{\bmu}\rmdef
\ldots,\triupjktarg{0}{k_u}{t_u}{\bmu},\triupjktarg{0}{k_u'}{t_u'}{\bmu},\ldots
\ee
indexed by paired sequence $(\bmk_u,\bmt_u)=\ldots,(k_u,t_u),\ldots,(k_u',t_u'),\ldots$.
Define $\blktriupktarg{\bmk_u}{\bmt_u}{\calu}$ to be the set of sequences 
$\{\blktriupktarg{\bmk_u}{\bmt_u}{\bmu}:  \bmu\in\calu\}$.

Define any paired sequence $(\bmk,\bmt)$.  Consider the product
\be
\label{eqn17}
\prod_{t\in\bmt} A^{[t,t+k]}
\ee
for ordered pairs $(k,t)$ in $(\bmk,\bmt)$.  Each group $A^{[t,t+k]}$ is a normal
subgroup of $A$.  Then the product (\ref{eqn17}) is a normal subgroup of $A$.
We have seen the image of $A^{[t,t+k]}$ in $A$ is the lower elementary group
$\grplwjktargnop{0}{k}{t}{\call}$ in $(\calu,\circ)$, which is normal in $(\calu,\circ)$.
Then the image of the product (\ref{eqn17}) in $A$ is the product
\be
\label{eqn18}
(\blktrilwktargnop{\bmk}{\bmt}{\call},\circ)\rmdef\prod_{t\in\bmt} \grplwjktargnop{0}{k}{t}{\call}
\ee
in $(\calu,\circ)$, for ordered pairs $(k,t)$ in $(\bmk,\bmt)$.  Each group
$\grplwjktargnop{0}{k}{t}{\call}$ is a normal subgroup of $(\calu,\circ)$.  Then the product
(\ref{eqn18}) is a normal subgroup of $(\calu,\circ)$.
Note that $\blktrilwktargnop{\bmk}{\bmt}{\call}$
is a subset of $\calu$, whereas $\blktriupktarg{\bmk_u}{\bmt_u}{\bmu}$ is a sequence
of triangles of $\bmu\in\calu$ as in (\ref{seq4}).

\begin{lem}
\label{nsubg}
For any paired sequence $(\bmk,\bmt)$, $(\blktrilwktargnop{\bmk}{\bmt}{\call},\circ)$ is a 
normal subgroup of $(\calu,\circ)$.
\end{lem}

Consider the single normal subgroup $\grplwjktargnop{0}{k}{t}{\call}$ of $(\calu,\circ)$.
Let $(\bmk^*,\bmt^*)$ be the paired sequence given by the single ordered pair $(k,t)$.
The complementary paired sequence $(\bmk_u^*,\bmt_u^*)$ of $(\bmk^*,\bmt^*)$ is the paired sequence
$$
(\bmk_u^*,\bmt_u^*)=\ldots,(0,t+k+3),(0,t+k+2),(0,t+k+1),\ldots,(0,t-1),(0,t-2),(0,t-3),\ldots.
$$
We know $\trilwjktargnop{0}{k}{t}{\call}$ is the subset of all $\bmu\in\calu$ 
with trivial entries except for those in $\trilwjktarg{0}{k}{t}{\bmu}$.
Then $\trilwjktargnop{0}{k}{t}{\call}$ is the subset of all $\bmu\in\calu$ 
which are the identity on $\blktriupktarg{\bmk_u^*}{\bmt_u^*}{\calu}$, or
$\{\bmu\in\calu:  \blktriupktarg{\bmk_u^*}{\bmt_u^*}{\bmu}=\blktriupktarg{\bmk_u^*}{\bmt_u^*}{\bmu_\bone}\}$,
where $\bmu_\bone$ is the identity of $(\calu,\circ)$.  Now let $(\bmk,\bmt)$ be the paired sequence
used to define $(\blktrilwktargnop{\bmk}{\bmt}{\call},\circ)$.  Then we see that
$\blktrilwktargnop{\bmk}{\bmt}{\call}$ is the subset of all $\bmu\in\calu$ 
which are the identity on $\blktriupktarg{\bmk_u}{\bmt_u}{\calu}$, or
$\{\bmu\in\calu:  \blktriupktarg{\bmk_u}{\bmt_u}{\bmu}=\blktriupktarg{\bmk_u}{\bmt_u}{\bmu_\bone}\}$,
where $(\bmk_u,\bmt_u)$ is the complementary paired sequence of $(\bmk,\bmt)$.  This gives the following.

\begin{thm}
\label{thm48}
The normal subgroup $(\blktrilwktargnop{\bmk}{\bmt}{\call},\circ)$ of $(\calu,\circ)$
defined by paired sequence $(\bmk,\bmt)$ is the normal subgroup formed by the set of all
$\bmu\in\calu$ which are the identity on $\blktriupktarg{\bmk_u}{\bmt_u}{\calu}$,
where $(\bmk_u,\bmt_u)$ is the complementary paired sequence of $(\bmk,\bmt)$.
\end{thm}

We have just seen that the generator group 
$(\calu,\circ)$ of any \ellctl\ complete group system $A$ is a collection of 
elementary groups $\grpupjktarg{0}{k}{t}{\calu}{\ccirc}$ of $(\calu,\circ)$
for each $t\in\bmcpz$, for $0\le k\le\ell$.  In the remainder of this subsection, we show that a
finite or infinite set of elementary groups in the generator group also forms a group,
with properties similar to an elementary group.

Since $(\blktrilwktargnop{\bmk}{\bmt}{\call},\circ)$ is normal in $(\calu,\circ)$, there is
a quotient group $(\calu,\circ)/(\blktrilwktargnop{\bmk}{\bmt}{\call},\circ)$.
The projection of the cosets in $(\calu,\circ)/(\blktrilwktargnop{\bmk}{\bmt}{\call},\circ)$ on set
$\blktriupktarg{\bmk_u}{\bmt_u}{\calu}$ is $\blktriupktarg{\bmk_u}{\bmt_u}{\calu}$.
We now define a group on $\blktriupktarg{\bmk_u}{\bmt_u}{\calu}$.

Let $(\bmk_u,\bmt_u)$ be any paired sequence.  Let 
$\blktriupktarg{\bmk_u}{\bmt_u}{\bmudt},\blktriupktarg{\bmk_u}{\bmt_u}{\bmuddt}\in\blktriupktarg{\bmk_u}{\bmt_u}{\calu}$.
Define an operation $\proditarg{\bmk_u}{\bmt_u}{\oplus}$ on 
$\blktriupktarg{\bmk_u}{\bmt_u}{\calu}$ by
\be
\label{defn2}
\blktriupktarg{\bmk_u}{\bmt_u}{\bmudt}\proditarg{\bmk_u}{\bmt_u}{\oplus}\blktriupktarg{\bmk_u}{\bmt_u}{\bmuddt}
\rmdef\blktriupktarg{\bmk_u}{\bmt_u}{\bmudt\circ\bmuddt}.
\ee
Note that the operation $\proditarg{\bmk_u}{\bmt_u}{\oplus}$ in (\ref{defn2}) can be evaluated as
\begin{align}
\nonumber
\blktriupktarg{\bmk_u}{\bmt_u}{\bmudt}\proditarg{\bmk_u}{\bmt_u}{\oplus}\blktriupktarg{\bmk_u}{\bmt_u}{\bmuddt}
&=\blktriupktarg{\bmk_u}{\bmt_u}{\bmudt\circ\bmuddt} \\
\nonumber
&=\ldots,\triupjktarg{0}{k}{t}{\bmudt\circ\bmuddt},\triupjktarg{0}{k'}{t'}{\bmudt\circ\bmuddt},\ldots \\
\label{evaln2}
&=\ldots,\triupjktarg{0}{k}{t}{\bmudt}\prodjktarg{0}{k}{t}{\ccirc}\triupjktarg{0}{k}{t}{\bmuddt},
\triupjktarg{0}{k'}{t'}{\bmudt}\prodjktarg{0}{k'}{t'}{\ccirc}\triupjktarg{0}{k'}{t'}{\bmuddt},\ldots
\end{align}
for each $\blktriupktarg{\bmk_u}{\bmt_u}{\bmudt},\blktriupktarg{\bmk_u}{\bmt_u}{\bmuddt}\in\blktriupktarg{\bmk_u}{\bmt_u}{\calu}$,
and the last line is easy to evaluate using groups $\grpupjktarg{0}{k}{t}{\calu}{\ccirc}$ isomorphic 
to elementary groups.

\begin{lem}
The operation $\proditarg{\bmk_u}{\bmt_u}{\oplus}$ is well defined.
\end{lem}

\begin{prf}
Let $\blktriupktarg{\bmk_u}{\bmt_u}{\bmudt},\blktriupktarg{\bmk_u}{\bmt_u}{\bmuddt}\in\blktriupktarg{\bmk_u}{\bmt_u}{\calu}$.
Let $\blktriupktarg{\bmk_u}{\bmt_u}{\bmugr},\blktriupktarg{\bmk_u}{\bmt_u}{\bmuac}\in\blktriupktarg{\bmk_u}{\bmt_u}{\calu}$
such that $\blktriupktarg{\bmk_u}{\bmt_u}{\bmugr}=\blktriupktarg{\bmk_u}{\bmt_u}{\bmudt}$ and 
$\blktriupktarg{\bmk_u}{\bmt_u}{\bmuac}=\blktriupktarg{\bmk_u}{\bmt_u}{\bmuddt}$.
To show the operation is well defined, we need to show
$$
\blktriupktarg{\bmk_u}{\bmt_u}{\bmugr}\proditarg{\bmk_u}{\bmt_u}{\oplus}\blktriupktarg{\bmk_u}{\bmt_u}{\bmuac}
=\blktriupktarg{\bmk_u}{\bmt_u}{\bmudt}\proditarg{\bmk_u}{\bmt_u}{\oplus}\blktriupktarg{\bmk_u}{\bmt_u}{\bmuddt},
$$
or $\blktriupktarg{\bmk_u}{\bmt_u}{\bmugr\circ\bmuac}=\blktriupktarg{\bmk_u}{\bmt_u}{\bmudt\circ\bmuddt}$.
From (\ref{evaln2}), this is the same as showing
\begin{multline*}
\ldots,\triupjktarg{0}{k}{t}{\bmugr}\prodjktarg{0}{k}{t}{\ccirc}\triupjktarg{0}{k}{t}{\bmuac},
\triupjktarg{0}{k'}{t'}{\bmugr}\prodjktarg{0}{k'}{t'}{\ccirc}\triupjktarg{0}{k'}{t'}{\bmuac},\ldots \\
=\ldots,\triupjktarg{0}{k}{t}{\bmudt}\prodjktarg{0}{k}{t}{\ccirc}\triupjktarg{0}{k}{t}{\bmuddt},
\triupjktarg{0}{k'}{t'}{\bmudt}\prodjktarg{0}{k'}{t'}{\ccirc}\triupjktarg{0}{k'}{t'}{\bmuddt},\ldots,
\end{multline*}
or
\begin{multline}
\ldots,\triupjktarg{0}{k}{t}{\bmugr}\prodjktarg{0}{k}{t}{\ccirc}\triupjktarg{0}{k}{t}{\bmuac}
=\triupjktarg{0}{k}{t}{\bmudt}\prodjktarg{0}{k}{t}{\ccirc}\triupjktarg{0}{k}{t}{\bmuddt}, \\
\label{evaln3}
\triupjktarg{0}{k'}{t'}{\bmugr}\prodjktarg{0}{k'}{t'}{\ccirc}\triupjktarg{0}{k'}{t'}{\bmuac}
=\triupjktarg{0}{k'}{t'}{\bmudt}\prodjktarg{0}{k'}{t'}{\ccirc}\triupjktarg{0}{k'}{t'}{\bmuddt},\ldots.
\end{multline}
But if $\blktriupktarg{\bmk_u}{\bmt_u}{\bmugr}=\blktriupktarg{\bmk_u}{\bmt_u}{\bmudt}$, then
$$
\ldots,\triupjktarg{0}{k}{t}{\bmugr},\triupjktarg{0}{k'}{t'}{\bmugr},\ldots=
\ldots,\triupjktarg{0}{k}{t}{\bmudt},\triupjktarg{0}{k'}{t'}{\bmudt},\ldots,
$$
or
\be
\label{evaln4}
\ldots,\triupjktarg{0}{k}{t}{\bmugr}=\triupjktarg{0}{k}{t}{\bmudt},
\triupjktarg{0}{k'}{t'}{\bmugr}=\triupjktarg{0}{k'}{t'}{\bmudt},\ldots.
\ee
Similarly, if $\blktriupktarg{\bmk_u}{\bmt_u}{\bmuac}=\blktriupktarg{\bmk_u}{\bmt_u}{\bmuddt}$,
then
\be
\label{evaln5}
\ldots,\triupjktarg{0}{k}{t}{\bmuac}=\triupjktarg{0}{k}{t}{\bmuddt},
\triupjktarg{0}{k'}{t'}{\bmuac}=\triupjktarg{0}{k'}{t'}{\bmuddt},\ldots.
\ee
From Lemma \ref{lem74a}, the operation $\prodjktarg{0}{k}{t}{\ccirc}$ 
on $\triupjktarg{0}{k}{t}{\calu}$ is well defined.  Then using
(\ref{evaln4}) and (\ref{evaln5}), we have verified (\ref{evaln3}).
\end{prf}

\begin{thm}
\label{thm92}
The set $\blktriupktarg{\bmk_u}{\bmt_u}{\calu}$ with operation $\proditarg{\bmk_u}{\bmt_u}{\oplus}$
forms a group $\grpblktriupktarg{\bmk_u}{\bmt_u}{\calu}{\oplus}$.
\end{thm}

\begin{prf}
The proof is essentially the same as Theorem \ref{thm41}.  We recapitulate this proof now.
First we show the operation $\proditarg{\bmk_u}{\bmt_u}{\oplus}$ is associative.
Let $\blktriupktarg{\bmk_u}{\bmt_u}{\bmu},\blktriupktarg{\bmk_u}{\bmt_u}{\bmudt},\blktriupktarg{\bmk_u}{\bmt_u}{\bmuddt}\in\blktriupktarg{\bmk_u}{\bmt_u}{\calu}$.
We need to show
$$
(\blktriupktarg{\bmk_u}{\bmt_u}{\bmu}\proditarg{\bmk_u}{\bmt_u}{\oplus}\blktriupktarg{\bmk_u}{\bmt_u}{\bmudt})\proditarg{\bmk_u}{\bmt_u}{\oplus}\blktriupktarg{\bmk_u}{\bmt_u}{\bmuddt}
=\blktriupktarg{\bmk_u}{\bmt_u}{\bmu}\proditarg{\bmk_u}{\bmt_u}{\oplus}(\blktriupktarg{\bmk_u}{\bmt_u}{\bmudt}\proditarg{\bmk_u}{\bmt_u}{\oplus}\blktriupktarg{\bmk_u}{\bmt_u}{\bmuddt}).
$$
But using (\ref{defx}) we have
\begin{align*}
(\blktriupktarg{\bmk_u}{\bmt_u}{\bmu}\proditarg{\bmk_u}{\bmt_u}{\oplus}\blktriupktarg{\bmk_u}{\bmt_u}{\bmudt})\proditarg{\bmk_u}{\bmt_u}{\oplus}\blktriupktarg{\bmk_u}{\bmt_u}{\bmuddt}
&=\blktriupktarg{\bmk_u}{\bmt_u}{\bmu\circ\bmudt}\proditarg{\bmk_u}{\bmt_u}{\oplus}\blktriupktarg{\bmk_u}{\bmt_u}{\bmuddt} \\
&=\blktriupktarg{\bmk_u}{\bmt_u}{(\bmu\circ\bmudt)\circ\,\bmuddt},
\end{align*}
and
\begin{align*}
\blktriupktarg{\bmk_u}{\bmt_u}{\bmu}\proditarg{\bmk_u}{\bmt_u}{\oplus}(\blktriupktarg{\bmk_u}{\bmt_u}{\bmudt}\proditarg{\bmk_u}{\bmt_u}{\oplus}\blktriupktarg{\bmk_u}{\bmt_u}{\bmuddt})
&=\blktriupktarg{\bmk_u}{\bmt_u}{\bmu}\proditarg{\bmk_u}{\bmt_u}{\oplus}\blktriupktarg{\bmk_u}{\bmt_u}{\bmudt\circ\bmuddt} \\
&=\blktriupktarg{\bmk_u}{\bmt_u}{\bmu\circ\,(\bmudt\circ\bmuddt)}.
\end{align*}
Therefore the operation $\proditarg{\bmk_u}{\bmt_u}{\oplus}$ is associative since the operation $\circ$ in 
group $(\calu,\circ)$ is associative.

Let $\bmu_\bone$ be the identity of $(\calu,\circ)$.  We show
$\blktriupktarg{\bmk_u}{\bmt_u}{\bmu_\bone}$ is the identity of $\grpblktriupktarg{\bmk_u}{\bmt_u}{\calu}{\oplus}$.
Let $\bmu\in\calu$ and $\blktriupktarg{\bmk_u}{\bmt_u}{\bmu}\in\blktriupktarg{\bmk_u}{\bmt_u}{\calu}$.  
But using (\ref{defx}) we have
\begin{align*}
\blktriupktarg{\bmk_u}{\bmt_u}{\bmu_\bone}\proditarg{\bmk_u}{\bmt_u}{\oplus}\blktriupktarg{\bmk_u}{\bmt_u}{\bmu} 
&=\blktriupktarg{\bmk_u}{\bmt_u}{\bmu_\bone\circ\bmu} \\
&=\blktriupktarg{\bmk_u}{\bmt_u}{\bmu}
\end{align*}
and
\begin{align*}
\blktriupktarg{\bmk_u}{\bmt_u}{\bmu}\proditarg{\bmk_u}{\bmt_u}{\oplus}\blktriupktarg{\bmk_u}{\bmt_u}{\bmu_\bone} 
&=\blktriupktarg{\bmk_u}{\bmt_u}{\bmu\circ\bmu_\bone} \\
&=\blktriupktarg{\bmk_u}{\bmt_u}{\bmu}.
\end{align*}

Fix $\bmu\in\calu$.  Let $\bmubr$ be the inverse of $\bmu$ in $(\calu,\circ)$.
We show $\blktriupktarg{\bmk_u}{\bmt_u}{\bmubr}$ is the inverse of $\blktriupktarg{\bmk_u}{\bmt_u}{\bmu}$
in $\grpblktriupktarg{\bmk_u}{\bmt_u}{\calu}{\oplus}$.  But using (\ref{defx}) we have
\begin{align*}
\blktriupktarg{\bmk_u}{\bmt_u}{\bmubr}\proditarg{\bmk_u}{\bmt_u}{\oplus}\blktriupktarg{\bmk_u}{\bmt_u}{\bmu} 
&=\blktriupktarg{\bmk_u}{\bmt_u}{\bmubr\circ\bmu} \\
&=\blktriupktarg{\bmk_u}{\bmt_u}{\bmu_\bone}
\end{align*}
and
\begin{align*}
\blktriupktarg{\bmk_u}{\bmt_u}{\bmu}\proditarg{\bmk_u}{\bmt_u}{\oplus}\blktriupktarg{\bmk_u}{\bmt_u}{\bmubr} 
&=\blktriupktarg{\bmk_u}{\bmt_u}{\bmu\circ\bmubr} \\
&=\blktriupktarg{\bmk_u}{\bmt_u}{\bmu_\bone}.
\end{align*}
Together these results show $\grpblktriupktarg{\bmk_u}{\bmt_u}{\calu}{\oplus}$ is a group.
\end{prf}

We have the following analogue of Theorem \ref{homo3}.

\begin{thm}
\label{homo5}
There is a homomorphism from $(\calu,\circ)$ to $\grpblktriupktarg{\bmk_u}{\bmt_u}{\calu}{\oplus}$
given by the map $\omega_{\bmk_u}^{\bmt_u}:  \calu\ra\blktriupktarg{\bmk_u}{\bmt_u}{\calu}$ with assignment
$\omega_{\bmk_u}^{\bmt_u}:  \bmu\mapsto\blktriupktarg{\bmk_u}{\bmt_u}{\bmu}$.
\end{thm}

\begin{prf}
This follows immediately from the definition of operation $\proditarg{\bmk_u}{\bmt_u}{\oplus}$.
\end{prf}

\begin{cor}
For any paired sequence $(\bmk,\bmt)$ and its
complementary paired sequence $(\bmk_u,\bmt_u)$ of $(\bmk,\bmt)$, we have
$(\calu,\circ)/(\blktrilwktargnop{\bmk}{\bmt}{\call},\circ)\simeq\grpblktriupktarg{\bmk_u}{\bmt_u}{\calu}{\oplus}$.
\end{cor}

\begin{prf}
Note that $(\blktrilwktargnop{\bmk}{\bmt}{\call},\circ)$ is the kernel of the 
homomorphism $\omega_{\bmk_u}^{\bmt_u}$.  Now the result follows from the first homomorphism theorem.
\end{prf}

We now show there is a homomorphism from $A$ to $\grpblktriupktarg{\bmk_u}{\bmt_u}{\calu}{\oplus}$.
Combining Theorem \ref{homo7} and Theorem \ref{homo5} gives the following.

\begin{thm}
\label{homo9}
There is a homomorphism from $A$ to $\grpblktriupktarg{\bmk_u}{\bmt_u}{\calu}{\oplus}$
given by the composition map
$\omega_{\bmk_u}^{\bmt_u}\bullet\xi:  A\ra\blktriupktarg{\bmk_u}{\bmt_u}{\calu}$,
where isomorphism $A\simeq(\calu,\circ)$ is given by the bijection 
$\xi:  A\ra\calu$ in Theorem \ref{homo7}, and the
homomorphism from $(\calu,\circ)$ to $\grpblktriupktarg{\bmk_u}{\bmt_u}{\calu}{\oplus}$
is given by the map $\omega_{\bmk_u}^{\bmt_u}:  \calu\ra\blktriupktarg{\bmk_u}{\bmt_u}{\calu}$
in Theorem \ref{homo5}.
\end{thm}

\vspace{3mm}
{\bf 5.3  Normal chains of the generator group (harmonic theory)}%hhh
\vspace{3mm}

A {\it filling sequence} $\bmf$ of $\bmn$ is a walk of ordered pairs in $\bmn$ in (\ref{top}) 
that includes all ordered pairs in $\bmn$ once and only once,
$$
\bmf\rmdef(k',t'),(k'',t''),\ldots,(k''',t'''),\ldots.
$$
The ordered pairs $(k',t')$ in $\bmf$ are called {\it filled ordered pairs} in $\bmn$.
Clearly, the filling sequence $\bmf$ is specified by a sequence of filled ordered pairs.
The filled ordered pairs $(k',t')$ in $\bmf$ can be in any time order.

For each time epoch $\tau$, $\tau>0$, we let $\bmf^\tau$ be the {\it filling subsequence} 
of $\bmf$ of the $\tau$ ordered pairs at the start of the filling sequence.
Then $\bmf^1=(k',t')$, $\bmf^2=(k',t'),(k'',t'')$, and so on.
Fix time epoch $\tau$, $\tau>0$, and consider any filling subsequence $\bmf^\tau$ of $\bmf$
at time epoch $\tau$,
\be
\label{ufop11}
\bmf^\tau=(k',t'),(k'',t''),\ldots.
\ee
The ordered pairs in $\bmf^\tau$ are filled ordered pairs in $\bmn$ 
at time epoch $\tau$, and the ordered pairs not in $\bmf^\tau$ are called
{\it unfilled ordered pairs} in $\bmn$ at time epoch $\tau$.
The ordered pairs in $\bmf^{\tau+1}$ are the filled ordered pairs in $\bmf^\tau$
and one unfilled ordered pair in $\bmn$ at time epoch $\tau$.

We now subject the filling subsequence $\bmf^\tau$ of $\bmf$
at time epoch $\tau$ to another constraint.  We require that the filled ordered pairs
in $\bmf^\tau$ form a sequence $\blktrilwktarg{\bmk^\tau}{\bmt^\tau}{\bmn}$ of lower elementary triangles
\be
\label{seqtri1}
\blktrilwktarg{\bmk^\tau}{\bmt^\tau}{\bmn}\rmdef
\ldots,\trilwjktarg{0}{k''}{t''}{\bmn},\ldots,\trilwjktarg{0}{k'}{t'}{\bmn},\ldots,
\ee
indexed by a paired sequence
$$
(\bmk^\tau,\bmt^\tau)=\ldots,(k'',t''),\ldots,(k',t'),\ldots.
$$
We call a filling subsequence $\bmf^\tau$ with this property a {\it normal filling subsequence} of $\bmf$.
If $\bmf^\tau$ is a normal filling subsequence of $\bmf$ for all $\tau>0$, we say
$\bmf$ is a {\it normal filling sequence} of $\bmn$.  If
$\bmf$ is a normal filling sequence of $\bmn$, the filled ordered pairs
in $\bmf^\tau$ must form a sequence $\blktrilwktarg{\bmk^\tau}{\bmt^\tau}{\bmn}$ 
of lower elementary triangles, and the filled ordered pairs
in $\bmf^{\tau+1}$ must form a sequence $\blktrilwktarg{\bmk^{\tau+1}}{\bmt^{\tau+1}}{\bmn}$ 
of lower elementary triangles.  This means the sequence
of lower elementary triangles $\blktrilwktarg{\bmk^{\tau+1}}{\bmt^{\tau+1}}{\bmn}$ is 
formed from the sequence of lower elementary triangles $\blktrilwktarg{\bmk^\tau}{\bmt^\tau}{\bmn}$
and the unfilled ordered pair in $\bmn$ at time epoch $\tau$ that is in $\bmf^{\tau+1}$
but not $\bmf^\tau$.

\begin{thm}
Let $\bmf$ be a normal filling sequence of $\bmn$.  For each $\tau>0$, let $\bmf^\tau$ be a 
normal filling subsequence of $\bmf$ with a paired sequence $(\bmk^\tau,\bmt^\tau)$.
For each $\tau>0$, $(\blktrilwktargnop{\bmk^\tau}{\bmt^\tau}{\call},\circ)$ 
is a normal subgroup of $(\calu,\circ)$.
\end{thm}

\begin{prf}
This follows from Lemma \ref{nsubg}.
\end{prf}

Suppose $(k^+,t^+)$ is the filled ordered pair of $\bmf^{\tau+1}$ that is not filled in $\bmf^\tau$.
Then we must have a filled triangle $\trilwjktarg{0}{k^+}{t^+}{\bmn}$ in 
$\blktrilwktarg{\bmk^{\tau+1}}{\bmt^{\tau+1}}{\bmn}$.
If we are to have a filled triangle $\trilwjktarg{0}{k^+}{t^+}{\bmn}$ in $\blktrilwktarg{\bmk^{\tau+1}}{\bmt^{\tau+1}}{\bmn}$,
then triangles $\trilwjktarg{0}{k^+-1}{t^++1}{\bmn}$ and $\trilwjktarg{0}{k^+-1}{t^+}{\bmn}$
must be filled in $\blktrilwktarg{\bmk^\tau}{\bmt^\tau}{\bmn}$.  Triangle $\trilwjktarg{0}{k^+-1}{t^++1}{\bmn}$
cannot be a subtriangle of a larger triangle in $\blktrilwktarg{\bmk^\tau}{\bmt^\tau}{\bmn}$ because $(k^+,t^+)$ is unfilled
in $\blktrilwktarg{\bmk^\tau}{\bmt^\tau}{\bmn}$.  However triangle $\trilwjktarg{0}{k^+-1}{t^+}{\bmn}$ can be a subtriangle
of a larger triangle $\trilwjktarg{0}{k^+-1+m^+}{t^+-m^+}{\bmn}$ in $\blktrilwktarg{\bmk^\tau}{\bmt^\tau}{\bmn}$, where $m^+>0$ and
$k^+-1+m^+\le\ell$.  Then $\blktrilwktarg{\bmk^\tau}{\bmt^\tau}{\bmn}$ must be the sequence of triangles
\be
\label{seqtri3}
\blktrilwktarg{\bmk^\tau}{\bmt^\tau}{\bmn}=
\ldots,\trilwjktarg{0}{k''}{t''}{\bmn},\ldots,\trilwjktarg{0}{k^+-1}{t^++1}{\bmn},\trilwjktarg{0}{k^+-1+m^+}{t^+-m^+}{\bmn},\trilwjktarg{0}{k^\gamma}{t^\gamma}{\bmn},\ldots,\trilwjktarg{0}{k'}{t'}{\bmn},\ldots,
\ee
indexed by paired sequence
$$
(\bmk^\tau,\bmt^\tau)=\ldots,(k'',t''),\ldots,(k^+-1,t^++1),(k^+-1+m^+,t^+-m^+),(k^\gamma,t^\gamma),\ldots,(k',t'),\ldots,
$$
where $m^+\ge 0$ and $k^+-1+m^+\le\ell$, and $t^\gamma<t^+-m^+$.  If $m^+=0$,   
then $\blktrilwktarg{\bmk^{\tau+1}}{\bmt^{\tau+1}}{\bmn}$ is given by the sequence of triangles
\be
\label{seqtri2a}
\blktrilwktarg{\bmk^{\tau+1}}{\bmt^{\tau+1}}{\bmn}=
\ldots,\trilwjktarg{0}{k''}{t''}{\bmn},\ldots,\trilwjktarg{0}{k^+}{t^+}{\bmn},\trilwjktarg{0}{k^\gamma}{t^\gamma}{\bmn},\ldots,\trilwjktarg{0}{k'}{t'}{\bmn},\ldots,
\ee
indexed by paired sequence
$$
(\bmk^{\tau+1},\bmt^{\tau+1})=\ldots,(k'',t''),\ldots,(k^+,t^+),(k^\gamma,t^\gamma),\ldots,(k',t'),\ldots.
$$
If $m^+>0$,   
then $\blktrilwktarg{\bmk^{\tau+1}}{\bmt^{\tau+1}}{\bmn}$ is given by the sequence of triangles
\be
\label{seqtri2b}
\blktrilwktarg{\bmk^{\tau+1}}{\bmt^{\tau+1}}{\bmn}=
\ldots,\trilwjktarg{0}{k''}{t''}{\bmn},\ldots,\trilwjktarg{0}{k^+}{t^+}{\bmn},\trilwjktarg{0}{k^+-1+m^+}{t^+-m^+}{\bmn},\trilwjktarg{0}{k^\gamma}{t^\gamma}{\bmn},\ldots,\trilwjktarg{0}{k'}{t'}{\bmn},\ldots,
\ee
indexed by paired sequence
$$
(\bmk^{\tau+1},\bmt^{\tau+1})=\ldots,(k'',t''),\ldots,(k^+-1,t^++1),(k^+-1+m^+,t^+-m^+),(k^\gamma,t^\gamma),\ldots,(k',t'),\ldots.
$$
In either case, we see that the sequence of triangles 
$\blktrilwktarg{\bmk^\tau}{\bmt^\tau}{\bmn}$ in (\ref{seqtri3}) is contained in the sequence of triangles 
$\blktrilwktarg{\bmk^{\tau+1}}{\bmt^{\tau+1}}{\bmn}$ in (\ref{seqtri2a}) or  (\ref{seqtri2b}), which we write as
$\blktrilwktarg{\bmk^\tau}{\bmt^\tau}{\bmn}\subset\blktrilwktarg{\bmk^{\tau+1}}{\bmt^{\tau+1}}{\bmn}$.

\begin{lem}
\label{lem81}
Let $\bmf$ be a normal filling sequence of $\bmn$.  For each $\tau>0$, let $\bmf^\tau$ be a 
normal filling subsequence of $\bmf$ with a paired sequence $(\bmk^\tau,\bmt^\tau)$.
At each time epoch $\tau>0$, we have
$(\blktrilwktexpargnop{\bmk}{\bmt}{\call}{\tau},\circ)
\subset(\blktrilwktexpargnop{\bmk}{\bmt}{\call}{\tau+1},\circ)$.  In addition, we have
$\bmu_\bone\subset(\blktrilwktexpargnop{\bmk}{\bmt}{\call}{1},\circ)$.
\end{lem}

\begin{prf}
The filled ordered pairs in $\bmn$ at time epoch $\tau$ are the sequence of 
triangles $\blktrilwktarg{\bmk^\tau}{\bmt^\tau}{\bmn}$, and the filled ordered pairs in $\bmn$
at time epoch $\tau+1$ are the sequence of triangles
$\blktrilwktarg{\bmk^{\tau+1}}{\bmt^{\tau+1}}{\bmn}$.
The sequence of triangles $\blktrilwktarg{\bmk^\tau}{\bmt^\tau}{\bmn}$
is contained in the sequence of triangles $\blktrilwktarg{\bmk^{\tau+1}}{\bmt^{\tau+1}}{\bmn}$, or
$\blktrilwktarg{\bmk^\tau}{\bmt^\tau}{\bmn}\subset\blktrilwktarg{\bmk^{\tau+1}}{\bmt^{\tau+1}}{\bmn}$.
This means $(\blktrilwktexpargnop{\bmk}{\bmt}{\call}{\tau},\circ)
\subset(\blktrilwktexpargnop{\bmk}{\bmt}{\call}{\tau+1},\circ)$.
In a similar way, we have $\bmn\subset\blktrilwktarg{\bmk^1}{\bmt^1}{\bmn}$ and
then $\bmu_\bone\subset(\blktrilwktexpargnop{\bmk}{\bmt}{\call}{1},\circ)$.
\end{prf}

Since $(\blktrilwktexpargnop{\bmk}{\bmt}{\call}{\tau},\circ)$ is a normal subroup of $(\calu,\circ)$ 
for any paired sequence $(\bmk^\tau,\bmt^\tau)$ at any time epoch $\tau$, 
then from Lemma \ref{lem81}, for any normal filling sequence $\bmf$ of $\bmn$ 
we may construct a normal chain
\be
\label{nc1}
\bmu_\bone\subset(\blktrilwktexpargnop{\bmk}{\bmt}{\call}{1},\circ)
\subset\cdots\subset(\blktrilwktexpargnop{\bmk}{\bmt}{\call}{\tau},\circ)
\subset(\blktrilwktexpargnop{\bmk}{\bmt}{\call}{\tau+1},\circ)\subset\cdots\subset(\calu,\circ).
\ee
We say (\ref{nc1}) is the {\it normal chain of a normal filling sequence} $\bmf$ of $\bmn$.
This gives the following.

\begin{thm}
Any normal filling sequence $\bmf$ of $\bmn$ gives a normal chain (\ref{nc1}) of $(\calu,\circ)$.
\end{thm}

Each normal chain (\ref{nc1}) of $(\calu,\circ)$ gives a \cdc\ of $(\calu,\circ)$.
We now find \creps\ of the \cdc.  First we find \creps\ of the quotient group 
$(\blktrilwktexpargnop{\bmk}{\bmt}{\call}{\tau+1},\circ)/(\blktrilwktexpargnop{\bmk}{\bmt}{\call}{\tau},\circ)$.
A \crep\ of 
$(\blktrilwktexpargnop{\bmk}{\bmt}{\call}{\tau+1},\circ)/(\blktrilwktexpargnop{\bmk}{\bmt}{\call}{\tau},\circ)$ 
is an element of $(\blktrilwktexpargnop{\bmk}{\bmt}{\call}{\tau+1},\circ)$ that is not an element 
of $(\blktrilwktexpargnop{\bmk}{\bmt}{\call}{\tau},\circ)$.
Suppose $(k^+,t^+)$ is the filled ordered pair
of $\bmf^{\tau+1}$ that is not filled in $\bmf^\tau$.  
Clearly generator $\bmu_{g,k^+}^{t^+}$ of $(\calu,\circ)$ is an element of 
$(\blktrilwktexpargnop{\bmk}{\bmt}{\call}{\tau+1},\circ)$ that is not an element of 
$(\blktrilwktexpargnop{\bmk}{\bmt}{\call}{\tau},\circ)$.
Then the set of generators $\{\bmu_{g,k^+}^{t^+}:  g_{k^+}^{t^+}\in G_{k^+}^{t^+}\}$
are elements of 
$(\blktrilwktexpargnop{\bmk}{\bmt}{\call}{\tau+1},\circ)$ that are not elements of 
$(\blktrilwktexpargnop{\bmk}{\bmt}{\call}{\tau},\circ)$.
For any $g_{k^+}^{t^+}\in G_{k^+}^{t^+}$, we show that
$\bmu_{g,k^+}^{t^+}\circ(\blktrilwktexpargnop{\bmk}{\bmt}{\call}{\tau},\circ)$
is a coset of $(\blktrilwktexpargnop{\bmk}{\bmt}{\call}{\tau},\circ)$ in
$(\blktrilwktexpargnop{\bmk}{\bmt}{\call}{\tau+1},\circ)$.
First, it is clear that $\bmu_{g,k^+}^{t^+}\circ(\blktrilwktexpargnop{\bmk}{\bmt}{\call}{\tau},\circ)$
is a coset of $(\blktrilwktexpargnop{\bmk}{\bmt}{\call}{\tau},\circ)$.
Next we show the coset is contained in
$(\blktrilwktexpargnop{\bmk}{\bmt}{\call}{\tau+1},\circ)$.

\begin{lem}
\label{lem60}
We have that $\bmu_{g,k^+}^{t^+}\circ(\blktrilwktargnop{\bmk^\tau}{\bmt^\tau}{\call},\circ)$
is a coset of $(\blktrilwktargnop{\bmk^\tau}{\bmt^\tau}{\call},\circ)$.
The coset $\bmu_{g,k^+}^{t^+}\circ(\blktrilwktargnop{\bmk^\tau}{\bmt^\tau}{\call},\circ)$ is contained in
$(\blktrilwktargnop{\bmk^{\tau+1}}{\bmt^{\tau+1}}{\call},\circ)$, and the cosets
$\bmu_{g,k^+}^{t^+}\circ(\blktrilwktargnop{\bmk^\tau}{\bmt^\tau}{\call},\circ)$ for each
$g_{k^+}^{t^+}\in G_{k^+}^{t^+}$ are disjoint.
\end{lem}

\begin{prf}
Let $\bmu\in(\blktrilwktargnop{\bmk^\tau}{\bmt^\tau}{\call},\circ)$.
We have seen above that if $(k^+,t^+)$ is the filled ordered pair of $\bmf^{\tau+1}$ 
that is not filled in $\bmf^\tau$, then
$\blktrilwktarg{\bmk^\tau}{\bmt^\tau}{\bmn}\subset\blktrilwktarg{\bmk^{\tau+1}}{\bmt^{\tau+1}}{\bmn}$.
This means that the sequence of triangles $\blktrilwktarg{\bmk^\tau}{\bmt^\tau}{\bmu}$
is contained in the sequence of triangles $\blktrilwktarg{\bmk^{\tau+1}}{\bmt^{\tau+1}}{\bmu}$.
In other words, we have $\bmu\in(\blktrilwktargnop{\bmk^{\tau+1}}{\bmt^{\tau+1}}{\call},\circ)$.
Clearly $\bmu_{g,k^+}^{t^+}\in(\blktrilwktargnop{\bmk^{\tau+1}}{\bmt^{\tau+1}}{\call},\circ)$.
Then $\bmu_{g,k^+}^{t^+}\circ\bmu$ is in $(\blktrilwktargnop{\bmk^{\tau+1}}{\bmt^{\tau+1}}{\call},\circ)$.
Since $\bmu_{g,k^+}^{t^+}\circ\bmu\in(\blktrilwktargnop{\bmk^{\tau+1}}{\bmt^{\tau+1}}{\call},\circ)$,
the coset $\bmu_{g,k^+}^{t^+}\circ(\blktrilwktargnop{\bmk^\tau}{\bmt^\tau}{\call},\circ)$ is contained in
$(\blktrilwktargnop{\bmk^{\tau+1}}{\bmt^{\tau+1}}{\call},\circ)$. It is clear the cosets
$\bmu_{g,k^+}^{t^+}\circ(\blktrilwktargnop{\bmk^\tau}{\bmt^\tau}{\call},\circ)$ are disjoint for each
$g_{k^+}^{t^+}\in G_{k^+}^{t^+}$.
\end{prf}

\begin{lem}
\label{lem61}
The cosets $\bmu_{g,k^+}^{t^+}\circ(\blktrilwktargnop{\bmk^\tau}{\bmt^\tau}{\call},\circ)$
for each $g_{k^+}^{t^+}\in G_{k^+}^{t^+}$ are all the cosets of 
$(\blktrilwktargnop{\bmk^\tau}{\bmt^\tau}{\call},\circ)$
in $(\blktrilwktargnop{\bmk^{\tau+1}}{\bmt^{\tau+1}}{\call},\circ)$.
\end{lem}

\begin{prf}
The cardinality of $(\blktrilwktargnop{\bmk^{\tau+1}}{\bmt^{\tau+1}}{\call},\circ)$
is $|G_k^t|\times |(\blktrilwktargnop{\bmk^\tau}{\bmt^\tau}{\call},\circ)|$.
But this is the same as the cardinality of all the cosets
$\bmu_{g,k^+}^{t^+}\circ(\blktrilwktargnop{\bmk^\tau}{\bmt^\tau}{\call},\circ)$ for each
$g_{k^+}^{t^+}\in G_{k^+}^{t^+}$. 
\end{prf}

Similarly the \creps\ of $(\blktrilwktexpargnop{\bmk}{\bmt}{\call}{1},\circ)/\bmu_\bone$
are the generators $\{\bmu_{g,k^1}^{t^1}:  g_{k^1}^{t^1}\in G_{k^1}^{t^1}\}$
where $(k^1,t^1)$ is the filled ordered pair in $\bmf^1$.
Combining Lemmas \ref{lem60} and \ref{lem61} gives the following.

\begin{thm}
The generators of $(\calu,\circ)$ form a \crepc\ of any \cdc\ of $(\calu,\circ)$
formed by the normal chain (\ref{nc1}) of $(\calu,\circ)$ of any normal filling 
sequence $\bmf$ of $\bmn$.
\end{thm}

Under the bijection $\beta^{-1}:  \calu\ra\calr$ in Theorem \ref{thm88a},
a generator $\bmu_{g,k}^t$ in $(\calu,\circ)$ is taken to a generator $\bmr_g^{[t,t+k]}$ 
in $(\calr,*)$, with assignment $\beta^{-1}:  \bmu_{g,k}^t\mapsto\bmr_g^{[t,t+k]}$.  
Under the bijection $\alpha:  \calr\ra A$ in Theorem \ref{thm31}, a generator $\bmr_g^{[t,t+k]}$ 
in $(\calr,*)$ is taken to a generator $\bmg^{[t,t+k]}$ in $A$, with assignment
$\alpha:  \bmr_g^{[t,t+k]}\mapsto\bmg^{[t,t+k]}$.
Thus the bijection $\alpha\bullet\beta^{-1}:  \calu\ra A$ takes the set of generators 
$\bmu_{g,k}^t$ in $(\calu,\circ)$ to the set of generators $\bmg^{[t,t+k]}$ in $A$.
Furthermore, the bijection $\alpha\bullet\beta^{-1}$ is an isomorphism between
$(\calu,\circ)$ and $A$.  Therefore a normal chain of $(\calu,\circ)$
is taken to a normal chain of $A$, and a \crep\ in $(\calu,\circ)$ is taken to
a \crep\ in $A$.  Then we can reconstruct the set of elements $A$ using the \creps\ of 
the \cdc\ of $(\calu,\circ)$ obtained from any normal filling sequence of $\bmn$.
This gives the following.

\begin{thm}
\label{recon}
We can reconstruct $A$ using the generators of $(\calu,\circ)$ as
the \creps\ of the \cdc\ of $(\calu,\circ)$ obtained from any normal filling sequence of $\bmn$.
\end{thm}
We now give some examples of this result.

In general there are many normal filling sequences $\bmf$ of $\bmn$
and therefore many normal chains of $(\calu,\circ)$.  We now give 4 
important examples of normal filling sequences of $\bmn$.  We specify the filling
sequence by the sequence of filled ordered pairs.  If we go left to right
in (\ref{top}), the time index decreases.  We call this {\it reverse time}.
If we go right to left, the time index increases.  We call this {\it forward time}.
For the first example of a sequence of filled ordered pairs, we go up the columns in (\ref{top}),
from left to right.  This gives the sequence of filled ordered pairs $(k,t)$,
\be
\label{tdndsbt}
\ldots,(0,t),(1,t),(2,t),\ldots,(j,t),\ldots,(\ell,t),
              (0,t-1),(1,t-1),(2,t-1),\ldots,(j,t-1),\ldots,(\ell,t-1),\ldots.
\ee
It can be easily verified that this gives a normal filling sequence.
We call this the {\it time domain normal filling sequence of $\bmn$} in {\it reverse time}.
For the second example of a sequence of filled ordered pairs,
we go up the diagonals in (\ref{top}), from right to left.
This gives the sequence of filled ordered pairs $(k,t)$,
$$
\ldots,(0,t),(1,t-1),(2,t-2),\ldots,(j,t-j),\ldots,(\ell,t-\ell),
              (0,t+1),(1,t),(2,t-1),\ldots,(j,t-j+1),\ldots,(\ell,t-\ell+1),\ldots.
$$
We call this the time domain normal filling sequence of $\bmn$ in {\it forward time}.
Note that we go up the columns in reverse time and up the diagonals in forward time
because of the triangular shape of $\triupjktarg{0}{0}{t}{\bmu}$.
Next we go row by row in (\ref{top}), from bottom row to top row and left to right.
This gives the sequence of filled ordered pairs $(k,t)$,
\be
\label{sdndsbt}
\ldots,(0,t),(0,t-1),\ldots,\ldots,(1,t),(1,t-1),\ldots,(2,t),(2,t-1),
\ldots,(j,t),(j,t-1),\ldots,(\ell,t),(\ell,t-1),\ldots.
\ee
We call this the {\it spectral domain normal filling sequence of $\bmn$} in {\it reverse time}.
Finally we may go row by row in (\ref{top}), from bottom row to top row and right to left.
This gives the sequence of filled ordered pairs $(k,t)$,
$$
\ldots,(0,t-1),(0,t),\ldots,(1,t-1),(1,t),\ldots,(2,t-1),(2,t),
\ldots,(j,t-1),(j,t),\ldots,(\ell,t-1),(\ell,t),\ldots.
$$
We call this the spectral domain normal filling sequence of $\bmn$ in {\it forward time}.
The examples given here can be viewed as limiting cases; in general a normal filling
sequence need not be time ordered in forward or reverse time.

If we use generators of $(\calu,\circ)$ obtained from the normal chain of the time domain
normal filling sequence of $\bmn$ in reverse time (\ref{tdndsbt}), 
we obtain the time domain encoder of $A$ 
in (\ref{enctdc}).  If we use generators of $(\calu,\circ)$ obtained from the normal chain of the 
spectral domain normal filling sequence of $\bmn$ in reverse time (\ref{sdndsbt}), 
we obtain the Forney-Trott spectral domain encoder of $A$ 
in (\ref{encftd}).  Thus the time domain encoder of $A$ discussed in Section 3
and the spectral domain encoder of $A$ in \cite{FT} can be more easily
obtained just from the normal chain of their respective normal filling sequence of $\bmn$.
In addition, there are many other encoders and \crepc s of $A$ that can be obtained from
the normal chain of many other normal filling sequences of $\bmn$.

Let $(\bmk_u,\bmt_u)$ be any paired sequence.
Let group $\grpblktriupktarg{\bmk_u}{\bmt_u}{\calu}{\oplus}$
be the set of elementary groups formed from paired sequence $(\bmk_u,\bmt_u)$ as in Subsection 5.2.
We now find a normal chain of the group $\grpblktriupktarg{\bmk_u}{\bmt_u}{\calu}{\oplus}$
in much the same way as we just found normal chains of $(\calu,\circ)$ using the
normal filling sequence.

From Lemma \ref{lem47},
for any paired sequence $(\bmk_u,\bmt_u)$, there is a complementary paired sequence 
$(\bmk',\bmt')$ of $(\bmk_u,\bmt_u)$, and $\blktrilwktarg{\bmk'}{\bmt'}{\bmn}$ and 
$\blktriupktarg{\bmk_u}{\bmt_u}{\bmn}$ partition $\bmn$ into two sawtooth patterns.  
A filling sequence $\bmf$ of $\bmn$ is a walk of ordered pairs in $\bmn$
that includes all ordered pairs in $\bmn$ once and only once.  To study
$\grpblktriupktarg{\bmk_u}{\bmt_u}{\calu}{\oplus}$, we consider a filling
sequence $\bmf_\blktridn$ of $\bmn$,
$$
\bmf_\blktridn\rmdef[\blktrilwktarg{\bmk'}{\bmt'}{\bmn}],(k',t'),(k'',t''),\ldots,(k''',t'''),\ldots,
$$
where $[\blktrilwktarg{\bmk'}{\bmt'}{\bmn}]$ is any sequence of all the ordered pairs
in $\blktrilwktarg{\bmk'}{\bmt'}{\bmn}$, and the remaining sequence
$(k',t'),(k'',t''),\ldots,(k''',t'''),\ldots$ are all the ordered pairs
in $\blktriupktarg{\bmk_u}{\bmt_u}{\bmn}$.

For each time epoch $\tau$, $\tau>0$, we let $\bmf_\blktridn^\tau$ be the {\it filling subsequence} 
of $\bmf_\blktridn$ of the ordered pairs in $[\blktrilwktarg{\bmk'}{\bmt'}{\bmn}]$ 
and the first $\tau$ ordered pairs after 
the ordered pairs in $[\blktrilwktarg{\bmk'}{\bmt'}{\bmn}]$.
Then $\bmf_\blktridn^1=[\blktrilwktarg{\bmk'}{\bmt'}{\bmn}],(k',t')$, 
$\bmf_\blktridn^2=[\blktrilwktarg{\bmk'}{\bmt'}{\bmn}],(k',t'),(k'',t'')$, and so on.
Fix time epoch $\tau$, $\tau>0$, and consider any filling subsequence $\bmf_\blktridn^\tau$ 
of $\bmf_\blktridn$ at time epoch $\tau$,
\be
%\label{ufop11}
\bmf_\blktridn^\tau=[\blktrilwktarg{\bmk'}{\bmt'}{\bmn}],(k',t'),(k'',t''),\ldots.
\ee
The ordered pairs in $\bmf_\blktridn^\tau$ are filled ordered pairs in
$\blktrilwktarg{\bmk'}{\bmt'}{\bmn}$ and $\blktriupktarg{\bmk_u}{\bmt_u}{\bmn}$
at time epoch $\tau$, and the ordered pairs not in $\bmf_\blktridn^\tau$ are
unfilled ordered pairs in $\blktriupktarg{\bmk_u}{\bmt_u}{\bmn}$ at time epoch $\tau$.

We now subject the filling subsequence $\bmf_\blktridn^\tau$ of $\bmf_\blktridn$
at time epoch $\tau$ to another constraint.  We require that the filled ordered pairs
in $\bmf_\blktridn^\tau$ form a sequence $\blktrilwktarg{\bmk_\blktridn^\tau}{\bmt_\blktridn^\tau}{\bmn}$ 
of lower elementary triangles
\be
%\label{seqtri1}
\blktrilwktarg{\bmk_\blktridn^\tau}{\bmt_\blktridn^\tau}{\bmn}\rmdef
\ldots,\trilwjktarg{0}{k''}{t''}{\bmn},\ldots,\trilwjktarg{0}{k^\epsilon}{t^\epsilon}{\bmn},\ldots,\trilwjktarg{0}{k'}{t'}{\bmn},\ldots,
\ee
indexed by a paired sequence
\be
\label{seq5}
(\bmk_\blktridn^\tau,\bmt_\blktridn^\tau)=\ldots,(k'',t''),\ldots,(k^\epsilon,t^\epsilon),\ldots,(k',t'),\ldots.
\ee
In (\ref{seq5}), the ordered pair $(k^\epsilon,t^\epsilon)$ is in
$(\bmk',\bmt')$, and the ordered pairs $(k'',t''),(k',t')$ are not in $(\bmk',\bmt')$.
Note that the ordered pairs in $[\blktrilwktarg{\bmk'}{\bmt'}{\bmn}]$ already form a 
sequence of lower elementary triangles $\blktrilwktarg{\bmk'}{\bmt'}{\bmn}$
indexed by $(\bmk',\bmt')$, and so we just need
to ensure that we can use $(\bmk',\bmt')$ and the additional ordered pairs 
$(k',t'),(k'',t''),\ldots$ in $\bmf_\blktridn^\tau$ to 
form a paired sequence $(\bmk_\blktridn^\tau,\bmt_\blktridn^\tau)$
and a sequence $\blktrilwktarg{\bmk_\blktridn^\tau}{\bmt_\blktridn^\tau}{\bmn}$ of lower elementary triangles.
We call a filling subsequence $\bmf_\blktridn^\tau$ 
with this property a {\it normal filling subsequence} of $\bmf_\blktridn$.
If $\bmf_\blktridn^\tau$ is a normal filling subsequence of $\bmf_\blktridn$ for all $\tau>0$, we say
$\bmf_\blktridn$ is a {\it normal filling sequence} of $\bmn$.

\begin{thm}
Let $\bmf_\blktridn$ be a normal filling sequence of $\bmn$.
For each $\tau>0$, let $\bmf_\blktridn^\tau$ be a 
normal filling subsequence of $\bmf_\blktridn$ with a paired sequence $(\bmk_\blktridn^\tau,\bmt_\blktridn^\tau)$.
For each $\tau>0$, $(\blktrilwktargnop{\bmk_\blktridn^\tau}{\bmt_\blktridn^\tau}{\call},\circ)$ 
is a normal subgroup of $(\calu,\circ)$.
\end{thm}

\begin{prf}
This follows from Lemma \ref{nsubg}.
\end{prf}

Previously we have seen that if
$(k^+,t^+)$ is the filled ordered pair of $\bmf^{\tau+1}$ that is not filled in $\bmf^\tau$,
then $\blktrilwktarg{\bmk^\tau}{\bmt^\tau}{\bmn}\subset\blktrilwktarg{\bmk^{\tau+1}}{\bmt^{\tau+1}}{\bmn}$,
where $(\bmk^\tau,\bmt^\tau)$ is the paired sequence of $\bmf^\tau$
and $(\bmk^{\tau+1},\bmt^{\tau+1})$ is the paired sequence of $\bmf^{\tau+1}$.
Now suppose $(k^+,t^+)$ is the filled ordered pair of $\bmf_\blktridn^{\tau+1}$ that is not 
filled in $\bmf_\blktridn^\tau$.  Then in a similar way, we have  
$\blktrilwktarg{\bmk_\blktridn^\tau}{\bmt_\blktridn^\tau}{\bmn}\subset\blktrilwktarg{\bmk_\blktridn^{\tau+1}}{\bmt_\blktridn^{\tau+1}}{\bmn}$,
where $(\bmk_\blktridn^\tau,\bmt_\blktridn^\tau)$ is the paired sequence of $\bmf_\blktridn^\tau$
and $(\bmk_\blktridn^{\tau+1},\bmt_\blktridn^{\tau+1})$ is the paired sequence of $\bmf_\blktridn^{\tau+1}$.
This gives the following.

\begin{lem}
\label{lem81a}
Let $\bmf_\blktridn$ be a normal filling sequence of $\bmn$.
For each $\tau>0$, let $\bmf_\blktridn^\tau$ be a normal filling subsequence of 
$\bmf_\blktridn$ with a paired sequence $(\bmk_\blktridn^\tau,\bmt_\blktridn^\tau)$.
At each time epoch $\tau>0$, we have
$(\blktrilwktexpargnop{\bmk_\blktridn}{\bmt_\blktridn}{\call}{\tau},\circ)
\subset(\blktrilwktexpargnop{\bmk_\blktridn}{\bmt_\blktridn}{\call}{\tau+1},\circ)$.  In addition, we have
$(\blktrilwktargnop{\bmk'}{\bmt'}{\call},\circ)
\subset(\blktrilwktexpargnop{\bmk_\blktridn}{\bmt_\blktridn}{\call}{1},\circ)$.
\end{lem}

\begin{prf}
The proof is similar to Lemma \ref{lem81}.
\end{prf}

We know that $(\blktrilwktargnop{\bmk'}{\bmt'}{\call},\circ)$ is normal in $(\calu,\circ)$.
Since $(\blktrilwktexpargnop{\bmk_\blktridn}{\bmt_\blktridn}{\call}{\tau},\circ)$ is a normal subroup of $(\calu,\circ)$ 
for any paired sequence $(\bmk_\blktridn^\tau,\bmt_\blktridn^\tau)$ at any time epoch $\tau$, 
then from Lemma \ref{lem81a}, for any normal filling sequence $\bmf_\blktridn$ of 
$\bmn$ we may construct a normal chain
\be
\label{nc10}
(\blktrilwktargnop{\bmk'}{\bmt'}{\call},\circ)
\subset(\blktrilwktexpargnop{\bmk_\blktridn}{\bmt_\blktridn}{\call}{1},\circ)
\subset\cdots\subset(\blktrilwktexpargnop{\bmk_\blktridn}{\bmt_\blktridn}{\call}{\tau},\circ)
\subset(\blktrilwktexpargnop{\bmk_\blktridn}{\bmt_\blktridn}{\call}{\tau+1},\circ)\subset\cdots\subset(\calu,\circ).
\ee
We say (\ref{nc10}) is the {\it normal chain of a normal filling sequence} $\bmf_\blktridn$
of $\bmn$.  This gives the following.

\begin{thm}
Any normal filling sequence $\bmf_\blktridn$ of $\bmn$
gives a normal chain (\ref{nc10}) of $(\calu,\circ)$.
\end{thm}

From Lemmas \ref{lem60} and \ref{lem61}, we have seen 
the cosets $\bmu_{g,k^+}^{t^+}\circ(\blktrilwktexpargnop{\bmk}{\bmt}{\call}{\tau},\circ)$
for each $g_{k^+}^{t^+}\in G_{k^+}^{t^+}$ are all the cosets of 
$(\blktrilwktexpargnop{\bmk}{\bmt}{\call}{\tau},\circ)$ in
$(\blktrilwktexpargnop{\bmk}{\bmt}{\call}{\tau+1},\circ)$, where
$(k^+,t^+)$ is the filled ordered pair of $\bmf^{\tau+1}$ that is not filled in $\bmf^\tau$,  
$\bmu_{g,k^+}^{t^+}$ is a generator of $(\calu,\circ)$,
$(\bmk^\tau,\bmt^\tau)$ is the paired sequence of $\bmf^\tau$,
and $(\bmk^{\tau+1},\bmt^{\tau+1})$ is the paired sequence of $\bmf^{\tau+1}$.  By the same reasoning,
the cosets $\bmu_{g,k^+}^{t^+}\circ(\blktrilwktexpargnop{\bmk_\blktridn}{\bmt_\blktridn}{\call}{\tau},\circ)$
for each $g_{k^+}^{t^+}\in G_{k^+}^{t^+}$ are all the cosets of 
$(\blktrilwktexpargnop{\bmk_\blktridn}{\bmt_\blktridn}{\call}{\tau},\circ)$ in
$(\blktrilwktexpargnop{\bmk_\blktridn}{\bmt_\blktridn}{\call}{\tau+1},\circ)$, where
$(k^+,t^+)$ is the filled ordered pair of $\bmf_\blktridn^{\tau+1}$ that is not filled in $\bmf_\blktridn^\tau$, 
$\bmu_{g,k^+}^{t^+}$ is a generator of $(\calu,\circ)$ where $(k^+,t^+)$ is in
$\blktriupktarg{\bmk_u}{\bmt_u}{\bmn}$,
$(\bmk^\tau,\bmt^\tau)$ is the paired sequence of $\bmf_\blktridn^\tau$,
and $(\bmk^{\tau+1},\bmt^{\tau+1})$ is the paired sequence of $\bmf_\blktridn^{\tau+1}$.
Similarly the \creps\ of 
$(\blktrilwktexpargnop{\bmk_\blktridn}{\bmt_\blktridn}{\call}{1},\circ)/(\blktrilwktargnop{\bmk'}{\bmt'}{\call},\circ)$
are the generators $\{\bmu_{g,k^1}^{t^1}:  g_{k^1}^{t^1}\in G_{k^1}^{t^1}\}$
where $(k^1,t^1)$ is the filled ordered pair in $\bmf_\blktridn^1$.
This gives the following.

\begin{thm}
The generators $\bmu_{g,k^+}^{t^+}$ of $(\calu,\circ)$ where $(k^+,t^+)$ is in
$\blktriupktarg{\bmk_u}{\bmt_u}{\bmn}$, form a \crepc\ of any \cdc\ of $(\calu,\circ)$
formed by the normal chain (\ref{nc10}) of $(\calu,\circ)$ of any normal filling 
sequence $\bmf_\blktridn$ of $\bmn$.
\end{thm}

We know that $(\blktrilwktexpargnop{\bmk_\blktridn}{\bmt_\blktridn}{\call}{\tau},\circ)$ 
is a normal subgroup of $(\calu,\circ)$.
There is a homomorphism from $(\calu,\circ)$ to $\grpblktriupktarg{\bmk_u}{\bmt_u}{\calu}{\oplus}$
given by the map $\omega_{\bmk_u}^{\bmt_u}:  \calu\ra\blktriupktarg{\bmk_u}{\bmt_u}{\calu}$ with assignment
$\omega_{\bmk_u}^{\bmt_u}:  \bmu\mapsto\blktriupktarg{\bmk_u}{\bmt_u}{\bmu}$.  Then 
$\omega_{\bmk_u}^{\bmt_u}(\blktrilwktexpargnop{\bmk_\blktridn}{\bmt_\blktridn}{\call}{\tau},\circ)$ is a normal subgroup of 
$\omega_{\bmk_u}^{\bmt_u}(\calu,\circ)=\grpblktriupktarg{\bmk_u}{\bmt_u}{\calu}{\oplus}$.  Further if 
$(\blktrilwktexpargnop{\bmk_\blktridn}{\bmt_\blktridn}{\call}{\tau},\circ)
\subset(\blktrilwktexpargnop{\bmk_\blktridn}{\bmt_\blktridn}{\call}{\tau+1},\circ)$ 
at each time epoch $\tau>0$, then
$\omega_{\bmk_u}^{\bmt_u}(\blktrilwktexpargnop{\bmk_\blktridn}{\bmt_\blktridn}{\call}{\tau},\circ)
\subset\omega_{\bmk_u}^{\bmt_u}(\blktrilwktexpargnop{\bmk_\blktridn}{\bmt_\blktridn}{\call}{\tau+1},\circ)$
at each time epoch $\tau>0$.  Finally we know that 
$\omega_{\bmk_u}^{\bmt_u}(\blktrilwktargnop{\bmk'}{\bmt'}{\call},\circ)
=\grpblktriupktarg{\bmk_u}{\bmt_u}{\bmu_\bone}{\oplus}$.
Then using normal chain (\ref{nc10}), we have the
normal chain
\begin{multline}
\label{nc11}
\grpblktriupktarg{\bmk_u}{\bmt_u}{\bmu_\bone}{\oplus}
=\omega_{\bmk_u}^{\bmt_u}(\blktrilwktargnop{\bmk'}{\bmt'}{\call},\circ)
\subset\omega_{\bmk_u}^{\bmt_u}(\blktrilwktexpargnop{\bmk_\blktridn}{\bmt_\blktridn}{\call}{1},\circ)\subset\cdots
\subset\omega_{\bmk_u}^{\bmt_u}(\blktrilwktexpargnop{\bmk_\blktridn}{\bmt_\blktridn}{\call}{\tau},\circ)
\subset\omega_{\bmk_u}^{\bmt_u}(\blktrilwktexpargnop{\bmk_\blktridn}{\bmt_\blktridn}{\call}{\tau+1},\circ)\subset  \\
\cdots\subset\omega_{\bmk_u}^{\bmt_u}(\calu,\circ)=\grpblktriupktarg{\bmk_u}{\bmt_u}{\calu}{\oplus}
\end{multline}
of $\grpblktriupktarg{\bmk_u}{\bmt_u}{\calu}{\oplus}$.

We have just seen that the cosets 
$\bmu_{g,k^+}^{t^+}\circ(\blktrilwktexpargnop{\bmk_\blktridn}{\bmt_\blktridn}{\call}{\tau},\circ)$
for each $g_{k^+}^{t^+}\in G_{k^+}^{t^+}$ are all the cosets of 
$(\blktrilwktexpargnop{\bmk_\blktridn}{\bmt_\blktridn}{\call}{\tau},\circ)$ in
$(\blktrilwktexpargnop{\bmk_\blktridn}{\bmt_\blktridn}{\call}{\tau+1},\circ)$, where
$(k^+,t^+)$ is the filled ordered pair of $\bmf_\blktridn^{\tau+1}$ that is not filled in $\bmf_\blktridn^\tau$, 
and $\bmu_{g,k^+}^{t^+}$ is a generator of $(\calu,\circ)$ where $(k^+,t^+)$ is in
$\blktriupktarg{\bmk_u}{\bmt_u}{\bmn}$.  It follows that the cosets of
$\omega_{\bmk_u}^{\bmt_u}(\blktrilwktexpargnop{\bmk_\blktridn}{\bmt_\blktridn}{\call}{\tau},\circ)$ in
$\omega_{\bmk_u}^{\bmt_u}(\blktrilwktexpargnop{\bmk_\blktridn}{\bmt_\blktridn}{\call}{\tau+1},\circ)$ are
$\omega_{\bmk_u}^{\bmt_u}(\bmu_{g,k^+}^{t^+}\circ(\blktrilwktexpargnop{\bmk_\blktridn}{\bmt_\blktridn}{\call}{\tau},\circ))$, where
$(k^+,t^+)$ is the filled ordered pair of $\bmf_\blktridn^{\tau+1}$ that is not filled in $\bmf_\blktridn^\tau$, 
and $\bmu_{g,k^+}^{t^+}$ is a generator of $(\calu,\circ)$ where $(k^+,t^+)$ is in
$\blktriupktarg{\bmk_u}{\bmt_u}{\bmn}$.  We have
$$
\omega_{\bmk_u}^{\bmt_u}(\bmu_{g,k^+}^{t^+}\circ(\blktrilwktexpargnop{\bmk_\blktridn}{\bmt_\blktridn}{\call}{\tau},\circ))
=\omega_{\bmk_u}^{\bmt_u}(\bmu_{g,k^+}^{t^+})\proditarg{\bmk_u}{\bmt_u}{\oplus}
\omega_{\bmk_u}^{\bmt_u}((\blktrilwktexpargnop{\bmk_\blktridn}{\bmt_\blktridn}{\call}{\tau},\circ)).
$$
We say the projection $\omega_{\bmk_u}^{\bmt_u}(\bmu_{g,k^+}^{t^+})$
of the generator $\bmu_{g,k^+}^{t^+}$ of $(\calu,\circ)$ is a {\it generator} of
$\grpblktriupktarg{\bmk_u}{\bmt_u}{\calu}{\oplus}$.  This gives the following.

\begin{thm}
The generators of $\grpblktriupktarg{\bmk_u}{\bmt_u}{\calu}{\oplus}$ form a \crepc\ of any \cdc\ of
$\grpblktriupktarg{\bmk_u}{\bmt_u}{\calu}{\oplus}$ formed by the normal chain (\ref{nc11}) of
$\grpblktriupktarg{\bmk_u}{\bmt_u}{\calu}{\oplus}$ of any normal filling sequence $\bmf_\blktridn$
of $\bmn$.
\end{thm}

Note that a nontrivial generator $\omega_{\bmk_u}^{\bmt_u}(\bmu_{g,k^+}^{t^+})$
of $\grpblktriupktarg{\bmk_u}{\bmt_u}{\calu}{\oplus}$ has all trivial entries
except for $g_{k^+}^{t^+}$.  A special case of group $\grpblktriupktarg{\bmk_u}{\bmt_u}{\calu}{\oplus}$
is group $\grpupjktarg{0}{0}{t}{\calu}{\ccirc}$.  In this case, a nontrivial generator of 
$\grpupjktarg{0}{0}{t}{\calu}{\ccirc}$ has all trivial entries
except for one entry.  We call this an {\it eigentriangle} of
$\grpupjktarg{0}{0}{t}{\calu}{\ccirc}$; we say the identity generator
is also an eigentriangle.  Then group $\grpupjktarg{0}{0}{t}{\calu}{\ccirc}$
has an expansion in terms of a \crepc\ of eigentriangles.
From Corollary \ref{cor65}, there is an isomorphism from $\grpupjktarg{0}{0}{t}{\calr}{\cast}$
to $\grpupjktarg{0}{0}{t}{\calu}{\ccirc}$
with assignment $\beta^t:  \triupjktarg{0}{0}{t}{\bmr}\mapsto\triupjktarg{0}{0}{t}{\bmu}$, 
for each $t\in\bmcpz$, if $\beta:  \bmr\ra\bmu$
is the assignment of the isomorphism $(\calr,*)\simeq(\calu,\circ)$.
Then the assignment $\beta^t:  \triupjktarg{0}{0}{t}{\bmr}\mapsto\triupjktarg{0}{0}{t}{\bmu}$
shows that an eigentriangle of $\grpupjktarg{0}{0}{t}{\calu}{\ccirc}$
is an eigentriangle of $\grpupjktarg{0}{0}{t}{\calr}{\cast}$.  Therefore the
group $\grpupjktarg{0}{0}{t}{\calr}{\cast}$ also has an expansion in terms of eigentriangles,
which are generators of $\grpupjktarg{0}{0}{t}{\calr}{\cast}$.

Any linear block code $B$ of length $L$ \cite{MS} is isomorphic to a strongly controllable 
complete group system $A_B$ which is nontrivial on the time interval $[0,L-1]$ and
trivial elsewhere.  The generator group $(\calu,\circ)$ of $A_B$ is isomorphic to a 
direct product group on the sets of generators in $A_B$, whose nontrivial portions
on the interval $[0,L-1]$ are the sets of generators in $B$.  
In the typical case, $B$ has a generator of length $L$.
Then $A_B$ is \ellctl\ where $\ell=L-1$.  In this case, the generator group of $A_B$
is simply the lower elementary group $\grplwjktargnop{0}{\ell}{t}{\call}$,
where $t=0$.  Then we can study the block code $B$ using the 
lower elementary group $\grplwjktargnop{0}{\ell}{t}{\call}$ for $t=0$.

Note that we can construct a normal filling sequence $\bmf$ of $\bmn$ which
first fills the single index triangle $\trilwjktarg{0}{\ell}{t}{\bmn}$ with a subsequence 
$\bmf^\tau$ before filling the rest of $\bmn$.  Let $\bmf^\tau$ have
paired sequence $(\bmk^\tau,\bmt^\tau)$.  Then $(\blktrilwktargnop{\bmk^\tau}{\bmt^\tau}{\call},\circ)$ 
is a normal subgroup of $(\calu,\circ)$, and in fact,
$(\blktrilwktargnop{\bmk^\tau}{\bmt^\tau}{\call},\circ)=\grplwjktargnop{0}{\ell}{t}{\call}$.
Then the normal subchain
$$
\bmu_\bone\subset(\blktrilwktexpargnop{\bmk}{\bmt}{\call}{1},\circ)
\subset\cdots\subset(\blktrilwktexpargnop{\bmk}{\bmt}{\call}{\tau},\circ)
$$
of normal chain (\ref{nc1}) is a normal chain of $\grplwjktargnop{0}{\ell}{t}{\call}$,
and therefore of block code $B$.
It is clear that $\tau=(\ell+1)!$.  In general, if $\ell$ is large, 
there are many ways to fill the index triangle $\trilwjktarg{0}{\ell}{t}{\bmn}$
that give a subsequence $\bmf^\tau$ of a normal filling 
sequence $\bmf$ of $\bmn$, and therefore there are many normal chains of $B$.

In addition, we can study the structure of the block code $B$ using
sets of elementary groups $\grpblktriupktarg{\bmk_u}{\bmt_u}{\calu}{\oplus}$
discussed in Subsection 5.2,
the normal chains of $\grpblktriupktarg{\bmk_u}{\bmt_u}{\calu}{\oplus}$ just discussed
above, and the homomorphism in Theorem \ref{homo9}.  More details and examples
can be found in \cite{KM6} v9-v12.

\newpage
{\bf 6.  THE ELEMENTARY SYSTEM}%hhh
\vspace{3mm}

\vspace{3mm}
{\bf 6.1  The elementary system}
\vspace{3mm}

Fix $k$ such that $0\le k\le\ell$.  For each $t\in\bmcpz$,
let set $V_k^t$ be a collection of elements $v_{k}^t$, which may be
integers or any arbitrary objects; besides this, there is no requirement
on set $V_k^t$.  Consider the set $\calv$, which is the double Cartesian product
\be
\label{dprod1}
\bigotimes_{t=+\infty}^{t=-\infty} \bigotimes_{0\le k\le\ell} V_k^t.
\ee
We call $\calv$ the {\it elementary set}.  A row in $\calv$ is
assumed to be written in time reverse order as $\ldots,v_k^t,v_k^{t-1},\ldots$.

Let $\bmv$ be an element in $\calv$.  Because $\bmv$ has the same form as $\bmu$ in (\ref{ugttf}),
we can use the same triangle notation $\triupjktarg{0}{k}{t}{\bmv}$ and
$\triupjktarg{0}{k}{t}{\calv}$ for $\bmv$ and $\calv$, for $0\le k\le\ell$, 
for $t\in\bmcpz$, as $\triupjktarg{0}{k}{t}{\bmu}$ and 
$\triupjktarg{0}{k}{t}{\calu}$ for $\bmu$ and $\calu$.
Note that the entry in index position $(0,k)$ in $\triupjktarg{0}{k}{t}{\bmv}$
is $v_{k}^t$.

The {\it elementary list} $\call$ is an infinite collection of groups defined 
on triangular subsets $\triupjktarg{0}{k}{t}{\calv}$ of the elementary set $\calv$. 
The groups in the elementary list $\call$ are
$$
\{\grpupjktarg{0}{k}{t}{\calv}{\odot}:  0\le k\le\ell,t\in\bmcpz\}.
$$
The {\it $(\ell+1)$-depth elementary system} $\cale$ is an elementary set $\calv$,
the elementary list $\call$, and a homomorphism condition.
The homomorphism condition is that for each $k$ such that $0\le k<\ell$, for each
$t\in\bmcpz$, there is a homomorphism from $\grpupjktarg{0}{k}{t}{\calv}{\odot}$ to 
$\grpupjktarg{0}{k+1}{t}{\calv}{\odot}$ and $\grpupjktarg{0}{k+1}{t-1}{\calv}{\odot}$
under the projection map from set $\triupjktarg{0}{k}{t}{\calv}$ to set
$\triupjktarg{0}{k+1}{t}{\calv}$ and set $\triupjktarg{0}{k+1}{t-1}{\calv}$
given by the assignment $\triupjktarg{0}{k}{t}{\bmv}\mapsto\triupjktarg{0}{k+1}{t}{\bmv}$
and $\triupjktarg{0}{k}{t}{\bmv}\mapsto\triupjktarg{0}{k+1}{t-1}{\bmv}$.

The $(\ell+1)$-depth elementary system $\cale$ is nested.  For example,
the list $\{\grpupjktarg{0}{k}{t}{\calv}{\odot}:  k=\ell,t\in\bmcpz\}$
forms a 1-depth elementary system $\cale_\ell$.  For this trivial case,
the homomorphism condition is vacuous.  The list
$\{\grpupjktarg{0}{k}{t}{\calv}{\odot}:  \ell-1\le k\le\ell,t\in\bmcpz\}$
forms a 2-depth elementary system $\cale_{\ell-1}$.  In general the following holds.
 
\begin{thm}
The list $\{\grpupjktarg{0}{k}{t}{\calv}{\odot}:  \ell-m+1\le k\le\ell,t\in\bmcpz\}$
forms an $m$-depth elementary system $\cale_{\ell-m+1}$.
\end{thm}

In Subsection 6.2 we show that we can form an $(\ell+1)$-depth global group $(\calv,\cdot)$ 
from any $(\ell+1)$-depth elementary system $\cale$ with elementary set $\calv$.  
Then in Subsection 6.3, we show that we can start with the $(\ell+1)$-depth elementary system 
$\cale_A$ of any \ellctl\ complete group system $A$ and use the global group formed from 
$\cale_A$ to obtain an isomorphic copy of the generator group $(\calu,\circ)$ of $A$ and 
hence recover $A$.  In fact, we show that we can recover an \ellctl\ complete group system $A$
from any $(\ell+1)$-depth elementary system $\cale$.  Therefore the study of
\ellctl\ complete group systems $A$ is essentially the study of 
$(\ell+1)$-depth elementary systems $\cale$.
Finally in Subsection 6.4, we give a brief discussion of how to construct all $(\ell+1)$-depth 
elementary systems $\cale$.

\vspace{3mm}
{\bf 6.2  The global group $(\calv,\cdot)$}
\vspace{3mm}

We now define a global operation $\cdot$ on the elementary set
$\calv$ using the infinite collection of groups in the elementary system $\cale$.
We show this forms a group $(\calv,\cdot)$.  We say $(\calv,\cdot)$ is the {\it global group}
of elementary system $\cale$.

We define the product operation $\bmvdt\cdot\bmvddt$ to be the element $\bmvbr\in\calv$ given by
\be
\label{eq100}
\triupjktarg{0}{0}{t}{\bmvbr}\rmdef
\triupjktarg{0}{0}{t}{\bmvdt}\prodjktarg{0}{0}{t}{\odot}\triupjktarg{0}{0}{t}{\bmvddt}.
\ee
for $t=+\infty$ to $-\infty$.

\begin{lem}
The product (\ref{eq100}) for $t=+\infty$ to $-\infty$ gives an element in $\calv$.
\end{lem}

\begin{prf}
To show that the product (\ref{eq100}) for $t=+\infty$ to $-\infty$ gives an element
$\bmvbr$ in $\calv$, we first show that if $\triupjktarg{0}{0}{t}{\bmvbr}$ and 
$\triupjktarg{0}{0}{t'}{\bmvbr}$ intersect, they are the same on their intersection.
Assume without loss in generality that $t'<t$.  Then the intersection is a
triangle $\triupjktarg{0}{k}{t'}{\bmvbr}$ for some $k<\ell$.
But $\triupjktarg{0}{0}{t}{\bmvbr}=\triupjktarg{0}{0}{t}{\bmvdt}\prodjktarg{0}{0}{t}{\odot}\triupjktarg{0}{0}{t}{\bmvddt}$
and $\triupjktarg{0}{0}{t'}{\bmvbr}=\triupjktarg{0}{0}{t'}{\bmvdt}\prodjktarg{0}{0}{t'}{\odot}\triupjktarg{0}{0}{t'}{\bmvddt}$.
But by finite induction the homomorphism condition ensures there is a homomorphism from 
$\grpupjktarg{0}{0}{t}{\calv}{\odot}$ to $\grpupjktarg{0}{k}{t'}{\calv}{\odot}$, 
and from $\grpupjktarg{0}{0}{t'}{\calv}{\odot}$ to $\grpupjktarg{0}{k}{t'}{\calv}{\odot}$.
Therefore $\triupjktarg{0}{0}{t}{\bmvdt}\prodjktarg{0}{0}{t}{\odot}\triupjktarg{0}{0}{t}{\bmvddt}$
and $\triupjktarg{0}{0}{t'}{\bmvdt}\prodjktarg{0}{0}{t'}{\odot}\triupjktarg{0}{0}{t'}{\bmvddt}$
give the same result on their intersection 
$\triupjktarg{0}{k}{t'}{\bmvdt}\prodjktarg{0}{k}{t'}{\odot}\triupjktarg{0}{k}{t'}{\bmvddt}$.
Therefore $\triupjktarg{0}{0}{t}{\bmvbr}$ and $\triupjktarg{0}{0}{t'}{\bmvbr}$
are the same on their intersection $\triupjktarg{0}{k}{t'}{\bmvbr}$.  But if
the products in (\ref{eq100}) for $t=+\infty$ to $-\infty$ are consistent on their
intersections, then it is clear that $\bmvbr$ is just some element
of the double Cartesian product $\calv$.
\end{prf}

\vspace{4mm}
\begin{lem}
\label{lem94}
We have that $\bmvdt\cdot\bmvddt\in\calv$ and the operation $\bmvdt\cdot\bmvddt$ 
is well defined.
\end{lem}

\begin{prf}
We have just shown $\bmvdt\cdot\bmvddt\in\calv$.
We only need to show $\bmvdt\cdot\bmvddt$ is well defined.
Let $\bmvgr,\bmvac\in\calv$ and suppose $\bmvgr=\bmvdt$ and $\bmvac=\bmvddt$.  But then
$$
\triupjktarg{0}{0}{t}{\bmvdt}\prodjktarg{0}{0}{t}{\odot}\triupjktarg{0}{0}{t}{\bmvddt}
=\triupjktarg{0}{0}{t}{\bmvgr}\prodjktarg{0}{0}{t}{\odot}\triupjktarg{0}{0}{t}{\bmvac}
$$
for $t=+\infty$ to $-\infty$.  This means $\bmvdt\cdot\bmvddt=\bmvgr\cdot\bmvac$.
\end{prf}

\begin{thm}
The set $\calv$ with operation $\cdot$ forms a group $(\calv,\cdot)$.
\end{thm}

\begin{prf}
First we show the operation $\cdot$ is associative.
Let $\bmv,\bmvdt,\bmvddt\in\calv$.  We need to show
\be
\label{eva00}
(\bmv\cdot\bmvdt)\cdot\bmvddt
=\bmv\cdot(\bmvdt\cdot\bmvddt).
\ee
To find the left hand side of (\ref{eva00}), we evaluate
\be
\label{eval11}
(\triupjktarg{0}{0}{t}{\bmv}\prodjktarg{0}{0}{t}{\odot}\triupjktarg{0}{0}{t}{\bmvdt})\prodjktarg{0}{0}{t}{\odot}\triupjktarg{0}{0}{t}{\bmvddt}
\ee
for $t=+\infty$ to $-\infty$.
And to find the right hand side of (\ref{eva00}), we evaluate
\be
\label{eval12}
\triupjktarg{0}{0}{t}{\bmv}\prodjktarg{0}{0}{t}{\odot}(\triupjktarg{0}{0}{t}{\bmvdt}\prodjktarg{0}{0}{t}{\odot}\triupjktarg{0}{0}{t}{\bmvddt})
\ee
for $t=+\infty$ to $-\infty$.  But we know group
$\grpupjktarg{0}{0}{t}{\calv}{\odot}$ is associative so (\ref{eval11}) is the 
same as (\ref{eval12}).  This means the left hand side and right hand side
of (\ref{eva00}) evaluate to the same element in $\calv$.

We show $\bmv_\bone$ is the identity of $(\calv,\cdot)$.  Let $\bmv\in\calv$.  
We need to show $\bmv_\bone\cdot\bmv=\bmv$ and $\bmv\cdot\bmv_\bone=\bmv$.
First, to find $\bmv_\bone\cdot\bmv$ we evaluate
\be
\label{eval15}
\triupjktarg{0}{0}{t}{\bmv_\bone}\prodjktarg{0}{0}{t}{\odot}\triupjktarg{0}{0}{t}{\bmv}
\ee
for $t=+\infty$ to $-\infty$.  But we know that
$\triupjktarg{0}{0}{t}{\bmv_\bone}$ is the identity of group
$\grpupjktarg{0}{0}{t}{\calv}{\odot}$.  Then (\ref{eval15}) reduces to
\be
\label{eval16}
\triupjktarg{0}{0}{t}{\bmv_\bone}\prodjktarg{0}{0}{t}{\odot}\triupjktarg{0}{0}{t}{\bmv}
=\triupjktarg{0}{0}{t}{\bmv}
\ee
for $t=+\infty$ to $-\infty$.  Then $\bmv_\bone\cdot\bmv=\bmv$.  A similar argument 
shows that $\bmv\cdot\bmv_\bone=\bmv$.

Let $\bmv\in\calv$.  We show $\bmv^{-1}$ is the inverse of $\bmv$ in $(\calv,\cdot)$.
We need to show $\bmv^{-1}\cdot\bmv=\bmv_\bone$ and $\bmv\cdot\bmv^{-1}=\bmv_\bone$.
First, to find $\bmv^{-1}\cdot\bmv$ we evaluate
\be
\label{eval17}
\triupjktarg{0}{0}{t}{\bmv^{-1}}\prodjktarg{0}{0}{t}{\odot}\triupjktarg{0}{0}{t}{\bmv}
\ee
for $t=+\infty$ to $-\infty$.  But we know that
$\triupjktarg{0}{0}{t}{\bmv^{-1}}$ is the inverse of
$\triupjktarg{0}{0}{t}{\bmv}$ in $\grpupjktarg{0}{0}{t}{\calv}{\odot}$.
Then (\ref{eval17}) reduces to
\be
\label{eval18}
\triupjktarg{0}{0}{t}{\bmv^{-1}}\prodjktarg{0}{0}{t}{\odot}\triupjktarg{0}{0}{t}{\bmv}
=\triupjktarg{0}{0}{t}{\bmv_\bone}
\ee
for $t=+\infty$ to $-\infty$.  
Then $\bmv^{-1}\cdot\bmv=\bmv_\bone$.  A similar argument shows that $\bmv\cdot\bmv^{-1}=\bmv_\bone$.

Together these results show $(\calv,\cdot)$ is a group.
\end{prf}

We call the group $(\calv,\cdot)$ formed from an $(\ell+1)$-depth elementary system
$\cale$ an {\it $(\ell+1)$-depth global group}.
The groups $\grpupjktarg{0}{k}{t}{\calv}{\odot}$ in the elementary system are
the {\it elementary groups} of $(\calv,\cdot)$.  We have just shown the following.

\begin{thm}
Any $(\ell+1)$-depth elementary system $\cale$ forms an $(\ell+1)$-depth 
global group $(\calv,\cdot)$ by the procedure just described.
\end{thm}

\begin{thm}
\label{thm113}
The $(\ell+1)$-depth global group $(\calv,\cdot)$ formed from the $(\ell+1)$-depth elementary system
$\cale$ is uniquely determined by $\cale$.
\end{thm}

\begin{prf}
By definition, the global operation $\cdot$ in $(\calv,\cdot)$ is uniquely determined
by the elementary groups $\{\grpupjktarg{0}{0}{t}{\calu}{\ccirc}:  t\in\bmcpz\}$ in the elementary
list of $\cale$.
\end{prf}

The $(\ell+1)$-depth global group is nested.  For example,
the top row of an $(\ell+1)$-depth global group $(\calv,\cdot)$ is a
1-depth global group $(\calv,\cdot)_\ell$.  The top two rows
of an $(\ell+1)$-depth global group $(\calv,\cdot)$ is a
2-depth global group $(\calv,\cdot)_{\ell-1}$.
In general the following holds.

\begin{thm}
The top $m$ rows of an $(\ell+1)$-depth global group $(\calv,\cdot)$ form an
$m$-depth global group $(\calv,\cdot)_{\ell-m+1}$.
\end{thm}

To summarize, in Subsection 6.2 we have constructed the chain
\be
\cale\ra(\calv,\cdot),
\ee
where $\cale$ is an $(\ell+1)$-depth elementary system, and $(\calv,\cdot)$ is an
$(\ell+1)$-depth global group formed from $\cale$.

\vspace{3mm}
{\bf 6.3  Construction of all \ellctl\ complete group systems $A$
from the elementary system}
\vspace{3mm}

We now show that any \ellctl\ complete group system $A$ can be reduced 
to an elementary system $\cale_A$.

\begin{thm}
\label{thm99}
The $(\ell+1)$-depth generator group $(\calu,\circ)$ of any \ellctl\ complete group system $A$ contains
a unique $(\ell+1)$-depth elementary system $\cale_A$ with elementary set $\calu$ and elementary list
$\{\grpupjktarg{0}{k}{t}{\calu}{\ccirc}:  0\le k\le\ell,t\in\bmcpz\}$.
\end{thm}

\begin{prf}
First we show that $\calu$ can be considered to be an elementary set.
We have seen that $\calu$ is just the double Cartesian product 
(\ref{input4}).  Comparing set $\calu$ in (\ref{input4}) to set $\calv$
in (\ref{dprod1}), we see there is a
bijection between the two sets provided there is a bijection between the
sets $G_k^t$ and $V_k^t$, for $0\le k\le\ell$, for each $t\in\bmcpz$.
Then $\calu$ is an elementary set.  Lastly, from Theorem \ref{thm41b}, 
the upper elementary groups $\grpupjktarg{0}{k}{t}{\calu}{\ccirc}$ of $(\calu,\circ)$, 
for $0\le k\le\ell$ and $t\in\bmcpz$, satisfy the homomorphism condition of groups in an 
$(\ell+1)$-depth elementary system.  We define the elementary system $\cale_A$
to be the elementary set $\calu$ and elementary list
$\{\grpupjktarg{0}{k}{t}{\calu}{\ccirc}:  0\le k\le\ell,t\in\bmcpz\}$.
The elementary system $\cale_A$ is unique from Theorem \ref{thm41c}.
\end{prf}
We call $\cale_A$ the {\it elementary system of} $A$.
Essentially the elementary system $\cale_A$ of $A$ is just the generator group $(\calu,\circ)$ of $A$
stripped of its global operation $\circ$.
Then we can summarize the results of this paper so far by the chain
%%%%%%%%%%%%%%%%%%%%%%%%%%%%%%%%%%%%%%%%%
\be
\label{chain8a}
\begin{array}{lllllll}
 \da & \la                       & \la         & {\stackrel{=}{\la}}      & \imfu         &      & \\
 \da &                           &             &                          & \ua\,f_u      &      & \\
 A   & {\stackrel{\simeq}{\ra}}  & (\calr,*)   & {\stackrel{\simeq}{\ra}} & (\calu,\circ) & \ra  & \cale_A        
\end{array}
\ee
%%%%%%%%%%%%%%%%%%%%%%%%%%%%%%%%%%%%%%%%%
where $(\calr,*)$ is a decomposition group; $(\calu,\circ)$ is a generator group; and $\cale_A$ is an 
elementary system.  In the remainder of this subsection, we ask whether we can reverse 
the chain in (\ref{chain8a}), i.e., can we recover $A$ from $\cale_A$.
Then we define a new notion of isomorphism for group systems and show how to obtain
all \ellctl\ complete group systems $C$ up to this new isomorphism from the set of all
$(\ell+1)$-depth elementary systems $\cale$. 

\begin{thm}
\label{thm120}
We know the generator group $(\calu,\circ)$ of any \ellctl\ complete group system $A$ forms an 
$(\ell+1)$-depth elementary system $\cale_A$ with elementary set $\calu$.
Form the $(\ell+1)$-depth global group $(\calu,\star)$ of $\cale_A$.
The global operation $\bmudt\star\bmuddt$ in $(\calu,\star)$ is the same as the global 
operation $\bmudt\circ\bmuddt$ in $(\calu,\circ)$.  Therefore there is an isomorphism 
$(\calu,\star)\simeq (\calu,\circ)$ under the 1-1 correspondence $\calu=\calu$ 
given by the assignment $\bmu=\bmu$.
\end{thm}

\begin{prf}
From Theorem \ref{thm99}, we have seen that $\cale_A$ is an elementary system
with elementary set $\calu$ and elementary groups $\grpupjktarg{0}{k}{t}{\calu}{\ccirc}$.
We use $\cale_A$ to define a global group $(\calu,\star)$.  Let $\bmudt,\bmuddt\in\calu$.  
By definition, the product operation $\bmudt\star\bmuddt$ in global group 
$(\calu,\star)$ is uniquely determined by the evaluation of
\be
\label{subop11}
\triupjktarg{0}{0}{t}{\bmudt}\prodjktarg{0}{0}{t}{\ccirc}\triupjktarg{0}{0}{t}{\bmuddt},
\ee
for $t=+\infty$ to $-\infty$.  From Lemma \ref{lem95a},
the product operation $\bmudt\circ\bmuddt$ in generator group $(\calu,\circ)$ 
is uniquely determined by the evaluation of
\be
\label{subop12}
\triupjktarg{0}{0}{t}{\bmudt}\prodjktarg{0}{0}{t}{\ccirc}\triupjktarg{0}{0}{t}{\bmuddt}
\ee
for $t=+\infty$ to $-\infty$.  We see that (\ref{subop11}) and 
(\ref{subop12}) are the same.  Therefore $\bmudt\star\bmuddt$ and $\bmudt\circ\bmuddt$ 
are the same.  Therefore there is an isomorphism $(\calu,\star)\simeq (\calu,\circ)$ under the
1-1 correspondence $\calu=\calu$ given by the assignment $\bmu=\bmu$.
\end{prf}

For $\bmudt,\bmuddt\in\calu$, the global operation $\bmudt\star\bmuddt$ in $(\calu,\star)$ 
is the same as the global operation $\bmudt\circ\bmuddt$ in $(\calu,\circ)$.  This is a
little stronger condition than isomorphism, and we say that $(\calu,\star)$ is
{\it essentially identical} to $(\calu,\circ)$, written $(\calu,\star)\eid (\calu,\circ)$.

\begin{lem}
The global group $(\calu,\star)$ of $\cale_A$ is essentially identical to the generator group 
$(\calu,\circ)$ of $A$, $(\calu,\star)\eid (\calu,\circ)$.
\end{lem}

We have just constructed the chain
\be
\label{chain8b}
\cale_A\ra(\calu,\star),
\ee
where $\cale_A$ is an $(\ell+1)$-depth elementary system, and $(\calu,\star)$ is an
$(\ell+1)$-depth global group.  We can incorporate the chain (\ref{chain8b}) into chain
(\ref{chain8a}) as follows.
%%%%%%%%%%%%%%%%%%%%%%%%%%%%%%%%%%%%%%%%%
\be
\label{chain8c}
\begin{array}{lllllll}
 \da & \la                        & \la         & {\stackrel{=}{\la}}      & \imfu         &     &             \\
 \da &                            &             &                          & \ua\,f_u      &     &             \\
 A   & {\stackrel{\simeq}{\ra}}   & (\calr,*)   & {\stackrel{\simeq}{\ra}} & (\calu,\circ) & \ra & \cale_A     \\       
     &                            &             &                          & \uda\,\eid    &     & \uda =      \\
     &                            &             &                          & (\calu,\star) & \la & \cale_A
\end{array}
\ee
%%%%%%%%%%%%%%%%%%%%%%%%%%%%%%%%%%%%%%%%%
This gives the following.
\begin{thm}
\label{thm116}
We may recover any \ellctl\ complete group system $A$ from the $(\ell+1)$-depth 
elementary system $\cale_A$ of $A$ using the chain (\ref{zchain8c}).
%%%%%%%%%%%%%%%%%%%%%%%%%%%%%%%%%%%%%%%%%
\be
\label{zchain8c}
\begin{array}{lllll}
\da & {\stackrel{=}{\la}}        & \imfu         &      &         \\
\da &                            & \ua\,f_u      &      &         \\
A   &                            & (\calu,\star) & \la  & \cale_A    
\end{array}
\ee
%%%%%%%%%%%%%%%%%%%%%%%%%%%%%%%%%%%%%%%%%
\end{thm}

\begin{prf}
Since $(\calu,\circ)\eid (\calu,\star)$, we can easily recover the generator group 
$(\calu,\circ)$ of $A$ from the global group $(\calu,\star)$ of $\cale_A$.
Then we can recover $A$ from $(\calu,\circ)$
using the homomorphism $f_u$ in the \fhgs\ as done previously in Subsection 6.5.
\end{prf}

In fact, we can recover $A$ directly from $(\calu,\star)$ using the \fhgs.
The homomorphism $f_u$ in the top half of chain (\ref{chain8c})
just uses the primary elementary groups $\grpupjktarg{0}{0}{t}{\calu}{\ccirc}$ of $(\calu,\circ)$
for each $t\in\bmcpz$.  But the latter groups are already available in $(\calu,\star)$.
In fact these groups are available in $\cale_A$, so $A$ can be recovered
directly from the elementary system $\cale_A$ as well.

So far we have shown that the set of all elementary systems $\cale_A$
of all \ellctl\ complete group systems $A$, or $\{\cale_A\}$, is contained in the set of all 
$(\ell+1)$-depth elementary systems $\cale$, $\{\cale\}$, or $\{\cale_A\}\subset\{\cale\}$.
We now show that we can construct at least one \ellctl\ complete group system $A$ from
any $(\ell+1)$-depth elementary system $\cale$.

Assume we are given any $(\ell+1)$-depth elementary system $\cale$.  First find the
$(\ell+1)$-depth global group $(\calv,\cdot)$ of $\cale$.  We now show that we can
always construct a special \ellctl\ complete group system $(V_s,\ovcdot)$ from any 
$(\ell+1)$-depth global group $(\calv,\cdot)$ of $\cale$.  Previously, we recovered
$A$ from $(\calu,\circ)$ using the \fhgs\ with $(\calu,\circ)$ as an input group.
The construction of $(V_s,\ovcdot)$ uses the \fhgs\ with $(\calv,\cdot)$ as an input group,
as summarized in chain (\ref{chain7f}).  We give this construction now.

For each $t\in\bmcpz$, define a map $\theta_v^t:  \calv\ra\triupjktarg{0}{0}{t}{\calv}$ with assignment
$\theta_v^t:  \bmv\ra\triupjktarg{0}{0}{t}{\bmv}$.  Using (\ref{eq100}), the map
$\theta_v^t$ is a homomorphism from $(\calv,\cdot)$ to $\grpupjktarg{0}{0}{t}{\calv}{\odot}$.  
Consider the Cartesian product 
$$
V_\amalg\rmdef\cdots\times\triupjktarg{0}{0}{t}{\calv}\times\triupjktarg{0}{0}{t+1}{\calv}\times\cdots.
$$
Define the direct product group $(V_\amalg,\ovcdot)$ by
$$
(V_\amalg,\ovcdot)\rmdef\cdots\times\grpupjktarg{0}{0}{t}{\calv}{\odot}\times\grpupjktarg{0}{0}{t+1}{\calv}{\odot}\times\cdots.
$$
Then from Theorem \ref{fhgs}, using $(\calv,\cdot)$ for information group $\msfcpg$
and the primary elementary group $\grpupjktarg{0}{0}{t}{\calv}{\odot}$ for $\msfcpg^t$, $t\in\bmcpz$, 
there is a homomorphism $\theta_v:  \calv\ra V_\amalg$, from $(\calv,\cdot)$
to the direct product group $(V_\amalg,\ovcdot)$, defined by
$$
\theta_v(\bmv)\rmdef\ldots,\theta_v^t(\bmv),\theta_v^{t+1}(\bmv),\ldots.
$$
Define
\begin{align*}
\bmv_s\rmdef &\ldots,\theta_v^t(\bmv),\theta_v^{t+1}(\bmv),\ldots \\
           = &\ldots,\triupjktarg{0}{0}{t}{\bmv},\triupjktarg{0}{0}{t+1}{\bmv},\ldots.
\end{align*}
Then $\theta_v:  \calv\ra V_\amalg$ with assignment $\theta_v:  \bmv\mapsto\bmv_s$.
We can think of $\bmv_s$ as the sequence of triangles $\triupjktarg{0}{0}{t}{\bmv}$ of $\bmv$, 
now written in conventional time order and not overlapped.  Then
$$
(\calv,\cdot)/(\calv,\cdot)_K\simeq\imtheta_v,
$$
where group $\imtheta_v$ is the image of the homomorphism $\theta_v$, and where 
$(\calv,\cdot)_K$ is the kernel of the homomorphism $\theta_v$.  Since group $\imtheta_v$ is a
subgroup of the direct product group $(V_\amalg,\ovcdot)$, then $\imtheta_v$ is a
group system where global operation $\ovcdot$ is defined by the componentwise 
operation $\prodjktarg{0}{0}{t}{\odot}$ in group $\grpupjktarg{0}{0}{t}{\calv}{\odot}$ 
for each $t\in\bmcpz$.  We denote group $\imtheta_v$ by $(V_s,\ovcdot)$,
where $V_s$ is the subset of the Cartesian product $\calv_\amalg$ determined by $\bmv\in\calv$, or
equivalently the subset of $\calv_\amalg$ defined by $\imtheta_v$.
We call $(V_s,\ovcdot)$ the {\it global group system} of $(\calv,\cdot)$.  
We can summarize the preceding discussion as follows.

\begin{thm}
\label{thm83}
The \fhgs\ constructs a group system $(V_s,\ovcdot)$ with component group
$\grpupjktarg{0}{0}{t}{\calv}{\odot}$ from an $(\ell+1)$-depth generator group $(\calv,\cdot)$ 
using a homomorphism $\theta_v$, where $\theta_v=\ldots,\theta_v^t,\theta_v^{t+1},\ldots$,
and $\theta_v^t$ is a homomorphism 
$\theta_v^t:  (\calv,\cdot)\mapsto\grpupjktarg{0}{0}{t}{\calv}{\odot}$, 
for each $t\in\bmcpz$.  The homomorphism $\theta_v$ is a bijection.  
We have $(\calv,\cdot)\simeq\imtheta_v=(V_s,\ovcdot)$ under the assignment
$\theta_v: \bmv\mapsto\bmv_s$ given by the bijection $\theta_v: \calv\ra V_s$.
\end{thm}

\begin{prf}
For each $\bmv_s\in V_s$, there can be only one
$\bmv\in\calv$ such that $\theta_v:  \bmv\ra\bmv_s$ because the sequence
$\bmv_s=\ldots,\triupjktarg{0}{0}{t}{\bmv},\triupjktarg{0}{0}{t+1}{\bmv},\ldots$ defines a unique $\bmv$.  Then
$\theta_v:  \calv\ra V_s$ given by the assignment $\theta_v:  \bmv\ra\bmv_s$ is a bijection.
Then the kernel $(\calv,\cdot)_K$ is the identity and $(\calv,\cdot)\simeq\imtheta_v=(V_s,\ovcdot)$.
\end{prf}
We can summarize the construction in Theorem \ref{thm83} as shown in chain (\ref{chain7f}).
%%%%%%%%%%%%%%%%%%%%%%%%%%%%%%%%%%%%%%%%%
\be
\label{chain7f}
\begin{array}{lllll}
\da             & {\stackrel{=}{\la}} & \imtheta_v                                             &      &        \\
\da             &                     & \ua\,\theta_v=\ldots,\theta_v^t,\theta_v^{t+1},\ldots  &      &        \\
(V_s,\ovcdot)   &                     & (\calv,\cdot)                                          & \la  & \cale    
\end{array}
\ee
%%%%%%%%%%%%%%%%%%%%%%%%%%%%%%%%%%%%%%%%%
The chain (\ref{chain7f}) forms a linear system with input group $(\calv,\cdot)$,
homorphism $\theta_v$, and output group $(V_s,\ovcdot)$.  We say a linear system is {\it invertible} 
if the homorphism is a bijection.  Since the homorphism $\theta_v$
in (\ref{chain7f}) is a bijection, then the linear 
system in (\ref{chain7f}) is invertible.  If the linear system in (\ref{chain7f}) is
invertible, then the input $\bmv\in\calv$ can be discovered from the output
$\bmv_s\in V_s$.

We say an element $\bmv\in\calv$ is a nonrivial {\it generator} 
$\bmv_{g,k}^t$ of $(\calv,\cdot)$ if $\bmv_{g,k}^t$
contains one and only one nontrivial element $v_k^t$ for some $k$ such that $0\le k\le\ell$ 
and some time $t\in\bmcpz$.  For each $k$ such that $0\le k\le\ell$ 
and each $t\in\bmcpz$, we always assume there is a trivial generator $\bmv_{g,k}^t$
of $(\calv,\cdot)$ which is the identity $\bmv_\bone$ of $(\calv,\cdot)$.

\begin{lem}
\label{lem98}
Let $\Xi\rmdef\{\bmv_{g,k}^t:  0\le k\le\ell,t\in\bmcpz\}$ 
be the set of generators in $(\calv,\cdot)$.
The sequences in set $\theta_v(\Xi)$ are generators of $(V_s,\ovcdot)$ and 
form a basis $\calb$ of $(V_s,\ovcdot)$.
\end{lem}

\begin{prf}
Let $\bmv$ be any element in $(\calv,\cdot)$.  Then $\theta_v:  \bmv\mapsto\bmv_s$,
where $\bmv_s$ is an element in $(V_s,\ovcdot)$.  For $\bmv_{g,k}^t\in\Xi$,
let $\theta_v:  \bmv_{g,k}^t\mapsto\bmv_s^{[t,t+k]}$, where 
$\bmv_s^{[t,t+k]}\in(V_s,\ovcdot)$.

We now show any element in $\theta_v(\Xi)$ is a generator of $(V_s,\ovcdot)$.  
Fix $k$ such that $0\le k\le\ell$.  Fix nontrivial generator $\bmv_{g,k}^t$ in $\Xi$.
Under the bijection $\theta_v$ of the \fhgs\, shown in (\ref{chain7f}), 
nontrivial element $v_k^t$ in $\bmv_{g,k}^t$ lies in $(k+1)$ primary elementary groups 
$\grpupjktarg{0}{0}{t}{\calv}{\odot},\grpupjktarg{0}{0}{t+1}{\calv}{\odot},\ldots,\grpupjktarg{0}{0}{t+k}{\calv}{\odot}$,
and therefore generator $\bmv_{g,k}^t$ becomes a sequence $\bmv_s^{[t,t+k]}$ of span $k+1$ in $V_s$.
To show $\bmv_s^{[t,t+k]}$ is a generator, we have to show $\bmv_s^{[t,t+k]}$ is a \crep\
of the time domain granule (\ref{qgx}).  It is sufficient to show $\bmv_s^{[t,t+k]}$ is a \crep\
of (\ref{qgx1}).  But $\bmv_s^{[t,t+k]}$ is a sequence of span $k+1$ in $V_s$.  Therefore it is a member
of the numerator of (\ref{qgx1}) but not of either term in the denominator.  Therefore $\bmv_s^{[t,t+k]}$ 
is a \crep\ of (\ref{qgx1}).  Clearly, $\theta_v:  \bmv_\bone\mapsto\bmv_{s,\bone}$ 
where $\bmv_{s,\bone}$ is the identity of $(V_s,\ovcdot)$.  

We have shown $\theta_v(\Xi)$ is a generator of $(V_s,\ovcdot)$ for $0\le k\le\ell$ and $t\in\bmcpz$.
Therefore $\theta_v(\Xi)$ is a basis $\calb$ of $(V_s,\ovcdot)$.
\end{prf}

We see that the \fhgs\ constructs generators in $(V_s,\ovcdot)$ from generators in $(\calv,\cdot)$.
There is a bijection between the generators of group system $(V_s,\ovcdot)$
and the generators in $(\calv,\cdot)$ given by the restriction of $\theta_v$ to the
generators of $(\calv,\cdot)$.

\begin{thm}
\label{thm81}
The group system $(V_s,\ovcdot)=\imtheta$ constructed by Theorem \ref{thm83} is \ellctl\ and complete.
\end{thm}

\begin{prf}
Since the generator group $(\calv,\cdot)$ is $(\ell+1)$-depth, there is at least one
generator $\bmv_{g,\ell}^t$ in $(\calv,\cdot)$ which has a nontrivial label\ $v_\ell^t$.  
Under the bijection $\theta_v$ of the \fhgs\, shown in (\ref{chain7f}),
$v_\ell^t$ lies in $(\ell+1)$ primary elementary groups 
$\grpupjktarg{0}{0}{t}{\calv}{\odot},\grpupjktarg{0}{0}{t+1}{\calv}{\odot},\ldots,\grpupjktarg{0}{0}{t+\ell}{\calv}{\odot}$,
and therefore generator $\bmv_{g,\ell}^t$ becomes a sequence $\bmv_s^{[t,t+\ell]}$ of span
$\ell+1$ in $V_s$.  From the same argument used to show $\bmv_s^{[t,t+k]}$ is a generator
of span $k+1$ in the proof of Lemma \ref{lem98}, we know $\bmv_s^{[t,t+\ell]}$ is a generator 
of span $\ell+1$.  Therefore $(V_s,\ovcdot)$ is \ellctl.

The group system $(V_s,\ovcdot)$ is complete since it is determined by component groups
$\grpupjktarg{0}{0}{t}{\calv}{\odot}$ in $(\calv,\cdot)$, which have no global constraints.
\end{prf}

\begin{lem}
\label{lem95b}
The generator group of $(V_s,\ovcdot)$ is $(\calv,\cdot)$.
\end{lem}

\begin{prf}
Under the bijection $\theta_v^{-1}: V_s\ra\calv$, the group system $(V_s,\ovcdot)$
collapses to the generator group, which is $(\calv,\cdot)$.
\end{prf}

Theorem \ref{thm83} shows that we can always construct a special \ellctl\ complete group system 
from any $(\ell+1)$-depth generator group $(\calv,\cdot)$, namely $(V_s,\ovcdot)$.
We can summarize these results as follows.

\begin{thm}
\label{thm115}
Given any $(\ell+1)$-depth elementary system $\cale$, we may always use the chain (\ref{chain7f})
to construct an \ellctl\ complete group system $(V_s,\ovcdot)$ from the $(\ell+1)$-depth 
global group $(\calv,\cdot)$ of $\cale$.
The homomorphism $\theta_v$ is a bijection, and we have $(\calv,\cdot)\simeq\imtheta_v=(V_s,\ovcdot)$.
The generator group of $(V_s,\ovcdot)$ is $(\calv,\cdot)$,
and the elementary system $\cale_{(V_s,\ovcdot)}$ of $(V_s,\ovcdot)$ is $\cale$.  
\end{thm}

Taken together, Theorems \ref{thm116} and \ref{thm115} give the following.

\begin{thm}
\label{thm119}
The set of all elementary systems $\cale_A$
of all \ellctl\ complete group systems $A$, or $\{\cale_A\}$, is the same as the set of all 
$(\ell+1)$-depth elementary systems $\cale$, $\{\cale\}$, or $\{\cale_A\}=\{\cale\}$.
\end{thm}

\begin{prf}
Theorem \ref{thm116} shows that $\{\cale_A\}\subset\{\cale\}$.  The group system
$(V_s,\ovcdot)$ of Theorem \ref{thm115} is an \ellctl\ complete group system $A$.
Then Theorem \ref{thm115} shows that given any $(\ell+1)$-depth elementary system $\cale$,
we can find an \ellctl\ complete group system $A$ whose elementary system 
$\cale_A$ is $\cale$; then $\{\cale\}\subset\{\cale_A\}$.
\end{prf}

We can use Theorem \ref{thm115} and chain (\ref{chain7f}) 
to find all \ellctl\ complete group systems $A$ up to isomorphism from
the set of all $(\ell+1)$-depth elementary systems $\cale$.  First find the set
of all $(\ell+1)$-depth global groups $(\calv,\cdot)$ from 
the set of all $(\ell+1)$-depth elementary systems $\cale$.  
To construct all \ellctl\ complete group systems 
$A$ up to isomorphism, we divide the the set of all $(\ell+1)$-depth global groups
into equivalence classes $[(\calv,\cdot)]$.  Pick one representative 
$(\calv,\cdot)$ from each equivalence class.
Construct the \ellctl\ complete global group system $(V_s,\ovcdot)$ of $(\calv,\cdot)$ 
using chain (\ref{chain7f}) of Theorem \ref{thm115}.
The set of all global group systems $(V_s,\ovcdot)$ obtained this way, one for each equivalence class
$[(\calv,\cdot)]$, is the set of all \ellctl\ complete group systems $A$ up to isomorphism
\cite {KM6} v9-v12.

In general, if we use the \fhgs\ with input group $(\calv,\cdot)$ and a
homomorphism from $(\calv,\cdot)$ to an alphabet group $A^t$ other than
$\grpupjktarg{0}{0}{t}{\calv}{\odot}$, for each $t\in\bmcpz$,
then we obtain an $l$-controllable group system, where $l$ may
be less than $\ell$.  And the $l$-controllable group system with $l<\ell$
can be isomorphic to $(V_s,\ovcdot)$ \cite {KM6} v9-v12.  This example shows
the definition of isomorphism of 
finite groups is somewhat defective for group systems; for one reason it
does not include the notion of time used in group systems.  
There is a more restrictive notion of isomorphism for the elementary system and global group,
called {\it list isomorphism} \cite{KM6} v11-v12.  Two group systems constructed from
list isomorphic elementary systems and global groups must be $l$-controllable for the same $l$.
We can find all \ellctl\ complete group systems $A$ up to list 
isomorphism from the set of all $(\ell+1)$-depth elementary systems $\cale$
up to list isomorphism in the same way as discussed just above for 
isomorphism \cite{KM6} v11-v12.

\vspace{3mm}
{\bf 6.4  Construction of any elementary system $\cale$}
\vspace{3mm}

In the previous subsection, we discussed how to construct all \ellctl\ complete group systems $C$
from all $(\ell+1)$-depth elementary systems $\cale$.
We now give a brief discussion of how to construct all $(\ell+1)$-depth elementary systems $\cale$.
We first discuss how to construct a single elementary system $\cale$.
Since an elementary system $\cale$ is nested, to construct an $(\ell+1)$-depth 
elementary system $\cale$, we first construct a 1-depth elementary system 
$\cale_\ell=\{\grpupjktarg{0}{k}{t}{\calv}{\odot}:  k=\ell,t\in\bmcpz\}$;
then a 2-depth elementary system 
$\cale_{\ell-1}=\{\grpupjktarg{0}{k}{t}{\calv}{\odot}:  \ell-1\le k\le\ell,t\in\bmcpz\}$,
where there is a homomorphism from groups $\grpupjktarg{0}{\ell-1}{t}{\calv}{\odot}$
in $\cale_{\ell-1}$ to groups
$\grpupjktarg{0}{\ell}{t}{\calv}{\odot}$ and $\grpupjktarg{0}{\ell}{t-1}{\calv}{\odot}$
in $\cale_\ell$ for each $t\in\bmcpz$; and continue on.  In this way,
we obtain a sequence of elementary systems
$\cale_\ell,\cale_{\ell-1},\ldots,\cale_m,\ldots,\cale_1,\cale_0=\cale$
which ends in $\cale_0=\cale$, where
$\cale_m=\{\grpupjktarg{0}{k}{t}{\calv}{\odot}:  m\le k\le\ell,t\in\bmcpz\}$.

We may construct the sequence 
$\cale_\ell,\cale_{\ell-1},\ldots,\cale_m,\ldots,\cale_1,\cale_0$
in the following way.  Assume we have found the partial sequence
$\cale_\ell,\cale_{\ell-1},\ldots,\cale_{m+1}$ for some $m$, $0<m\le\ell$.
We show how to find $\cale_m$.  To find $\cale_m$ we have to find groups
$\grpupjktarg{0}{m}{t}{\calv}{\odot}$ for each $t\in\bmcpz$, such that there is a 
homomorphism from $\grpupjktarg{0}{m}{t}{\calv}{\odot}$ to 
$\grpupjktarg{0}{m+1}{t}{\calv}{\odot}$ and $\grpupjktarg{0}{m+1}{t-1}{\calv}{\odot}$
in the elementary list, under the projection map from set $\triupjktarg{0}{m}{t}{\calv}$ 
to sets $\triupjktarg{0}{m+1}{t}{\calv}$ and $\triupjktarg{0}{m+1}{t-1}{\calv}$.  The elementary groups 
$\grpupjktarg{0}{m+1}{t}{\calv}{\odot}$ and $\grpupjktarg{0}{m+1}{t-1}{\calv}{\odot}$
intersect and form the subdirect product group \cite{KM6,MH}
\be
\label{tpg2}
\grpupjktarg{0}{m+1}{t}{\calv}{\odot}\Join\grpupjktarg{0}{m+1}{t-1}{\calv}{\odot}.
\ee
Note that set $\triupjktarg{0}{m}{t}{\calv}$ is the same as set 
$\triupjktarg{0}{m+1}{t}{\calv}\Join\triupjktarg{0}{m+1}{t-1}{\calv}$
except for the addition of entries $v_m^t$ at index position $(0,m)$
in $\triupjktarg{0}{m}{t}{\calv}$.
Since there must be a homomorphism from $\grpupjktarg{0}{m}{t}{\calv}{\odot}$ 
to each individual group $\grpupjktarg{0}{m+1}{t}{\calv}{\odot}$ and 
$\grpupjktarg{0}{m+1}{t-1}{\calv}{\odot}$, there must be a homomorphism 
to subdirect product group (\ref{tpg2}).  If $K$ is the kernel of this homomorphism,
then $\grpupjktarg{0}{m}{t}{\calv}{\odot}$ is an extension of (\ref{tpg2}) by $K$.
This approach gives the construction of group $\grpupjktarg{0}{m}{t}{\calv}{\odot}$.
Continuing in this way, we finally obtain $\cale_0=\cale$.   More construction details 
can be found in v2-v6 of \cite{KM6}.

To construct all $(\ell+1)$-depth elementary systems $\cale$, we just iterate
the above approach, first constructing all 1-depth elementary systems, then
all 2-depth elementary systems for each of the 1-depth elementary systems, and so on.
The approach in this subsection can also construct all linear block codes \cite{KM6} v9-v12.

\end{document}

===========================================================================
http://kb.mit.edu/confluence/pages/viewpage.action?pageId=3907196

How can I position figures and tables where I want them with LaTeX?
LaTeX uses specific rules to place floats (figures and tables).

You begin figures with \begin{figure}[loc] where loc is a sequence of 0 to 4 letters, each one 
specifying a location where the figure or table may be placed, as follows:

Code	Meaning
h	Here: at the position in the text where the environment appears.
t	Top: at the top of the next page.
b	Bottom: at the bottom of the next page.
p	Page of floats: on a seperate page containing no text, only figures and tables.
The default is tbp.

If you only specify h (here), and it doesn't fit there, it will float to the end. So, it is best to 
rank your choices (htp for example...)

The only things that you can be sure of, however, is that LaTeX will never place a float before it 
is referenced in the text. But it is possible that it will put it many pages AFTER the place where 
you actually placed it in the .tex file.

A good way to put the figures where you really want them is to wait until you're done with the paper. 
When all the text is ready, you can modify it so it puts the figures where you want them.

Look for the end of a page that you want your figure to come after. For example: if you want your 
figure to go on page 77, find the place in the text, where page 76 ends.

Insert a \clearpage command there. This command will force LaTeX to insert all the floats 
(figures and tables) that weren't printed yet to be printed before any other text is processed.